\documentclass[11pt,a4paper]{article}
\usepackage[dvips,breaklinks]{hyperref}
\usepackage[usenames,dvipsnames]{xcolor}
\hypersetup{colorlinks,urlcolor=blue,citecolor=blue,linkcolor=black,filecolor=black}
\hypersetup{pdfauthor={Kalliopi Petraki, Raymond R. Volkas},pdftitle={Review of asymmetric dark matter}}
\usepackage{breakurl}

\usepackage[sort&compress,numbers,merge]{natbib}
\usepackage{enumerate}
\usepackage[hang,flushmargin]{footmisc} 
\usepackage{amsfonts,amsmath,amssymb,amsthm}
\numberwithin{equation}{section}

\newcommand{\nc}{\newcommand}

\nc{\eq}[1]{Eq.~(\ref{#1})}
\nc{\eqs}[1]{Eqs.~(\ref{#1})}

\nc{\pd}{\partial}

\nc{\bea}{\begin{eqnarray}}
\nc{\eea}{\end{eqnarray}}
\nc{\bal}{\begin{alignedat}}
\nc{\eal}{\end{alignedat}}
\nc{\beq}{\begin{equation}}
\nc{\eeq}{\end{equation}}
\nc{\bit}{\begin{itemize}}
\nc{\eit}{\end{itemize}}
\nc{\benu}{\begin{enumerate}}
\nc{\eenu}{\end{enumerate}}
\nc{\bdes}{\begin{description}}
\nc{\edes}{\end{description}}

\nc{\nn}{\nonumber}

\nc{\hc}{\rm{h.c.}}
\nc{\cc}{\rm{c.c.}}

\nc{\sub}[1]{_{\rm{#1}}}
\nc{\ssub}[1]{_{_\rm{#1}}}
\nc{\super}[1]{^{\rm{#1}}}
\nc{\ssuper}[1]{^{^\rm{#1}}}

\nc{\pare}[1]{\left( #1 \right)}
\nc{\sqpare}[1]{\left[ #1 \right]}
\nc{\ang}[1]{\left\langle #1 \right\rangle}
\nc{\abs}[1]{\left| #1 \right|}

\def\g5{\gamma_{5}}

\def\keV{\: \text{keV}}
\def\MeV{\: \text{MeV}}
\def\GeV{\: \text{GeV}}
\def\TeV{\: \text{TeV}}

\def \cm{\: \text{cm}}

\def \km{\: \text{km}}

\def \snd{\: \text{s}}

\def \Gyr{\: \text{Gyr}}

\def\a{\alpha}
\def\b{\beta}
\def\g{\gamma}
\def\d{\delta}
\def\e{\epsilon}
\def\z{\zeta}
\def\h{\eta}
\def\th{\theta}

\def\k{\kappa}
\def\l{\lambda}
\def\m{\mu}
\def\n{\nu}
\def\ks{\xi}

\def\p{\pi}
\def\r{\rho}
\def\s{\sigma}
\def\t{\tau}

\def\f{\phi}
\def\x{\chi}

\def\G{\Gamma}
\def\D{\Delta}

\def\L{\Lambda}

\def\W{\Omega}

\def\DM  {_{_{\rm DM}}}
\def\BV  {B_{\rm V}}
\def\BD  {B_{\rm D}}
\def\BLV {(B-L)_{\rm V}}

\def\TV {T_{_{\rm V}}}
\def\TD {T_{_{\rm D}}}
\def\gV {g_{_{\rm V}}}
\def\gD {g_{_{\rm D}}}
\def\gVdec {g_{_{\rm V,dec}}}
\def\gDdec {g_{_{\rm D,dec}}}

\def\eD  {e_{_D}}
\def\pD  {p_{_D}}
\def\HD  {H_{_D}}
\def\gammaD  {\gamma_{_D}}
\def\aD {\a_{_{D}}}
\def\mD  {m_{_D}}
\def\muD {\mu_{_{D}}}
\def\ED {E_{_{D}}}

\def \ns{{_\text{NS}}}
\def \mp{M_\text{Pl}}

\def \refcite{\cite}

\begin{document}

\begin{titlepage}
\renewcommand*{\thefootnote}{\fnsymbol{footnote}}

\vspace*{-2cm}
\begin{flushright}
NIKHEF-2013-016
\end{flushright}

\vspace*{2cm}

\begin{center}
{\LARGE
\textbf{Review of asymmetric dark matter}\footnote{Invited review for the International Journal of Modern Physics A.}
}
\\[10mm]
{\large
Kalliopi Petraki~$^{a,}$\footnote{kpetraki@nikhef.nl}  and
Raymond R. Volkas~$^{b,}$\footnote{raymondv@unimelb.edu.au}
}
\\[5mm]
{\small\textit{
$^a$Nikhef, Science Park 105, 1098 XG Amsterdam, The Netherlands \\
$^b$ARC Centre of Excellence for Particle Physics at the Terascale,\\
School of Physics, The University of Melbourne, Victoria 3010, Australia\\
}}
\end{center}

\vspace*{0.5cm}
\date{\today}

\begin{abstract}
\noindent
Asymmetric dark matter models are based on the hypothesis that the present-day abundance of dark matter has the same origin as the abundance of ordinary or ``visible'' matter: an asymmetry in the number densities of particles and antiparticles.  They are largely motivated by the observed similarity in the mass densities of dark and visible matter, with the former observed to be about five times the latter.  This review discusses the construction of asymmetric dark matter models, summarizes cosmological and astrophysical implications and bounds, and touches on direct detection prospects and collider signatures.



\end{abstract}

\end{titlepage}

\renewcommand*{\thefootnote}{\arabic{footnote}}
\setcounter{footnote}{0}

\noindent\makebox[\linewidth]{\rule{\textwidth}{1pt}} 

\vspace*{-8pt}
\tableofcontents
\noindent\makebox[\linewidth]{\rule{\textwidth}{1pt}}

\section{Introduction}	
\label{sec:introduction}

The dark matter (DM) problem is an empirical proof of the reality of physics beyond the standard model (SM)~\cite{Bertone:2004pz}.  The most likely solution is the existence of one or more new particles that are stable on cosmological timescales, couple sufficiently weakly to ordinary or visible matter (VM), and have the correct properties to seed bottom-up large scale structure formation.  Solution via modified gravity seems unlikely, even if not completely ruled out, and in any case would constitute the discovery of new physics were it true.  The third generic solution has the DM being primordial black holes, perhaps formed entirely from the collapse of ordinary matter.  This solution also requires physics beyond the SM, such as appropriate density fluctuations seeded during inflation, or bubble-wall collisions, or whatever one can envisage to provide the required overdense regions.

This review focuses on the specific elementary particle proposal termed ``asymmetric dark matter (ADM)''.\footnote{For an earlier brief review of the subject, see Ref.~\refcite{Davoudiasl:2012uw}. }  
The motivation comes from the observation that the present-day mass density of DM is about a factor of five higher than the density of VM~\cite{Hinshaw:2012aka,Ade:2013zuv}
\begin{equation}
\Omega_{\rm DM} \simeq 5\, \Omega_{\rm VM}\ ,
\label{eq:factorof5}
\end{equation}
where $\Omega$ as usual denotes the mass density of a given component relative to the critical density.  The similarity in these observed densities suggests a common origin, some kind of a unification or strong connection between the physics and cosmological evolution of VM and DM.  The present-day density of VM has long been established as due to the baryon asymmetry of the universe: some time during the early universe, a tiny excess of baryons $B$ over antibaryons $\bar{B}$ evidently developed, parameterized by~\cite{Hinshaw:2012aka,Ade:2013zuv}
\begin{equation}
\eta(B) \equiv \frac{n_B - n_{\bar{B}}}{s} \simeq 10^{-10}\ ,
\end{equation}
where number densities are denoted $n$, and $s$ is entropy density.\footnote{The asymmetry in a charge $X$ is defined in general by $\eta(X) \equiv \sum_i X_i (n_{i} - n_{\bar{i}})/s$ where $i$ denotes a species carrying $X$-charge of $X_i$.  An asymmetry normalized in this way is useful because 
it remains constant during the isentropic expansion of the universe.} 
The baryons in the universe today constitute the excess remaining after all of the antibaryons annihilated with the corresponding number of baryons.  The ADM hypothesis simply states that the present-day DM density is similarly due to a DM particle-antiparticle asymmetry, and that these asymmetries are related due to certain processes that occurred rapidly during an early cosmological epoch but later decoupled.

Asymmetric DM may be contrasted with weakly-interacting massive particle or WIMP DM (for a recent review see, for example, Ref.~\refcite{Feng:2010gw}).  The latter postulates that the DM is a thermal, non-relativistic relic particle (usually self-conjugate) with mass in the GeV-TeV range that decouples when its weak-scale annihilations fall out of equilibrium due to the Boltzmann suppression of the WIMP population.  Famously, this cold DM (CDM) scenario ``miraculously'' provides about the correct DM mass density for generic weak-scale annihilation cross-section, with a specific value (weakly dependent on WIMP mass) derived by fitting the abundance exactly.  Furthermore, the idea fits in well with independent particle-physics motivations for new weak-regime physics such as supersymmetry.  However, in almost all WIMP scenarios the similarity of the DM and VM densities then must be taken to be a coincidence (with some attempts made to avoid this uncomfortable conclusion~\cite{McDonald:2011zz,McDonald:2011sv,Cui:2011ab}).  The pure form of the WIMP hypothesis is now also rather strongly constrained by direct and indirect DM detection bounds~\cite{Bai:2010hh,Buckley:2011kk,Fox:2012ee}, and at the time of writing there were no indications from the Large Hadron Collider (LHC) for WIMP production in $7-8$ TeV $pp$ collisions.

Asymmetric DM is one of a number of well-motivated alternatives to the WIMP solution -- other examples are keV-scale sterile neutrinos~\cite{Dodelson:1993je,Shi:1998km,Kusenko:2006rh,Petraki:2007gq,Petraki:2008ef,
Kusenko:2009up,Wu:2009yr,Shoemaker:2010fg,Kusenko:2010ik,
Canetti:2012vf,Canetti:2012kh,Merle:2013gea,Drewes:2013gca}, axions~\cite{Peccei:1977hh,Peccei:1977ur,Kim:1979if,Zhitnitsky:1980tq,Dine:1981rt,Asztalos:2009yp,Sikivie:2010bq,Sikivie:2009qn,Arik:2011rx}, and $Q$-balls\footnote{$Q$-balls are in fact an ADM candidate, as their production presupposes the existence of a net global charge in the universe~\cite{Kusenko:1997si}. However, the dynamics of $Q$-ball production differ from that of the ADM models covered in this review.}~\cite{Kusenko:1997si,Kasuya:1999wu,Demir:2000gj,Kusenko:2001vu,Kusenko:2004yw,Barnard:2011jw} -- which all deserve serious attention.  It does not have the (interesting) WIMP feature of a tight connection between the thermal freeze-out process, indirect detection, direct detection and collider production.  Indirect detection through dark particle-antiparticle annihilations is obviously irrelevant, because there are no DM antiparticles left to annihilate with (co-annihilations with, for example, nucleons are possible but not required).  Many models of asymmetric DM can be tested through direct detection and collider signatures, but the parameters involved typically include some that are independent of the physics behind the cosmological DM abundance.

In most ADM models, the number density of DM particles is comparable to that of baryons.  The observed relation of Eq.~\eqref{eq:factorof5} then requires the DM particle mass to be typically, though not exclusively, in the $1$-$15$ GeV regime, with the precise figure depending on details of the model, such as its DM baryonic charge, a concept to be introduced in the next section.  There is tantalizing support for a DM mass in this range from the DAMA+DAMA/LIBRA~\cite{Bernabei:2008yi,Bernabei:2010mq}, CoGeNT~\cite{Aalseth:2010vx,Aalseth:2011wp} and CRESST~\cite{Angloher:2011uu} experiments, and very recently from the observation of three candidate DM events by CDMS~\cite{Agnese:2013rvf}. The status of the first three of these results is at present unclear, with both systematic uncertainties and consistency or the lack thereof between them under active discussion. 
It has been shown, however, that these positive results may be reconciled with each other by constructing certain kinds of asymmetric DM sectors (albeit with some tension with null searches)~\cite{Foot:2003iv,Foot:2010hu,Foot:2011pi,Foot:2013msa,Foot:2012cs,Fornengo:2011sz}.  (Ref.~\refcite{Foot:2013msa} also discusses consistency with CDMS).  Two modifications from WIMP-style DM have been studied as a way to achieve reconciliation: the nature of the DM-nucleon interaction, and the DM velocity dispersion.  An altered velocity dispersion relation may be the result of non-negligible DM self-interactions. The latter can play another important role: affect gravitational clustering at subgalactic scales, and thereby resolve the small-scale structure problems of the collisionless $\Lambda$CDM paradigm (where $\L$ stands for a positive cosmological constant).  It is also very important to appreciate that a full ADM explanation for Eq.~\eqref{eq:factorof5} requires a microphysical justification for the few-GeV DM mass scale.  It is not difficult to produce a model where the baryon and DM \emph{number} densities are similar, but a compelling explanation for the required DM mass is more challenging.  Some possibilities for this will be discussed in the next section.

The idea that the universe may contain a relic, dark, particle--anti-particle asymmetric component has in one form or another been considered for a few decades.  One of the oldest ideas is that of mirror matter~\cite{Foot:1991bp,Foot:1991py,Foot:1995pa,Lee:1956qn,Kobzarev,Pavsic:1974rq,Blinnikov:1982eh,Blinnikov:1983gh,Berezhiani:2000gw,Ignatiev:2003js,Foot:2003jt,Foot:2004pq,Berezhiani:2003wj,Ciarcelluti:2004ik,Ciarcelluti:2004ip,Berezhiani:2005vv,Foot:2013msa}, where the dark sector has identical microphysics to the visible sector.  Though subsequent observational data have ruled out the specific scenarios considered in Refs.~\refcite{Blinnikov:1982eh} and \refcite{Blinnikov:1983gh}, these early papers contemplated a universe with equal amounts of (asymmetric) matter and mirror matter (modern and still viable incarnations will be discussed below).  Another relatively early idea was that DM stability may arise from an analog of baryon-number conservation in a technicolor sector, with the DM state being a neutral techni-baryon~\cite{Nussinov:1985xr,Barr:1990ca}.  The modern era of ADM research occurred after the observational result that the VM and DM densities were as close together as a factor of five was established.

The ADM idea is one well-motivated way to address the coincidence problem posed by Eq.~\eqref{eq:factorof5}.  As mentioned in passing above, there have been interesting attempts to connect the VM density to WIMP physics~\cite{McDonald:2011zz,McDonald:2011sv,Cui:2011ab}, and also to other ways of producing DM~\cite{Kusenko:1997si,Thomas:1995ze,Enqvist:1998en,Fujii:2002kr,Fujii:2002aj,
Enqvist:2003gh,Dine:2003ax,Roszkowski:2006kw,McDonald:2006if,Kitano:2008tk,
Shoemaker:2009kg,Higashi:2011qq,Doddato:2011fz,Mazumdar:2011zd,Kasuya:2011ix,
Doddato:2011hx,D'Eramo:2011ec,Gu:2010yf,Allahverdi:2010rh,McDonald:2012vw,Unwin:2012rp}.

In the next section the general features of ADM models are reviewed, and in Sec.~\ref{sec:models} some specific exemplar models are summarized.  Section \ref{sec:pheno} describes astrophysical and cosmological implications and bounds, and discusses various detection prospects.  Section~\ref{sec:conclusion} concludes this review.


\section{Asymmetric dark matter}
\label{sec:ADM}

The VM abundance in the universe today is made from a small number of SM fields: protons and bound neutrons (formed mainly from valance up and down quarks, and gluons), and electrons (with neutrinos and photons comprising the current radiation content of the VM).  These are the stable relics of a much larger SM particle content.  Since ADM models seek to draw a connection between DM and VM, it is natural to suppose that the DM may also be the stable member(s) of some relatively complicated gauge theory constituting a hidden sector.  In general, ADM models have gauge groups that contain the product structure
\begin{equation}
G_{\rm V} \times G_{\rm D}\ ,
\end{equation}
where the first factor is the SM gauge group or some extension thereof, and the second factor is a dark gauge group.  Some models have a gauge force that couples to both sectors, with an extended U(1)$_{B-L}$ being a common example.  The dark sector in general may have various fermion and scalar multiplets in representations of $G_{\rm D}$, and spontaneous gauge-symmetry breaking may occur.  Many models have the dark sector as simply just fermions or scalars or a mixture of the two.  Taking our cue once again from the visible world, it could well be that the DM is multi-component and that there is dark radiation (bosonic and/or fermionic) as well as dark matter.  A relatively complicated dark sector is not mandatory, but it is perfectly consistent with the ADM philosophy.  There are two features that are not optional: a conserved or approximately-conserved dark global quantum number so that a dark asymmetry can be defined in the first place, and an interaction that annihilates away the symmetric part of the dark  plasma, just as strong and electroweak interactions annihilate the symmetric component of the SM plasma into radiation.

Let us review why certain visible-sector particles are stable (or at least very long lived).  Protons are stable because they are the lightest particles carrying conserved baryon number $B$.  Neutrons are the next-to-lightest baryons, and they are unstable unless they are bound in appropriate nuclei.  Electrons are stable because they are the lightest electrically-charged particles, with electric-charge conservation mandated.  Note that the simultaneous existence of stable protons and electrons permits the universe to be electrically neutral.  The least massive neutrino mass eigenstate has its stability ensured through angular momentum conservation because it is the lightest fermion: its decay products would have to contain a half-integer spin particle.  Photons and gluons, being massless, are stable by kinematics (gluons are also stable because they are lightest colored states).  The reasons for stability are several: exact (or nearly exact) global symmetry, unbroken gauge symmetry, angular momentum conservation, kinematics, and bound state effects (nuclear physics).  

The diversity of causes for stability suggests that the dark sector may similarly contain a number of stable particles, as well as structures analogous to atoms or nuclei from unbroken dark-sector gauge interactions.  Rather than getting bogged down in possible complexities, we shall begin thinking about dark sectors by focusing simply on baryon number conservation.  In terms of mass, $B$ conservation is the most important of the stabilizing influences in the visible sector: the cosmological $B$ asymmetry (with the proton mass) determines the VM mass density.  Since we want to understand Eq.~\eqref{eq:factorof5}, it seems sensible to start by postulating a ``dark baryon number'' and establishing a relationship with ``visible'' baryon number.  Other considerations may be brought in as necessary. Denote the visible and dark baryon numbers by $B_{\rm V}$ and $B_{\rm D}$, respectively.  At low energies and temperatures, both of these must be conserved\footnote{Or extremely weakly broken; this qualification will be understood from now on.} to ensure the separate stability of VM and DM.  

\subsection{Symmetry structure}
\label{sec:symmetries}

In the very early universe, there are four generic possibilities for \emph{initial} asymmetry creation:
\begin{enumerate}
\setlength{\itemsep}{-1pt}
\item A non-trivial linear combination of $B_{\rm V}$ and $B_{\rm D}$ is exact, but a linearly-independent combination is broken.
\item $B_{\rm V}$ is broken, while $B_{\rm D}$ is not.
\item $B_{\rm D}$ is broken, while $B_{\rm V}$ is not.
\item Both are broken.
\end{enumerate}
The breaking of at least one baryon-number symmetry is required for creating an asymmetry.  Recall that the Sakharov conditions for generating an $X$-particle number asymmetry are: violation of $X$, $C$ and $CP$ through processes that occur out of equilibrium~\cite{Sakharov:1967dj}.\footnote{
For the first three cases, where there is one conserved baryon number combination, the requirement of conservation may be weakened to a violation of the Sakharov conditions.  For example, one could have $B_{\rm V}$ violated and the other Sakharov conditions obeyed, while $B_{\rm D}$ is also violated but always in a $CP$-conserving way.  This leads to the same outcome as case 2 above, in the sense that a $B_{\rm V}$ asymmetry is created, but $\eta(B_{\rm D}) = 0$.  Having acknowledged the caveat, it nevertheless seems nicer and more robust to mandate the conservation of a quantity if there is no wish to create an asymmetry in it.  }

Once the initial asymmetry creation has happened in cases 2-4, the ADM framework requires some new interactions to then become rapid so that chemical exchange between the two sectors relates the asymmetries.  For case 2, these interactions have to reprocess some VM asymmetry into a DM asymmetry~\cite{Hooper:2004dc,Kaplan:2009ag,Cohen:2009fz,Cai:2009ia,
Blennow:2010qp,Haba:2011uz,Servant:2013uwa}, while for case 3 the reverse must happen~\cite{Foot:2003jt,Foot:2004pq,Dutta:2006pt,
Buckley:2010ui,Dutta:2010va,Shelton:2010ta,Haba:2010bm,Feng:2013wn}.  For case 4~\cite{Berezhiani:2000gw,Berezhiani:2008zza,Falkowski:2011xh}, the initial asymmetries may be quite different, so subsequent interactions should drive the asymmetries towards some kind of equilibration to comply with the standard ADM philosophy.  If this does not happen, then while the DM is certainly asymmetric, the lack of relation between the asymmetries means that one is not addressing the primary motivation given by Eq.~\eqref{eq:factorof5}.  For mirror DM~\cite{Foot:1991bp,Foot:1991py,Foot:1995pa,Blinnikov:1982eh,Blinnikov:1983gh,Berezhiani:2000gw,Ignatiev:2003js,Foot:2003jt,Foot:2004pq,Berezhiani:2003wj,Ciarcelluti:2004ik,Ciarcelluti:2004ip,Berezhiani:2005vv,Berezhiani:2008zza,Foot:2013msa} with an exact mirror symmetry -- see below -- there is an interesting twist with respect to case-4 dynamics: the identical microphysics, but necessarily lower temperature, in the mirror sector means that whatever baryogenesis mechanism operates in the visible sector, the same will occur in the mirror sector, but in a different temperature and expansion rate regime~\cite{Berezhiani:2000gw,Berezhiani:2008zza}.  This makes asymmetry generation in the two sectors related, though in a different way from standard ADM scenarios.

Case 1 is qualitatively distinct from the others, because correlated asymmetries are created simultaneously in the visible and dark sectors via common interactions, with the universe always being symmetric in some linear combination of the visible and dark baryonic numbers.  
This very interesting scenario represents the unified generation of both visible and dark matter, and is arguably the most elegant implementation of ADM~\cite{Kuzmin:1996he,Kitano:2004sv,Kitano:2005ge,Gu:2009yy,Davoudiasl:2010am,
Gu:2010ft,Heckman:2011sw,Gelmini:1986zz,Oaknin:2003uv,Farrar:2005zd,Hall:2010jx,
Bell:2011tn,Cheung:2011if,Graesser:2011vj,vonHarling:2012yn,MarchRussell:2011fi,Petraki:2011mv}.  Because the universe is symmetric in one linear combination of baryon numbers, this scenario is also said to produce a ``baryon-symmetric universe''~\cite{Dodelson:1989ii,Dodelson:1989cq}.  Let us examine the symmetry structure in a little more detail.

Let the conserved and broken linear combinations be, respectively,
\begin{eqnarray}
B_{\rm con} & = & a\, B_{\rm V} + b\, B_{\rm D}\ , \nonumber \\
B_{\rm bro} & = & c\, B_{\rm V} + d\, B_{\rm D}\ , 
\label{eq:generalcombs}
\end{eqnarray}
where $a,b \neq 0$ to ensure that the conserved quantity is a non-trivial combination, and $ad - bc \neq 0$ to ensure that the two combinations are independent.  Abelian charges are of course defined up to a normalization convention, and $B_{\rm bro}$ may be redefined by the addition of a piece proportional to $B_{\rm con}$ because the result is still a broken charge.  This permits us to simplify the definitions of the conserved and broken baryon-number charges without loss of generality.  By scaling $B_{\rm con}$ we may set $a=1$ and by scaling $B_{\rm D}$ we may put $b = -1$.  Adding $(d-c) B_{\rm con}/2$ to $B_{\rm bro}$ and then setting $d+c = 2$ completes the simplification process.  The result is that Eq.~\eqref{eq:generalcombs} is equivalent to
\begin{eqnarray}
B_{\rm con} & = & B_{\rm V} - B_{\rm D}\ , \label{eq:Bcon} \\
B_{\rm bro} & = &  B_{\rm V} + B_{\rm D}\ . \label{eq:Bbro}
\end{eqnarray}
The Sakharov conditions can now be used to engineer dynamical schemes that give rise to an asymmetry in $B_{\rm bro}$ while maintaining $\eta(B_{\rm con}) = 0$ as a constraint, leading to
\begin{equation}
\eta(B_{\rm bro})  \equiv  \eta  
\ \Rightarrow \ 
\eta(B_{\rm V}) = \eta(B_{\rm D}) = \frac{\eta}{2}\ .
\label{eq:BV=BD=X/2--Bsym}
\end{equation}
Thus baryon-symmetric models can always be interpreted as implying that the asymmetries in the two sectors are equal, and that the DM is concealing a (generalized) baryon number that cancels the baryon asymmetry of VM (which is what $\eta(B_{\rm con}) = 0$ means).

The fundamental feature of having ${B_{\rm con}}$ always conserved now requires some additional discussion.  With $B_{\rm V}$ identified as visible-sector baryon number, the $B_{\rm con}$ defined in \eq{eq:Bcon} is anomalous and thus not actually conserved at the quantum level.  In the early-universe, an important aspect related to this is the reprocessing of $B_{\rm V}$ asymmetry by electroweak sphalerons into visible lepton number.  In order for $B_{\rm con}$ to also be quantally conserved, we must replace $B_{\rm V}$ with a suitable, related anomaly-free quantity.  The obvious choice is $(B-L)_{\rm V}$, and indeed this identification is made in many specific baryon-symmetric schemes.

The connection of the visible-sector generator within $B_{\rm con}$ to proton stability is then more subtle, since $(B-L)_V$ conservation alone cannot enforce it.  At the non-perturbative level, the sphaleron or related zero-temperature instanton process conserves a certain $\mathbb{Z}_3$ discrete symmetry that ensures the absolute stability of the proton within the SM, even in the face of the anomalous $(B+L)_{\rm V}$ violation.  At low energies and low temperatures, we must also arrange any model to produce the SM as the effective theory, and thus perforce have a conserved $B_{\rm V}$ at the perturbative level (in that limit).  Depending on how it is constructed, similar issues could arise in the dark sector, and it is worth noting that the absolute stability of DM (if that is what is desired) may be guaranteed by a discrete subgroup of U(1)$_{B_{\rm D}}$ rather than the full parent group.

With $B_{\rm con}$ now conserved at both the quantal and classical levels, one is free to gauge the associated abelian symmetry.  This further deepens the fundamental nature ascribed to this symmetry.  Because the gauged U(1) has to be spontaneously broken to ensure that the associated $Z'$ boson is sufficiently massive,\footnote{We are not considering the alternative possibility that the gauge boson is massless but the gauge coupling constant is tiny.} but we still want to have the global U(1), the gauged model has to be constructed in a particular way (see e.g. Refs.~\refcite{Petraki:2011mv,vonHarling:2012yn} for a full discussion).  Essentially, the enlarged symmetry should be the product of the gauged U(1) and the related global U(1).  The scalar field whose vacuum expectation value will spontaneously break the gauged U(1) must have zero charge under the global U(1) to ensure that the latter remains exact.  The $Z'$, which has decay channels into dark-sector particles and thus a substantial invisible width, is a generic feature of gauged baryon-symmetric models, and provides one very interesting way of searching for experimental evidence for at least that kind of ADM (see Sec.~\ref{sec:Zprime}).

In all four cases (as defined at the start of this subsection), depending on the context and the temperature regime of interest, the interactions that lead to a relation between the VM and DM asymmetries are either described by an explicit renormalizable theory (see e.g.\ Ref.~\refcite{Cui:2011qe}), or by effective operators of the form
\begin{equation}
{\cal O}_{\BLV}\, {\cal O}_{\BD}\ ,
\label{eq:sharing}
\end{equation}
where ${\cal O}_{\BLV}$ is formed from visible-sector fields in a combination that carries nonzero $\BLV$, while ${\cal O}_{B_{\rm D}}$ is a dark-sector analog (see e.g.\ Ref.~\refcite{Kaplan:2009ag}).  The interactions must preserve some linear combination of the $\BLV$ and $\BD$ numbers, otherwise they would wash out both asymmetries.  Some of these operators lead to interesting collider signatures if the effective scale is in the TeV regime (see Sec.~\ref{sec:monojets}).  In some models, non-perturbative sphaleron processes are used to relate the asymmetries~\cite{Barr:1991qn,Kribs:2009fy,Blennow:2010qp,Shelton:2010ta,Kang:2011wb}.

\subsection{Asymmetry generation}
\label{sec:AsymGen}

A full ADM theory should specify the dynamics of asymmetry generation, although a number of works simply assume an initial asymmetry was created by some means, and focus instead on how the asymmetry gets distributed between the visible sector and the dark sector.  The most common asymmetry creation scenarios are out-of-equilibrium decays~\cite{Weinberg:1979bt,Fukugita:1986hr}, Affleck-Dine dynamics~\cite{Affleck:1984fy,Dine:1995kz,Dine:2003ax,Enqvist:2003gh,Allahverdi:2012ju}, bubble nucleation during a first-order phase transition~\cite{Riotto:1999yt}, asymmetric freeze-out~\cite{Farrar:2005zd}, asymmetric freeze-in~\cite{Hall:2010jx}, and spontaneous genesis~\cite{Cohen:1987vi,Cohen:1988kt}.  We now briefly review the basic idea behind each of these.  We shall call the relevant particle number $X$, which may be $B_{\rm V}$, $B_{\rm D}$ or some linear combination depending on the model.\footnote{For all baryogenesis models, including those containing ADM, one must ensure that the baryon-asymmetry in the visible sector is not erased through the combined action of electroweak sphaleron transitions and $L$-violating Majorana neutrino masses.  See, for example, Ref.~\refcite{Blanchet:2008zg,Baldes:2013eva} for constraints on Majorana neutrino parameters from requiring that they not wash out a pre-existing baryon asymmetry.}

\setcounter{paragraph}{0}
\paragraph{Out-of-equilibrium decays of heavy particles.} 
This scenario~\cite{Dodelson:1991iv,Kuzmin:1996he,Berezhiani:2000gw,Agashe:2004bm,
Kitano:2004sv,Kitano:2005ge,Cosme:2005sb,Berezhiani:2008zza,Kaplan:2009de,Gu:2009yy,
An:2009vq,Davoudiasl:2010am,Gu:2010ft,Dulaney:2010dj,Haba:2010bm,Chun:2010hz,
Blennow:2010qp,Arina:2011cu,Barr:2011cz,Chun:2011cc,Falkowski:2011xh,Heckman:2011sw,
Blum:2012nf,Arina:2012fb,Arina:2012aj,Kuismanen:2012iz,Perez:2013nra,Feng:2013wn} 
is adapted from Fukugita-Yanagida-style leptogenesis~\cite{Fukugita:1986hr} as can occur in the type-I seesaw model of neutrino mass generation (the decay mechanism was earlier used in GUT baryogenesis~\cite{Ignatiev:1978uf,Weinberg:1979bt}).  The idea is that there is a massive unstable particle that decouples from the thermal plasma, and then decays through interactions that violate $X$, $C$ and $CP$.  Typically, the decaying particle $\psi$ is self-conjugate and the decay rates for the process $\psi \to x_1\, x_2\, \ldots$ and its charge-conjugate $\psi \to x_1^*\, x_2^*\, \ldots$ are unequal due to $CP$ violation, where $x_i$ denotes a particle whose $X$-charge equals $X_i$.  The unequal decay rates to final states of opposite $X$ create the asymmetry, and because the decays happen after the particle has lost thermal contact with its daughter particles, there is no wash out due to inverse decays.  As in standard leptogenesis, the decay amplitude must involve interference between at least two Feynman graphs in order for $CP$-violating phases to have physical consequences.

\paragraph{Affleck-Dine mechanism.}
Affleck-Dine (AD) dynamics~\cite{Affleck:1984fy,Dine:1995kz,Dine:2003ax,Enqvist:2003gh,Allahverdi:2012ju} is a very plausible mechanism in supersymmetric (SUSY) theories, and it is worth noting at the outset that for the purposes of AD asymmetry generation the supersymmetry breaking scale is allowed to be beyond the reach of the LHC.  Take a complex scalar field $\phi$ that carries nonzero $X$, and compute the Noether charge density
\begin{equation}
J^0 = i (\dot{\phi}^* \phi - \phi^* \dot{\phi}) = R^2 \dot{\theta}\ ,
\end{equation}
where $\phi \equiv (R/\sqrt{2})\exp(i\theta)$ is an amplitude and phase decomposition.  In the AD mechanism, suitable conditions for generating a time-dependent phase exist, thus creating $X$ charge carried by the coherent oscillations of the scalar field (which initially is a spatially-homogeneous condensate).  To create the correct amount of charge, the amplitude $R$ needs to be large to compensate for the fact that the violation of $X$-charge conservation must for phenomenological reasons be small.  This is assisted in supersymmetric theories by the generic existence of flat directions for renormalizable scalar potentials.  The flat directions are lifted by supersymmetry-breaking soft masses and by effective, non-renormalizable terms that also provide explicit $X$ violation.  The coupling of the AD field $\phi$ to the inflaton helps to set up an initial state of that field during inflation to be at the required high value, and it also implies that some of the parameters in the scalar potential change with time as the universe expands.  At those high field values, the effect of the small $X$-violating terms is amplified, and these terms together with $CP$-violating and time-dependent parameters kick the AD field in the angular direction and thereby create the $X$ charge or asymmetry.  The AD mechanism has been used in several ADM models~\cite{Suematsu:2005kp,Suematsu:2005zc,Bell:2011tn,Cheung:2011if,
Graesser:2011vj,vonHarling:2012yn,Kane:2011ih,Choi:2013fva}.

\paragraph{First-order phase transition.}
This is analogous to electroweak baryogenesis~\cite{Riotto:1999yt}.  The idea is that the phase transition from the symmetric phase of a gauge theory to the broken phase proceeds via bubble nucleation seeded by quantum tunneling through a potential barrier.  The Higgs field driving the phase transition Yukawa couples to fermions in a $CP$ violating way, and $X$ conservation is violated in the symmetric phase via rapid sphaleron transitions associated with a triangle anomaly between the $X$ current and the gauge fields.  The movement of the bubble walls creates departures from equilibrium that partner with the $X$- and $CP$-violating interactions to create an $X$ asymmetry carried by the fermions.  
In electroweak baryogenesis, the observed high value of the electroweak Higgs mass implies that the electroweak phase transition is not first-order (although it may be in the context of an extended Higgs sector).  
For the ADM application~\cite{Kaplan:1991ah,Dutta:2006pt,Shelton:2010ta,Dutta:2010va,Petraki:2011mv,Walker:2012ka} however, the phase transition in some models occurs for a new gauge force, either in the dark sector or in a third sector that mediates between the visible and dark worlds.  The resulting parameter freedom makes it trivial to arrange for the phase transition to be as strongly first order as desired.

\paragraph{Asymmetric freeze-out.}
It was stated above, and shall be discussed further below, that an interaction to annihilate away the symmetric part of the plasma is mandatory in ADM models.  The asymmetric freeze-out mechanism uses the same interactions to generate the asymmetry and to eliminate the symmetric part~\cite{Farrar:2005zd}.  Consider the DM particle $\chi$ and its antiparticle $\bar{\chi}$.  As well as self-annihilations of $\chi$ with $\bar{\chi}$ to SM states, $\chi$ and $\bar{\chi}$ may also experience $X$-violating co-annihilations with SM species.  Taking $\chi$ to be spin-1/2 and assigning $B_{\rm D}(\chi)=1$, the crossing-symmetry-related reactions
\begin{equation}
\chi + u_i \to \bar{d}_j + \bar{d}_k, \quad \chi + d_j \to \bar{u}_i + \bar{d}_k,
\label{eq:AsymCoann}
\end{equation}
where $i,j,k = 1,2,3$ are family indices, are the simplest co-annihilations involving quarks that conserve electric charge, color and $B_{\rm V} - B_{\rm D}$, but violate $B_{\rm V} + B_{\rm D}$.
They are generated through the ``neutron portal'' type of effective operator schematically written as $udd\chi$.  
By considering other gauge-invariant combinations of SM fields carrying nonzero-$B_{\rm V}$, 
more complicated co-annihilation reactions can be systematically identified.  Through $C$ and $CP$ violation, the rates for the $\chi$ and $\bar{\chi}$ co-annihilations can be unequal, and if the co-annihilation rates dominate over the self-annihilation rates, then $\chi$ and $\bar{\chi}$ will decouple at different temperatures and thus have exponentially different relic number densities.  This scenario gives rise to a baryon-symmetric universe when the asymmetry-creating co-annihilations preserve a linear combination of $B_{\rm V}$ and $B_{\rm D}$ as in the example of Eq.~\eqref{eq:AsymCoann}.  In the ADM context, for the cases where the DM mass is in the few-GeV regime, the asymmetric freeze-out co-annihilations must create an asymmetry in visible-baryon number directly, because no sphaleron reprocessing will be possible.

\paragraph{Asymmetric thermal production or asymmetric freeze-in.}
A particle that is too weakly coupled to ever attain thermal equilibrium with the cosmological bath can nevertheless be slowly produced from processes involving the bath particles (see Ref.~\refcite{McDonald:2001vt} for an application to the production of singet-scalar DM). This thermal production process has recently also been called ``thermal freeze-in''\cite{Hall:2009bx} and applied within the ADM paradigm~\cite{Hall:2010jx}.  Thermal freeze-in has been argued to be the inverse of thermal freeze-out.  Freeze-out occurs when a species $\chi$ that starts off in thermal equilibrium subsequently decouples from the bath.  Freeze-in sees the particle $\chi$ being so weakly coupled to the bath that while it is being slowly produced and heading towards thermal equilibrium, its co-moving number density becomes a constant -- freezes in -- before it is actually able to reach equilibrium.  The freeze-in happens when the temperature drops below the mass of the heaviest particle involved in the $\chi$ production process, thus Boltzmann suppressing that heaviest species.  If $\chi$ can be produced by the decays or inverse decays of bath particles, then that process will dominate over $\chi$ scattering processes.  Most of the $\chi$ production occurs just before freeze-in, partly because the characteristic Hubble time is longest at that point and often also because the microscopic process is most rapid then.  For example, if $\chi$ is produced through the decay of a particle of mass $m$, the decay rate is suppressed by an $m/T$ time dilation factor when $T \gg m$.  In this case the largest mass is $m$, and thus $T = m$ is the approximate freeze-in temperature (precise calculations show that $\chi$ is dominantly produced when $m/T$ is in the range $2$-$5$).  Asymmetric freeze-in simply means that $\chi$ and $\bar{\chi}$ freeze in with unequal co-moving number densities.  This is achieved in specific models through the decays of bath particles and antiparticles which violate $C$ and $CP$ and conserve only one linear combination of $B_{\rm V}$ and $B_{\rm D}$~\cite{Hall:2010jx}.  The daughter DM particles may be in thermal equilibrium with other species inhabiting the dark sector provided that the dark-bath temperature is lower than that of the visible bath.

\paragraph{Spontaneous genesis.}
The Sakharov conditions~\cite{Sakharov:1967dj} presuppose $CPT$ symmetry.  While this is rigorously a symmetry of all local, relativistic quantum field theory Lagrangians, it is spontaneously broken by the expanding universe solutions used in cosmology.  Through appropriate interactions, this can induce effective $CPT$ violation, together with $T$ violation, in the particle physics of the early universe.  Once $CPT$ invariance does not (effectively) hold, the Sakharov conditions need not all be obeyed in order to dynamically obtain a particle-number asymmetry~\cite{Cohen:1987vi,Cohen:1988kt}.  In ``spontaneous genesis'', the violation of particle number is, of course, still required, but the $CP$-violation and out-of-equilibrium conditions are not in general obeyed.  The basic mechanism requires an effective term of the form ${\cal L} \supset \partial_\mu \phi J^\mu_{X}/\Lambda$, where $J^\mu_X$ is the current corresponding to the particle number $X$ that will develop an asymmetry, $\phi$ is some scalar field, and $\Lambda$ is a high scale of new physics.  For spatially-homogeneous but time-dependent $\phi$ solutions, this term reduces to $(\dot{\phi}/\Lambda)(n_X - n_{\bar{X}})$.  Thus $\dot{\phi}/\Lambda$ is an effective chemical potential for $X$ number, leading to the approximate result $n_X - n_{\bar{X}} \sim T^2 \dot{\phi}/\Lambda$.  Specific models have to arrange for a suitable background scalar field configuration $\phi$ to develop.  Clearly, they must ensure that $\dot{\phi} \to 0$ at late times.  To stop wash out in this limit, the $X$-violating interactions have to decouple before the effective chemical potential becomes too small.  This mechanism has been used in the ADM models described in Refs.~\refcite{MarchRussell:2011fi,Banks:2006xr,Kamada:2012ht}.

\subsection{Freeze-out in the presence of an asymmetry}
\label{sec:freezeout}

The synergy of the symmetry structures described in Sec.~\ref{sec:symmetries} and an asymmetry-generation mechanism can dynamically relate the particle-antiparticle asymmetries in the visible and the dark sectors.  This translates into a tight relation between the relic number densities of VM and DM, provided that the excess of particles over antiparticles is the only relic from each sector contributing significantly to the total energy density. Indeed, in the visible sector, the strong baryon-antibaryon annihilation, via the $\BV$-preserving interactions of the SM, eventually drives the abundance of antibaryons effectively to zero, leaving only the excess of baryons present today. 
Similarly, the symmetric part of DM\footnote{In the absence of a more suitable term, by ``symmetric part of DM" we mean the non-excess part of the DM population, i.e.\ if particles are more numerous than antiparticles, the symmetric part is the antiparticles plus an equal number of particles.} is efficiently annihilated away in the early universe if there exist sufficiently strong $\BD$-preserving interactions that allow DM annihilation.

How large the DM annihilation cross-section has to be in successful ADM models is, of course, an important feature which determines their low-energy phenomenology. 
In the presence of a particle-number asymmetry, the freeze-out of annihilations in the primordial plasma differs from the case of symmetric species. In the latter case, the relic abundance is
\beq
\W\DM \simeq 0.2 \sqpare{ \ang{\s v}_{_{\rm WIMP}} / \ang{\s v} }  \  ,
\eeq
where $\ang{\s v}_{\rm _{WIMP}} \simeq 2.3 \times 10^{-26} \cm^3/\snd$ is the canonical value for the thermally averaged cross-section times velocity, which yields the observed DM abundance\footnote{For $m\DM \leqslant 10 \GeV$, $\ang{\s v}_{\rm _{WIMP}}$ depends noticeably on $m\DM$, and for $m\DM \leqslant {\rm few} \GeV$, $\ang{\s v}_{\rm _{WIMP}} \simeq (4.5 - 5) \times 10^{-26} \cm^3/\snd$~\cite{Steigman:2012nb}. }~\cite{Kolb:1990vq,Steigman:2012nb}.
Freeze-out in the presence of an asymmetry has been recently studied in Refs.~\refcite{Graesser:2011wi,Iminniyaz:2011yp}. Following Ref.~\refcite{Graesser:2011wi}, we define the fractional asymmetry of DM as
\beq
r \equiv \frac{n(\bar{\x})}{n(\x)} \ ,
\label{eq:r}
\eeq
where obviously $r = 0$ and $r=1$ correspond to the completely asymmetric and symmetric case respectively. The detailed Boltzmann-equation analysis shows that the late-time fractional asymmetry, $r_\infty,$ depends exponentially on the annihilation cross-section~\cite{Graesser:2011wi} 
\beq
r_\infty 
\approx \exp \sqpare{-2\pare{\frac{\s_{_0}}{ \s_{_\text{0,WIMP}}} } \pare{\frac{1-r_\infty}{1+r_\infty}}} 
\quad \xrightarrow{r_\infty \ll 1} \quad
\exp \sqpare{-2 \s_{_0} / \s_{_\text{0,WIMP}}}   
\  ,
\label{eq:FractionalAsymmetry}
\eeq
where $\s_{_0}$ is related to the thermally-averaged cross-section times velocity, 
\beq
\ang{\s v} = \s_{_0} (T/m\DM)^n\ , 
\eeq
with $n=0\text{ and }1$ for $s$-wave and $p$-wave annihilation respectively.
Because of this exponentially sensitive dependence,  $\s_{_0} \gtrsim 1.4 \, \s_{_\text{0,WIMP}}$ suffices to render $r_\infty \lesssim 0.1$. Thus, annihilating efficiently the symmetric part of DM in ADM models requires an annihilation cross-section which is larger than the canonical thermal-relic cross-section, albeit only by a factor of a few~\cite{Graesser:2011wi,Iminniyaz:2011yp},
\beq
\s \gtrsim \text{ few } \times \s_{\rm _{WIMP}} \  .
\label{eq:sigma ann}
\eeq

Equation~\eqref{eq:FractionalAsymmetry} does not alone determine the relic DM abundance. In the presence of an asymmetry, reproducing the observed DM density yields a prediction for the DM mass, as per
\beq
\frac{m\DM}{m_p} \frac{\eta (\BD) / q\DM}{\eta (\BV)} 
= \frac{1-r_\infty}{1+r_\infty} \frac{\W\DM}{\W_{\rm _{VM}}}
\ ,
\label{eq:ADM abundance}
\eeq
where $q\DM$ is the dark-baryonic charge of DM. We discuss the prediction of the DM mass in ADM models in Sec.~\ref{sec:DMmass}.

\medskip
\emph{Partially} asymmetric scenarios with $0 < r_\infty < 1$ arise if a dark asymmetry has been generated in the early universe, but one of the following is true:
\benu[(i)]
\item The DM annihilation cross-section is small (albeit still larger than the WIMP-miracle cross-section, in order to avoid overclosure)~\cite{Graesser:2011wi,Iminniyaz:2011yp}.
\item The dark baryon number is softly broken by a small (Majorana-type) mass term (see e.g. Ref.~\refcite{Okada:2012rm}), which results in dark particle-antiparticle oscillations after DM freeze-out~\cite{Buckley:2011ye,Cirelli:2011ac,Tulin:2012re}. The oscillations can regenerate a symmetric DM population, i.e. drive $r$ away from zero. This in turn can lead to recoupling of DM annihilations after the original asymmetric freeze-out has taken place, and thus washout of the DM density. The depletion of DM strongly constrains the soft breaking of the dark baryon number, and bounds the oscillation timescale to be larger than the age of the universe~\cite{Buckley:2011ye}.  Note that if the DM is stabilised by any discrete $Z_p\, (p>2)$ symmetry rather than a full U(1), then DM scattering processes can repopulate the universe with DM antiparticles and thus cause washout~\cite{Ivanov:2012hc}. This attests to the need in ADM models for a good low-energy global symmetry in the dark sector which ensures the conservation of the dark baryonic charge at late times, as already emphasized in Sec.~\ref{sec:symmetries}.
\item The cosmology of the early universe is non-standard.  Even if the DM annihilation cross-section satisfies the inequality \eqref{eq:sigma ann}, annihilations are less efficient if the expansion rate of the universe in the epoch before big-bang nucleosynthesis (BBN) is larger than in the standard cosmological model~\cite{Gelmini:2013awa}, such as in kination and in scalar-tensor theories. 
\eenu
While of course a continuous spectrum for the DM asymmetry is a priori possible, $0 \leqslant r_\infty \leqslant 1$, models with partial DM asymmetry do not feature a tight relation between the VM and DM relic abundances -- the primary motivation of the ADM scenario. This review concentrates on the $r_\infty \simeq 0$ case, with occasional comments on the implications of $r_\infty > 0$.

\subsection{Dark interactions}
\label{sec:DarkInter}

The need for efficient (or at least adequate) annihilation of DM in the early universe is a defining feature of both symmetric thermal-relic and asymmetric DM models.
This feature sets the expectations for experimental detection.
Interactions which allow DM to annihilate directly into SM particles have been probed in colliders, in DM direct-detection experiments, and by observations of galactic and extra-galactic radiation backgrounds.
Experiments have, in fact, closed-in on the parameter space of DM annihilation into SM degrees of freedom. While it still possible that DM annihilates into SM particles with cross-section $\s_{\x\bar{\x} \to {\rm SM}} \gtrsim \s_{\rm _{WIMP}}$, most effective operators which yield such processes are highly constrained~\cite{Bai:2010hh,Buckley:2011kk,Fox:2012ee,MarchRussell:2012hi}. The constraints are a bit more severe for asymmetric DM models, due to the somewhat larger annihilation cross-section required for efficient annihilation, in comparison to the symmetric DM case (c.f.\ \eq{eq:sigma ann}). The above considerations suggest that DM may in fact annihilate into SM particles via metastable mediators, or into dark-sector light species.

A minimal possibility for the efficient annihilation of (asymmetric) DM is a dark abelian gauge force, U(1)$_D$. This is a building block of many ADM models, and it appears automatically in mirror DM models as a consequence of the mirror symmetry (with U(1)$_D$ being the the mirror electromagnetism).  The dark gauge symmetry may be broken or unbroken. Dark matter charged under U(1)$_D$ can annihilate directly into dark photons, which in the following we will denote by $\g_{_D}$, as long as they are lighter than DM itself, $M_{_D} < m\DM$, where $M_{_D}$ is the dark-photon mass. Alternatively, DM can annihilate via the dark photon into other light species charged under U(1)$_D$.

If the lightest dark baryon -- which we shall now denote by $\pD$ -- is charged under a gauged U(1)$_D$, the dark baryonic asymmetry carried by $\pD$ amounts also to an asymmetry under the U(1)$_D$ gauge charge. If U(1)$_D$ is unbroken, or if its breaking leaves an unbroken global U(1) remnant under which the $\pD$ transforms\footnote{This is the case if the scalar field $\f_{_D}$ which breaks U(1)$_D$ has no Yukawa couplings to $\pD$, if $\pD$ is a fermion, or no scalar couplings other than those which depend on the mod of each field individually (i.e. $|\f_{_D}|^2 |\pD|^2$), if $\pD$ is a boson.}, the asymmetry carried by $\pD$ has to be compensated by an opposite asymmetry carried by some other species which also transforms under (the global remnant of) U(1)$_D$, and which we shall denote by $\eD$. Our notation of course alludes to the analogous situation in the visible sector, where the electromagnetic-charge asymmetry carried by the protons is compensated by an opposite asymmetry carried by the electrons. 
In this case, DM consists of equal amounts of dark protons, $\pD$, and dark electrons, $\eD$, which -- depending on the strength and the possible breaking of U(1)$_D$ -- may form bound neutral dark Hydrogen-like atoms, $\HD$.

A dark abelian force can mix kinetically with the hypercharge gauge boson~\cite{Holdom:1985ag,Foot:1991kb}
\beq
\d {\cal L} = \frac{\e}{2} \: F_{Y  \m\n} \, F_D^{\m\n}  \  .
\label{eq:kinetic mixing}
\eeq
This effectively imparts electromagnetic charges of order ${\cal O}(\e)$ to the dark-sector particles charged under U(1)$_D$. If the dark photon is massive, then it can decay via the kinetic mixing into SM fermions. Its  lifetime
\beq
\t_{_D} \approx 10^{-13}  \snd  
\pare{\frac{10^{-4}}{\e}}^2 
\pare{\frac{100 \MeV}{M_D}} \ ,
\label{eq:dark photon lifetime}
\eeq
is sufficiently short for a vast range of parameters to ensure that dark photons decay immediately after they decouple, leaving no extra radiation at late times. On the other hand, if the dark photon is massless, it will act as extra radiation during both the BBN epoch and during the decoupling of the cosmic microwave background (CMB).  The constraints shall be discussed fully in Sec.~\ref{sec:Neff}, but it is pertinent to note here that, within the observational errors, both the primordial element abundance measurements~\cite{Cyburt:2004yc,Mangano:2011ar}, and the CMB anisotropy results from WMAP~\cite{Hinshaw:2012aka} and Planck~\cite{Ade:2013zuv}, permit the existence of quite a significant amount of extra radiation.  Any such additional radiation gets red-shifted to insignificance in the late universe.  The massless dark photon scenario has the advantage of being more testable (and elegant), while massive dark photon models are safer from cosmological constraints.

Different possibilities for the efficient annihilation of DM into non-SM particles include Yukawa and scalar couplings to new light degrees of freedom. As is the case for the dark photon, these light species might subsequently decay into SM particles, or may be stable and contribute to the relativistic energy density of the universe at late times. As well as the extra radiation constraints covered in Sec.~\ref{sec:Neff}, we also discuss various other bounds and phenomenological implications of dark interactions in other parts of Sec.~\ref{sec:pheno}.

\subsection{The mass of the DM state}
\label{sec:DMmass}

The tight relation between the visible and the dark baryonic asymmetries established in ADM models translates into a relation between the \emph{number densities} of the VM and DM relic abundances. Assuming that the DM particle-antiparticle annihilations in the early universe were sufficiently efficient, as described in Sec.~\ref{sec:freezeout}, today's DM number density, $n\DM$, is found by
\beq
n[B_{\rm D}] = q\DM \, n\DM  \ ,
\label{eq:nDM}
\eeq
where $q\DM$ is the dark-baryonic charge of the DM particle and $n[B_{\rm D}]$ is the number density of dark baryons. 
Then, \eq{eq:factorof5} gives a prediction for the DM mass. The mass of the DM state is of course important, as it affects the direct and indirect detection expectations.

In baryon-symmetric models, $n[\BLV] = n[\BD]$, as per \eq{eq:BV=BD=X/2--Bsym}, and $n[\BV] = a_s n[\BLV]$, where $a_s \simeq 0.35$ or 1, depending on whether the generation of the asymmetries happened before or after the electroweak phase transition, respectively~\cite{Harvey:1990qw}. Then, from \eq{eq:ADM abundance} (with $r_\infty = 0$), the DM mass has to be $m\DM = a_s\, q\DM \pare{\W_{\rm DM} / \W_{\rm VM}} m_n$, or~\cite{Gu:2010ft,vonHarling:2012yn}
\beq
m\DM \simeq q\DM \times (1.6 - 5) \GeV \ .
\label{eq:mDM-Bsym}
\eeq
Note that if effects similar to the electroweak sphalerons are operative in the dark sector after the asymmetry generation has taken place, the above prediction may be modified by a factor of a few.

If the relation between the visible and dark asymmetries is established via chemical equilibrium, as in the cases 2 - 4 described in Sec.~\ref{sec:symmetries}, then the DM mass required to reproduce the correct DM abundance depends on the details of the chemical equilibrium. If DM is the only dark baryon participating in the chemical equilibrium, and the chemical decoupling of the two sectors occurs while DM is relativistic, then detailed calculations place the DM mass in the range~\cite{Ibe:2011hq}
\beq
m\DM \simeq q\DM^{-1} \times (5-7)  \GeV \ .
\label{eq:mDM-chem}
\eeq
The inverse proportionality to $q\DM$ is due to the ratio $\eta (\BD)/\eta (\BV)$ that appears in \eq{eq:ADM abundance} evaluating as proportional to $q^{-2}\DM$ from the chemical equilibrium equations and other constraints~\cite{Ibe:2011hq}.  However, if more dark-sector particles carrying $\BD$ participate in the chemical equilibrium (and subsequently decay and impart their dark baryonic charge to DM), then \eq{eq:mDM-chem} does not hold. 

It is also possible that the chemical decoupling of the two sectors occurs when DM is non-relativistic, but the SM quarks and leptons are still relativistic. While chemical equilibrium keeps the chemical potentials of the visible and dark sector particles at the same magnitude, the number density of the dark species is Boltzmann suppressed, and $n[\BD]/n[\BLV] \sim \exp (-m\DM/T)$. In this case a much larger DM mass is required in order to compensate for the thermal suppression of the DM number density, $m\DM \sim \pare{n[\BLV]/n[\BD]} \pare{\W_{\rm DM} / \W_{\rm VM}}$. While the exact value which reproduces the observed DM abundance depends on the details of the chemical equilibrium, in this regime the DM mass is expected to be $m\DM \sim \TeV$~\cite{Barr:1990ca,Buckley:2010ui}.

\medskip

Dark-matter particles may form bound states. In this case, the mass of the \emph{DM state} is different from the mass of the lightest dark baryon, and may deviate, sometimes significantly, from the predictions of \eqs{eq:mDM-Bsym} and \eqref{eq:mDM-chem}. We mention a few examples:
\bit
\item 
As discussed in Sec.~\ref{sec:DarkInter}, in many ADM models the lightest dark baryon is charged under a dark gauge force, similarly to the ordinary proton. If this dark force is unbroken, then the dark gauge-charge asymmetry carried by the dark proton has to be compensated by an opposite gauge-charge asymmetry carried by another particle, a dark electron. In this case, DM consists of dark atoms, bound states of the dark protons and dark electrons. The discussion above is still applicable, noting that the preceding equations give the mass of the DM state, rather than the mass of the lightest dark baryon.
\item
In the mirror DM scenario, dark baryons are bound in heavy mirror atoms due to the mirror nuclear interactions. The states expected to be detected in direct-detection experiments are primarily the mirror O, Ne, N, C (collectively called ``metals" in the astrophysics literature), with masses of a few tens of GeV~\cite{Foot:2003iv,Foot:2010hu,Foot:2011pi,Foot:2013msa}.
\item
Supersymmetric $Q$-balls are bound states of scalar fields carrying a conserved global charge~\cite{Kusenko:1997ad,Kusenko:1997zq,Kusenko:1997si}. If stable, they can make up the DM of the universe~\cite{Kusenko:1997si,Kasuya:1999wu,Demir:2000gj,Kusenko:2001vu,Kusenko:2004yw,Barnard:2011jw}.
In the minimal supersymmetric standard model (MSSM), stable $Q$-ball states arise as gauge-neutral linear combinations of squark fields. Their mass scales with their baryonic charge as 
$m_Q \sim \m Q^{3/4}$, where $\m$ is a mass parameter that relates to the attractive interaction responsible for the existence of $Q$-ball solutions.
For large enough $Q$, their mass-to-charge ratio is smaller than the mass of the proton, and baryonic $Q$-balls are energetically disallowed to decay. In typical scenarios, $\m \sim \TeV$, and baryonic DM $Q$-balls have masses $m_Q > 10^{12} \GeV$.
\eit

Typical ADM models, with mirror DM an exception, treat the DM mass as a parameter that is effectively measured through $\Omega_{\rm DM}$ in the context of a theory that predicts a specific DM number density.  These scenarios have to be considered incomplete.  As mentioned earlier, a full explanation of Eq.~\eqref{eq:factorof5} should have two characteristics: a relation between the visible and dark relic number densities, plus a good theoretical rationale for the DM mass scale.  Absent the latter, you effectively have just changed the mysterious coincidence of Eq.~\eqref{eq:factorof5} into the conundrum of why the DM mass scale happens to be in the few-GeV regime (for typical models).  A survey of the ADM literature reveals that most attention has been given to the number-density question, and rather less to the DM mass issue.  Nevertheless, some ideas have been proposed on how the required mass scale may be explained.

For the case of the typical few-GeV mass scale, it is very tempting to hypothesize a connection with the proton mass and hence the QCD confinement scale.  After all, the visible matter mass density depends exactly on this parameter.  The mirror matter model does this in the simplest possible way, as explained in more detail in the next section.  One simply supposes that the dark-sector microscopic theory is a copy of the visible-sector microphysics.  In that case there is a mirror QCD with exactly the same confinement scale as ordinary QCD.  Because of the plethora of stable nuclei there are several DM masses, but all of them are directly related to the QCD scale.  One may imagine that certain variations on the mirror matter idea might retain this feature, while allowing the other dark-sector microphysics to deviate from exact correspondence with the visible sector.

Another possibility is that the scale is similar to, say, the few-GeV bottom-quark mass.  It is due to the electroweak scale, but there is a small dimensionless parameter that reduces the mass appropriately.  This is certainly possible, but it should not be considered fully satisfactory.

A third speculation is that the DM mass scale is suppressed relative to the electroweak scale, or an even higher scale such as a hidden-sector spontaneous or dynamical supersymmetry-breaking scale, because of the weakness of messenger interactions that transmit the effects of symmetry breaking from one sector to another.  The suppression could be due to a high mass for the messengers, or a small kinetic-mixing parameter for gauge bosons~\cite{Cohen:2010kn}, or because of loop factors~\cite{Walker:2012ka}.  In a supersymmetric context, one could imagine that supersymmetry breaking is transmitted directly from the hidden sector to the visible sector through messengers, and then from the visible to the dark sector through other messengers: a cascade effect.  This speculation obviously fits in well with the idea that the DM is part of a relatively complex dark sector.  However, it seems difficult to get more than a qualitative explanation for the DM mass scale this way.

In technicolor models and other scenarios involving strong interactions~\cite{Nussinov:1985xr,Barr:1990ca,Khlopov:2005ew,Gudnason:2006ug,Gudnason:2006yj,Khlopov:2008ty,
Ryttov:2008xe,Foadi:2008qv,Kribs:2009fy,Frandsen:2009mi,Frandsen:2011kt,Khlopov:2011tn,Lewis:2011zb}, the mass of the DM state is sometimes determined by the confinement scale, but there are also other possibilities (e.g.~\refcite{Lewis:2011zb}).

Though it does not really solve the DM mass problem, it is appropriate to mention here that Ref.~\refcite{Barr:2011cz} is an interesting attempt to literally grand-unify the dark and visible sectors.


\section{Asymmetric dark matter models}
\label{sec:models}

We now survey a few specific models, chosen for their contrasting features.  Given the size of the ADM literature it is not possible to summarize or even mention most of the interesting proposals, though the reference list is a substantial compendium of the relevant papers.

\subsection{Mirror dark matter}
\label{sec:mirrorDM}

As stated above, mirror-matter models take the dark sector to be described by an exactly isomorphic gauge theory to the visible sector~\cite{Foot:1991bp,Foot:1991py,Foot:1995pa,Lee:1956qn,Kobzarev,Pavsic:1974rq,Blinnikov:1982eh,Blinnikov:1983gh,Berezhiani:2000gw,Ignatiev:2003js,Foot:2003jt,Foot:2004pq,Berezhiani:2003wj,Ciarcelluti:2004ik,Ciarcelluti:2004ip,Berezhiani:2005vv,Berezhiani:2008zza,Foot:2013msa}.  Restricting our scope to the visible sector being governed by the SM, the full gauge group is
\begin{equation}
G_{\rm SM} \times G'_{\rm SM}\ ,
\label{eq:GxG}
\end{equation}
and a discrete $Z_2$ symmetry interchanging the sectors is imposed~\cite{Foot:1991bp,Foot:1991py,Foot:1995pa}.  The name ``mirror matter'' derives from the usual choice that the discrete symmetry transforms visible-sector fermions into dark-sector fermions of opposite chirality, and vice-versa.  While the chirality-flipping feature is not mandatory, with this choice the full improper Lorentz group becomes a symmetry of nature, suggesting a fundamental reason for the existence of a hidden sector containing DM~\cite{Foot:1991bp}.  The symmetry-breaking sector of the minimal mirror-matter model contains simply the standard electroweak Higgs doublet and its mirror partner.  In this case, there are two possibilities for the mirror-parity symmetry depending on the Higgs potential parameter regime: unbroken with exactly the same electroweak vacuum expectation values (VEVs) in both sectors, or broken maximally in the sense that the mirror-doublet VEV is zero~\cite{Foot:2000tp}.  By adding extra scalar fields and/or softly breaking mirror parity, it is possible to have both VEVs nonzero and unequal~\cite{Berezhiani:1995yi,Cui:2011wk,Gu:2012fg,Gu:2013nya}.  Our focus is on the exact mirror-parity case in what follows.  

The two sectors interact gravitationally, through photon--mirror-photon kinetic mixing, Higgs mixing, and generally also through neutrino--mirror-neutrino mixing.  It is technically natural to take all of the non-gravitational interactions to be very weak.  Nevertheless, the kinetic mixing of the two photons is an important feature for both direct-detection phenomenology and the cosmological evolution of large-scale structure.

With mirror parity respected by the vacuum, the microphysics of the dark, mirror sector is identical to that of the visible: there are mirror protons, neutrons and electrons mass-degenerate with their visible counterparts, and there is a massless mirror photon.  Mirror matter thus consists of mirror atoms and mirror ions.  The few-to-10s of GeV ADM mass regime has a definitive explanation, as discussed earlier.

Despite the identical microphysics, the cosmological macrophysics must be different between the sectors, at least by the time that BBN takes place.  To meet the BBN upper bound on relativistic energy density during that epoch, and to ensure that mirror matter subsequently behaves sufficiently like CDM, the temperature of the mirror plasma $T'$ must be somewhat lower than that of the visible plasma $T$.  The ratio $T'/T$ should be of order $0.3$ or lower, but it need not be tiny~\cite{Berezhiani:2000gw,Ignatiev:2003js,Berezhiani:2003wj,Ciarcelluti:2004ik,Ciarcelluti:2004ip}.  The required temperature difference may reflect asymmetric reheating in the two sectors after inflation, or it may be due to a post-inflationary injection of entropy into the visible sector.  The temperature difference leads to a much higher mirror-He to mirror-H ratio than its visible analog, and this changes the star- and galaxy-formation outcomes in the mirror world (see, for example, Ref.~\refcite{Berezhiani:2005vv}).  Nonetheless, we expect stellar nucleosynthesis to take place and thus produce higher-mass isotopes such as mirror-O, -Ne, -N, -Fe, and so on.  The DM today should be in the form of mirror gas and ions plus compact stellar objects.  The spheroidal shape of observed DM haloes requires that a heating source counteract the dissipative nature of mirror DM, with visible-sector supernovae shown to provide about the correct heating power through the kinetic mixing of the photon and mirror photon~\cite{Foot:2004wz}.

Two possibilities for asymmetry generation have been studied in the literature.  One vision is that the visible and mirror plasmas are never in thermal contact~\cite{Berezhiani:2000gw}, motivated by the need to have $T'/T \lesssim 0.3$.  One chooses a baryogenesis mechanism in the visible sector with the knowledge that the mirror-parity symmetry forces exactly the same microphysical interactions in the mirror sector.  However, the temperature difference causes the asymmetries to be numerically different in the two worlds: the microscopic mirror-sector interaction rates depend on $T'$, while the expansion rate of the universe is driven by the larger visible-sector temperature $T$.  Reference~\refcite{Berezhiani:2000gw} studied asymmetry generation due to either out-of-equilibrium decays or bubble-nucleation during a first-order electroweak phase transition.  In both cases, it was found that the relic mirror-matter density was \emph{larger} than the relic visible-matter density for a significant region of parameter space.  This is a very interesting feature that correlates nicely with Eq.~\eqref{eq:factorof5} at the qualitative level.  Quantitatively, the precise ratio depends on unknown parameters and need not be near five.   As mentioned earlier, this is a case-4 scenario but without any chemical reprocessing between the sectors. 

The second approach falls under case 3: an asymmetry is generated in the mirror sector, and then reprocessed into the visible \cite{Foot:2003jt,Foot:2004pq}.  The analysis does not specify how the initial mirror-world asymmetry is created, focusing instead on the transfer mechanism.  The sequence of events is taken to be: 
\begin{enumerate}[(i)] 
\item The universe begins with $T' \gg T$, perhaps due to asymmetric reheating after inflation.
\item A mirror-lepton and/or mirror-baryon asymmetry subsequently develops.
\item The chemical reprocessing of the initial asymmetry includes transfer reactions to the visible sector through the dimension-five effective operators,
\begin{equation}
\left( \frac{1}{M_N} \right)_{ij}\, \overline{\ell}_{iL}\, \phi^c\, \ell'_{jR} \phi' + H.c.\ ,
\label{eq:mirrortransferops}
\end{equation}
where $\ell_{iL}$ is the usual lepton doublet with family index $i=1,2,3$,  $\phi$ is the electroweak Higgs doublet with $\phi^c \equiv i\tau_2 \phi^*$, and the primed fields are their mirror analogs.  These are the lowest-mass-dimension gauge-invariant effective operators, and the processes they induce are in equilibrium during the regime
\begin{equation}
10^{3}\, \left( \frac{M_N}{10^{10}\, {\rm GeV}} \right)^2 \lesssim \frac{T}{{\rm GeV}} \lesssim \frac{M_N}{{\rm GeV}}\ ,
\end{equation}
where $T$ is now the common temperature of the two sectors.  As is typical in scenarios of this kind, the asymmetry-transfer interactions bring the two sectors into kinetic equilibrium as well as performing the chemical reprocessing.
\item A subsequent injection of entropy causes $T > T'$.  The dynamics of this are left open.  A possibility is a second, very brief burst of inflation: if the initial $T' \gg T$ condition was due to mirror-inflaton-driven inflation, then the later stage could be induced by a visible-sector inflaton.
\end{enumerate}
The analysis requires solving a set of simultaneous chemical-potential equilibrium equations for all the processes that are faster than the expansion rate, with the initial asymmetries serving as conserved quantities.  The outcome depends on the initial asymmetries and the flavor content of the dominant operator in Eq.~\eqref{eq:mirrortransferops}.  The general results are catalogued in Refs.~\refcite{Foot:2003jt,Foot:2004pq}.  Examples of phenomenologically successful cases have the initial asymmetries 
\begin{equation}
\eta(L'_{\ell_{1}}) = \eta(L'_{\ell_{2}}) = \eta(L'_{\ell_{3}}) \neq 0,\quad \eta(B'_{u_{1R}}) = \eta(B'_{d_{1R}}) = \eta(B'_{d_{2R}}) \neq 0\ ,
\end{equation}
and one of the $i,j = 1,1$ or $2,2$ or $3,3$ effective operators of \eq{eq:mirrortransferops} dominating.  The results depend on assumptions about the final injection of entropy required to produce $T>T'$, with
\begin{equation}
\frac{\Omega_{\rm DM}}{\Omega_{\rm VM}} \simeq 4.6-5.0  \ ,
\end{equation}
amongst the possibilities.

\subsection{Dark(o)genesis}
\label{sec:dark(o)genesis}

Two examples of other models that have a dark asymmetry reprocessed into the visible sector are those named ``darkgenesis''~\cite{Haba:2010bm} and ``darkogenesis''~\cite{Shelton:2010ta}, which are similar in both name and construction.  Both are extensions of the next-to-MSSM (NMSSM), with darkgenesis employing the out-of-equilbrium-decay asymmetry generation mechanism, while darkogenesis uses a first-order phase transition.

In darkgenesis~\cite{Haba:2010bm}, one introduces lepton-number-carrying dark-sector chiral superfields $Y_i$, $X$ and $\bar{X}$, where the last two are conjugates of each other.  There must be at least two $Y_i$ fields ($i=1,2$) to ensure that there are physical $CP$-violating phases, and they have bare mass terms $Y_i Y_i$ in the superpotential that explicitly violate lepton-number conservation.  The $Y_1$ particle decays after decoupling from the bath into both $\bar{X}\bar{X}$ and $\bar{X}^* \bar{X}^*$ at different rates due to $CP$ violation.  The effective operator $\bar{X}^2 L H_u$ in the superpotential is used to transfer the asymmetry in the $\bar{X}$ sector into the visible sector, and it is the reason why $\bar{X}$, and hence also $Y_i$, carry (broken) lepton number.  Electroweak sphalerons then reprocess the visible-sector lepton asymmetry into a baryon asymmetry.  The DM is $\bar{X}$, with a mass of about 11 GeV.  
The fact that the model is based on the NMSSM rather than the MSSM allows the annihilation of the symmetric part to occur into the light pseudoscalar state allowed in this kind of scenario.

The dark sector of darkogenesis consists of a chiral gauge theory structured in a similar way to the electroweak sector of the SM.  There is a non-abelian dark gauge group, with SU(2)$_D$ being the minimal choice.  The dark fermions comprise at least two families which each have a left-handed doublet under SU(2)$_D$ and two right-handed singlets.  This structure means that the dark-fermion number $B_{\rm D}$ (called $D$ in Ref.~\refcite{Shelton:2010ta}) is an anomalous symmetry, just like $L_{\rm V}$ and $B_{\rm V}$ are in the SM.  Dark sphalerons then mediate rapid $B_{\rm D}$-violating processes above the critical temperature for the SU(2)$_D$ phase transition.  The required symmetry breaking proceeds via two Higgs doublets that also Yukawa couple to the fermions and provide them with mass in the broken phase.  The phase transition, which involves the $S$ field of the NMSSM as well as the two Higgs doublets, can be arranged to be strongly first-order, and a $B_{\rm D}$ asymmetry arises during it. The DM consists of the lightest fermion in this sector.  Two scenarios are considered for asymmetry transfer.  In the first, the dark phase transition occurs after the electroweak phase transition, so the dark asymmetry has to be processed directly into visible sector baryons.  This can be done, for example, through a ``neutron portal'' type of effective superpotential term, $X^2 u^c d^c d^c$, where the $X$ superfield appears quadratically to ensure that this interaction does not destabilize the DM.  In the second scenario, the dark phase transition occurs at a high temperature and the transfer can be accomplished by electroweak sphalerons.  This requires adding a messenger sector featuring fermions that carry both visible-sector weak isospin and $B_{\rm D}$.  In these models, the DM mass is in the 1-15 GeV range.  Annihilation of the symmetric part, depending on the precise model, proceeds either via a similar mechanism to that employed in darkgenesis, or into light fermions introduced for this purpose.

\subsection{Pangenesis}
\label{sec:pangenesis}

The word ``pangenesis'' has been coined to denote any theory that produces a baryon-symmetric universe through the Affleck-Dine mechanism in a SUSY context.  Recall that baryon-symmetric universes feature equal and opposite asymmetries in the visible and dark sectors under an always conserved symmetry, $B_{\rm con}$.  As we shall see, pangenesis works optimally when $B_{\rm con} = (B-L)_{\rm V} - B_{\rm D}$ is gauged, or more precisely, it is the global remnant of a gauge symmetry (which has to be broken to be consistent with observational constraints). The broken symmetry $B_{\rm bro}$ of Eq.~\eqref{eq:Bbro} is given by $(B-L)_{\rm V} + B_{\rm D}$ and renamed $X$. 

The specific model we review below is drawn from Ref.~\refcite{vonHarling:2012yn}; for earlier work see Refs.~\refcite{Bell:2011tn,Cheung:2011if}.  The visible sector is taken to be the MSSM plus right-handed neutrino chiral superfields to ensure the anomaly-freedom of ${(B-L)_{\rm V}}$.  The MSSM fields and the right-handed neutrinos are assigned gauge charges $B_{\rm con} = \BLV$. 
The dark sector consists of a supersymmetric U(1)$_{\rm D}$ gauge theory, with chiral superfields $\Delta$ and $\Lambda$ plus their opposite-chirality partners so that the fermionic components are vector-like.  The supermultiplets $\Delta$ and $\Lambda$ have gauge charges $[B_{\rm con}\, ,\, D]$ given by $[-q\DM\, ,\, 1]$ and $[0\, ,\, -1]$, respectively.  The gauge group of the model,  $G_{\rm SM} \times {\rm U}(1)_{B_{\rm con}} \times {\rm U}(1)_{\rm D}$, ensures that at the renormalizable level, there exist two accidental global symmetries, which can be taken to be $\BLV$, as usually defined in the visible sector, and $\BD$ in the dark sector, under which the superfields $\D$ and $\L$ have charges $q\DM$ and 0 respectively. 

The purpose of the U(1)$_{\rm D}$ gauge force is to annihilate the symmetric part of DM, as discussed in Sec.~\ref{sec:DarkInter}.  Recall that if the dark photon is massless, then its energy density contributes as dark radiation during BBN and at recombination.  If it is sufficiently massive it can decay into SM states such as electron-positron pairs through kinetic mixing, and there is no dark radiation (c.f. \eq{eq:dark photon lifetime}).  
The discussion below deals with the massless dark photon case only.  In that situation, $\Lambda$ is needed as well as $\Delta$ in order to have oppositely $D$-charged particles so that the universe can be neutral (they play the roles of the $\pD$ and $\eD$ particles introduced in Sec.~\ref{sec:DarkInter}).

As mentioned in Sec.~\ref{sec:AsymGen}, the AD mechanism relies on the existence of flat directions, carrying the charge one wishes to create, in supersymmetric scalar potentials.  These flat directions are associated with gauge-invariant monomials of the chiral superfields, and for the MSSM have been catalogued in Ref.~\refcite{Gherghetta:1995dv}.  For pangenesis, we need monomials that are invariant under both the visible- and dark-sector gauge groups, as well as $B_{\rm con}$, and carry nonzero $X$.  Examples are
\begin{equation}
(\Delta \Lambda)^2 u^c d^c d^c, \quad \Delta \Lambda (u^c d^c d^c)^2, \quad \Delta \Lambda\, d^c d^c d^c L L,
\end{equation}
which correspond to $q\DM = 1/2$, $2$ and $3$, respectively.  We now see that gauging $B_{\rm con}$ ensures that all directions of the scalar potential which transform under it, are not flat because they are lifted by the corresponding supersymmetric quartic $D$-terms. The AD mechanism is thus inoperative along such directions. In other words, as expected, gauging $B_{\rm con}$ ensures that no net  $B_{\rm con}$ is generated.

After choosing one of the flat directions, the associated monomial is added as a non-renormalizable term in the superpotential.  This serves both to explicitly break U(1)$_X$, as required for the creation of an $X$ asymmetry, and it lifts the flat direction at very high field values.  SUSY breaking is introduced in three ways into the scalar potential: as mediated from the SUSY-breaking hidden sector (including $X$-violating $A$-terms derived from the monomial), from the vacuum energy of the universe, and from the thermal bath.  The scalar potential includes various terms that violate both $X$ and $CP$, and are sensitive to the expansion of the universe through the Hubble parameter.  These features combine to see the Sakharov conditions satisfied, resulting in the creation of a condensate in the flat-direction field that carries nonzero $X$-charge.  This charge is subsequently transferred to visible- and dark-sector particles, assuming that stable $Q$-balls do not form instead.\footnote{For a parameter-space analysis of the mechanism in both gravity-mediated and gauge-mediated SUSY breaking, see Ref.~\refcite{vonHarling:2012yn}.}

The DM in this model is atomic.  The conserved $D$ and $B_{\rm D}$ charges\footnote{Recall  $\BLV$ and $B_{\rm D}$ are separately conserved in the low-energy and temperature limit.} imply the existence of two stable particles, which are the fermionic components $\delta$ (dark proton) and $\lambda$ (dark electron) of the supermultiplets $\Delta$ and $\Lambda$, respectively.  The DM today consists of hydrogen-like bound states of these particles.  The DM mass $m_{\delta} + m_{\lambda} \simeq q\DM\, (1.6 - 5)\ {\rm GeV}$ is as given by Eq.~\eqref{eq:mDM-Bsym}.
 
We finally note that, in supersymmetric theories, it is always possible to define a conserved discrete symmetry, an $R$-parity, which distinguishes between bosons and fermions carrying the same global charges. The inevitable conservation of $R$-parity in theories which feature global symmetries, such as theories of pangenesis, imples the stability of an additional particle. In the context of ADM, the stable lightest $R$-parity-odd particle is an unwanted relic. It is possible however to ensure that it does not contribute significantly to the energy density of the universe today~\cite{vonHarling:2012yn}.

\subsection{Asymmetric freeze-out model}
\label{sec:asymFOmodel}

As reviewed briefly above, asymmetric freeze-out requires the decoupling of DM particles and antiparticles to occur at different temperatures due to different co-annihilation rates driven by $CP$ violation.  A specific model~\cite{Farrar:2005zd} for a baryon-symmetric universe has been proposed based on the gauge-invariant effective interaction
\begin{equation}
\kappa \left[ \overline{(b_{L})^c} c_{L} \bar{X}_L d_{R}  
- \overline{(c_{L})^c} b_{L} \bar{X}_L d_{R} \right] + \hc
\label{eq:asymFOeffop}
\end{equation}
where the coupling strength is $\kappa = g^2/\Lambda^2$ with $\Lambda \lesssim 10$ TeV being the scale of new physics, $X$ is the DM fermion with mass about 5 GeV, and $d,c,b$ are the usual quarks (one must take account of the suppressed color index to see that this operator is nonzero).  Various diagrams involving this interaction plus electroweak effects featuring all three families produce different rates for the co-annihilation process $d + \bar{X} \to \bar{c} + \bar{b}$ and its charge conjugate (recall from Eq.~\eqref{eq:AsymCoann} that this neutron-portal co-annihilation channel is the simplest possible one).

A coupling strength of $\kappa \simeq 10^{-8}$ GeV$^{-2}$ is required to produce the correct asymmetric relic abundance of DM.  However, the same interaction, Eq.~\eqref{eq:asymFOeffop}, causes the DM to be unstable to decay into ordinary quarks.  Requiring the lifetime to be greater than the age of the universe is incompatible by many orders of magnitude with the $\kappa \simeq 10^{-8}$ GeV$^{-2}$ interaction strength.  
Dark-matter stability can still be ensured if $\kappa$ has a very strong temperature dependence which arises dynamically.  
One possibility is that \eq{eq:asymFOeffop} is mediated by a particle whose mass at the freeze-out temperature is moderate, but which acquires a much larger mass at zero temperature due to its coupling to a scalar field which undergoes a phase transition and obtains a very large vacuum expectation value.

The asymmetric freeze-out mechanism is elegant because the same co-annihilations that produce the asymmetry also co-annihilate away the symmetric part.  The fact that this is connected with unstable DM is fundamental:  any co-annihilation graph can, by crossing symmetry, be transformed into a DM decay graph.  This is in principle interesting from a phenomenological perspective.  Alternative ways of reconciling slow DM decay with sufficient asymmetric co-annihilation may also be interesting to pursue.

\subsection{Model with bosonic ADM}

Many ADM models have the DM particles as fermions.  But it is also interesting to consider the possibility of bosonic DM.  As we shall see in Sec.~\ref{sec:captureinstars}, fermionic and bosonic ADM have different phenomenological implications if they get captured in compact stellar objects.  We now review a class of bosonic ADM models that also allow the DM mass to be much higher than the few-GeV regime~\cite{Gu:2010ft}.  These models produce baryon-symmetric universes.

The DM particle is a complex scalar field $\chi$ that is a SM gauge singlet but carries dark baryon number.  Interactions between $\chi$ and SM fields are mediated by a number of other scalar fields, including two scalar leptoquarks, $\delta \sim (3,2,-1/3)(-2/3)$ and $\omega \sim (3,1,2/3)(-2/3)$, where the labels denote color, weak-isospin, hypercharge and visible baryon number, respectively.  The field $\delta$ couples to up and down quarks, while $\omega$ couples to pairs of down quarks.  In addition to the leptoquarks, there is either one complex SM-singlet scalar $X_1$, or a sequence of scalars $X_1, X_2, \ldots, X_n$.  The field $X_1$ couples to the leptoquarks via the scalar potential term $X_1 \omega \delta^2$, so it carries two units of visible baryon number.  In the case where only $X_1$ is introduced, it is required to couple to the DM field through $X_1^* \chi^b$, where $b = 2, 3$.  This term explicitly breaks the visible and dark baryon numbers to a linear combination, under which $X_1$ and $\chi$ carry $2$ and $2/b$ units of charge, respectively.  If there is sequence of $X_i$ scalars, then the baryon-number assignments are specified by the mixed interaction terms $X_2^* X_1^{a_{1}}$, $X_3^* X_2^{a_{2}}$, $\ldots$, $X_n^* X_{n-1}^{a_{n-1}}$ where the exponents $a_{1,2,\ldots,n-1} = 2, 3$.  The baryonic charges of the $X_i$ rise rapidly with increasing $i$.  For this model, the interaction with the DM field is taken to be $X_n^* \chi^b$ where $b = 2, 3 \neq a_{n-1}$.  The DM thus inherits the large baryonic charge of $X_n$.

Asymmetry generation proceeds through $CP$-violating, out-of-equilibrium decays of either $X_1$ or $X_n$, depending on the case.  The decay channels include both direct decay into $b$ DM particles, and through a cascade of virtual $X_i$ and leptoquark scalars to VM in the form of quarks and antiquarks.  By virtue of the conserved linear combination of visible and dark baryon numbers, the \emph{baryon} asymmetries are equal and opposite.  For the $X_1$-only model, the DM particle $\chi$ carries either $1$ or $2/3$ units of baryon number, leading to DM masses of about $1.6$ GeV (for $b=2$) and $1.1$ GeV (for $b=3$), where account has been taken of electroweak sphaleron reprocessing.  For the $X_n$ case, the high DM baryon-number assignment means that the DM \emph{number} density is unusually low, requiring a corresponding increase in the mass, as per \eq{eq:mDM-Bsym}.  For $n=7$, the mass creeps into the TeV scale.

Two possibilities are considered for annihilating the symmetric part.  One is direct annihilation into SM Higgs bosons through the $\chi^* \chi \phi^\dagger \phi$ term in the scalar potential.  Acknowledging that this possibility is constrained by direct-detection experiments, they also canvass the prospect of annihilation into the Goldstone bosons of an additional U(1) global symmetry that can be motivated through the Dirac see-saw mechanism.


\section{Phenomenology and bounds}
\label{sec:pheno}

The primary observational constraint discussed above for ADM was simply the motivating result of Eq.~\eqref{eq:factorof5}. But ADM, because of the potential richness of the hidden sector it inhabits, and of how that sector may interact with VM, holds the potential to reveal itself in many interesting ways in cosmology, astrophysics and through various kinds of terrestrial experiments.  We should start with the question:  Does ADM phenomenology \emph{have} to be unconventional?  The answer is no, as there are many ways to ensure that it behaves to a good approximation like standard collisionless CDM, with its true nature only accessible through very high-energy experiments.  However, it is extremely interesting to consider also the many ways that ADM could differ observationally and experimentally from that default possibility. Finding that the DM has some non-standard property would be a huge breakthrough, and we would do well to understand the range of possibilities for ADM and what constraints already exist.  
In fact, non-standard DM properties may help explain discrepancies that currently exist between the predictions of collisionless CDM and observations of the galactic structure, as well as provide a way to bring DM direct-detection experiments in better agreement. 
The question then shapes into this: How different from standard \emph{should} the properties of DM be, and does asymmetric DM provide a new DM paradigm which can successfully address existing observational anomalies?
The broad topics to be covered in this section are: dark radiation, galactic structure, direct and indirect detection, capture in stars, collider signatures and dark force constraints.

\subsection{Cosmology}
\label{sec:cosmobounds}

\subsubsection{Extra radiation}
\label{sec:Neff}

The fact that the symmetric part of the dark plasma has to be eliminated motivates serious consideration of the possible existence of dark radiation, such as a dark photon or other light (possibly massless) species, as discussed in Sec.~\ref{sec:DarkInter}. Dark radiation would give a non-standard contribution to the relativistic energy density of the universe that can have important consequences for BBN and CMB acoustic oscillations, and is already constrained by observations.  The latter fact in turn constrains ADM model building.

The interactions which relate the visible and dark baryonic asymmetries
are expected to bring the two sectors into thermal as well as chemical equilibrium at some point in the early universe. However, these interactions also have to eventually decouple to ensure the separate conservation of $B_{\rm V}$ and $B_{\rm D}$ at late times.  Let the kinetic decoupling temperature of the two sectors\footnote{Chemical decoupling occurs when the asymmetry-relating interactions fall out of equilibrium.  Kinetic decoupling may also occur at the same time, or later if there are additional VM-DM interactions that maintain kinetic-equilibrium without causing any chemical reprocessing.} 
be $T_{\rm dec}$, and suppose that the dark sector contains very light stable species.  In general, the dark sector will contain heavy unstable particles in addition to the DM and the dark radiation.  As these particles go non-relativistic, their contribution to the dark entropy density is transferred to the remaining states in the dark bath, which at sufficiently late times means the dark radiation.  This is what is often referred to as reheating.  A parallel process occurs in the visible sector.  The relative temperatures of the visible and dark radiation at late epochs such as BBN and recombination depend on the relative number of degrees of freedom that transfer their entropy to radiation after $T_{\rm dec}$.

It is customary to quantify extra radiation as an increase $\Delta N_{\rm eff}$ in the ``effective number of neutrino flavors'', defined through
\begin{equation}
\Delta \rho = \frac{7 \pi^2}{120} \left( \frac{4}{11} \right)^{4/3}  \Delta N_{\rm eff}\, \TV^4\ ,
\end{equation}
where $\Delta \rho$ is the energy density of the extra radiation and $\TV$ is the visible-photon temperature.  The energy density due to dark radiation is $\rho_{_{\rm D}} = \gD (\pi^2/30) \TD^4$, where $\TD$ is the temperature of the dark plasma, and $\gD$ is the temperature-dependent effective number of dark degrees of freedom, with a fermionic degree of freedom counting as $7/8$ that of a bosonic one (we assume for simplicity that the effective $g$ parameters for energy density and entropy are the same in the dark sector).  Using entropy conservation, we may write
\begin{equation}
\frac{\gV \TV^3}{\gD \TD^3} = \frac{\gVdec}{\gDdec}\ ,
\end{equation}
which relates the temperatures of the two sectors after their decoupling. Requiring $\rho_{_{\rm D}} \!<\!\Delta \rho$ leads to the bound
\begin{equation}
\gDdec <  
\gD^{1/4}\, \pare{ \frac{\gVdec}{\gV}} \frac{4}{11} \left( \frac{7}{4} \right)^{3/4}\,  \left (\Delta N_{\rm eff} \right)^{3/4} 
\: \simeq  \:
18 \left( \frac{\gD}{2} \right)^{1/4} \left( \frac{\gVdec}{106.75} \right) \left( \Delta N_{\rm eff} \right)^{3/4}
\end{equation}
on the size of the dark sector at decoupling from the visible sector.  The effective visible and dark degrees of freedom $g_{_{\rm V,D}}$ are to be evaluated either for the BBN or recombination epochs; in both cases, $\gV \simeq 3.91$~\cite{Kolb:1990vq}.

The systematic errors in the measurements of the primordial light-element abundances currently allow an energy density equivalent of up to about four effective neutrino species during BBN~\cite{Cyburt:2004yc,Mangano:2011ar}.  Taking as a plausible example that the dark radiation consists of just a dark photon, so that $\gD = 2$ at BBN, and that the visible plasma at $T_{\rm dec}$ has all SM species in thermal equilibrium, i.e. $T_{\rm dec} \gtrsim 100 \GeV$, we see that $\gDdec \lesssim 18$ is acceptable as far as successful BBN is concerned.  The full dark sector may contain additional degrees of freedom, but they should already have annihilated or decayed by $T_{\rm dec}$.  While this allows some leeway in the construction of hidden sectors, the constraint is reasonably significant.  The bound becomes looser if additional non-SM particles are admitted into the visible sector at $T_{\rm dec}$, and more stringent as $T_{\rm dec}$ decreases and with it also $\gVdec$. For example, for supersymmetric models and $T_{\rm dec}$ above the SUSY-breaking soft scale, the constraint becomes $\gDdec \lesssim 36$.

There are also very interesting constraints on $N_{\rm eff}$ during the period of (visible sector) recombination from CMB anisotropy measurements, especially from WMAP9~\cite{Hinshaw:2012aka}, ACT~\cite{Sievers:2013ica}, SPT~\cite{Hou:2012xq} and, most recently, Planck~\cite{Ade:2013zuv}.  These results can also be combined with baryon acoustic oscillation (BAO) measurements from the Sloan Digital Sky Survey~\cite{Padmanabhan:2012hf,Anderson:2012sa} and WiggleZ~\cite{Blake:2011wn}.  There has, for some time, been weak evidence for the existence of additional radiation.  For example, WMAP9, at fixed helium abundance and when combined with BAO and other observations, reported $N_{\rm eff} = 3.84 \pm 0.40$ where the error is $1\sigma$~\cite{Hinshaw:2012aka}.  Depending on which other data sets their results are combined with, or not, Planck has found values that range from about $2.7$ to a little over $4$ at $95$\% C.L.~\cite{Ade:2013zuv}.  While there is no strong evidence for additional radiation, current observations still cannot rule out a $\Delta N_{\rm eff}$ that is a substantial fraction of $1$.

Of course, all constraints from additional radiation are irrelevant if the symmetric part of the dark plasma annihilates into SM fields, either directly (which is constrained) or through intermediaries (which is much less restricted), as discussed in Sec.~\ref{sec:DarkInter}.

While the upper limit on $\D N_{\rm eff}$, and its possible non-zero value, obtained from BBN and CMB calculations do not constrain the nature of the extra radiation, it is possible CMB observations can distinguish dark radiation coupled to DM from extra relativistic neutrinos~\cite{Blennow:2012de,Foot:2012ai,CyrRacine:2012fz}. 
This may be the case if dark radiation forms a sizable fraction of the relativistic energy density at CMB, and also decouples from DM late, when Fourier modes relevant for the CMB are inside the horizon. The modes which enter the horizon before the onset of the dark-radiation free-streaming exhibit different behaviour from those which enter the horizon after the DM-dark radiation decoupling. Dark radiation could then produce non-uniform phase shifts and amplitude suppressions of the temperature anisotropy power spectrum~\cite{Foot:2012ai,CyrRacine:2012fz}.

\subsubsection{Structure formation and galactic dynamics}
\label{sec:structure}

The collisionless $\L$CDM paradigm has been extremely successful in reproducing the observed structure of our universe from supercluster scales down to galactic scales. However, there is a growing body of evidence that the rich structure predicted by collisionless CDM is in tension with the observed structure of the universe at galactic and sub-galactic scales. Indeed, simulations predict that galaxies have cuspy density inner profiles, while observations of the dwarf spheroidal galaxies are better fit with cored profiles~\cite{Gilmore:2006iy,Gilmore:2007fy,Gilmore:2008yp,deBlok:2009sp,Walker:2011zu,deNaray:2011hy}. 
Moreover, for a galaxy of the size of the Milky Way, simulations predict a much larger number of satellite galaxies than can be currently accounted for, giving rise to the ``missing satellite problem"~\cite{Klypin:1999uc,Moore:1999nt,Kravtsov:2009gi}. 
While arguments have been made that star formation may have failed in the smallest subhaloes, thus rendering them unobservable and alleviating the problem, recently it has been pointed out that, in addition to the smallest subhaloes, the most massive subhaloes predicted by simulations have no observed counterpart~\cite{BoylanKolchin:2011de}. To explain this with astrophysics, one has to accept that star formation, rather paradoxically, has failed where it is expected to have been most successful, namely in the most massive subhaloes. This is now known as the ``too big to fail" problem~\cite{BoylanKolchin:2011de}.

It is still plausible, albeit not straightforward or unambiguous, that these issues can be resolved within the collisionless CDM scenario by including baryonic physics in simulations and/or by improving observations (see e.g. \cite{MacLow:1998wv,Barkana:1999er,Shapiro:2003gxa,Hoeft:2005jn,
Guo:2009fn,Zwaan:2009dz,Pontzen:2011ty,Hopkins:2011rj,Breddels:2012cq,VeraCiro:2012na}). On the other hand, these discrepancies may be pointing towards a modified DM paradigm. Put differently, current observations can certainly accommodate and in fact favor non-standard DM properties. Of course, a new DM paradigm has to preserve the success of collisionless $\L$CDM in predicting the large-scale structure of the universe, while altering the predicted clustering patterns at small scales in a way that brings them in better agreement with observations. In this section we describe features arising in ADM models which may be able to fulfill this goal.

If DM couples directly to the light (or massless) degrees of freedom into which it annihilates, then in the early universe it must have been kinematically coupled to a thermal bath of dark radiation. If the kinetic decoupling of DM from dark radiation occurred late, it may have affected the growth of matter-density perturbations at small (i.e.\ dwarf-galaxy) scales. 
The kinetic decoupling of ADM from dark radiation may occur later than the kinetic decoupling of symmetric WIMP DM, since ADM must annihilate more efficiently
and is thus expected to interact more strongly with radiation.\footnote{However, if the temperature of the dark plasma is significantly lower than that of the SM particles at the time of the DM kinetic decoupling, the latter may occur at an earlier time than what is estimated for symmetric WIMP kinetic decoupling from SM particles. }
Similarly to what happens with ordinary matter and radiation, during the decoupling epoch both the damping of dark baryon acoustic oscillations~\cite{Loeb:2005pm,Bertschinger:2006nq} and dark-radiation diffusion (Silk damping~\cite{Silk:1967kq}) can reduce the amplitude of sub-horizon perturbations. This, in turn, suppresses the formation of structure at scales smaller than the damping horizon\footnote{The suppression of structure at small scales can also arise due to free-streaming of the DM particles if DM is warm~\cite{Bode:2000gq,Dalcanton:2000hn,Boyanovsky:2010pw}, e.g.\ in the form of sterile neutrinos~\cite{Dodelson:1993je,Shi:1998km,Kusenko:2006rh,Petraki:2007gq,Petraki:2008ef,
Kusenko:2009up,Wu:2009yr,Shoemaker:2010fg,Kusenko:2010ik,Canetti:2012vf,Canetti:2012kh,Merle:2013gea,Drewes:2013gca}. }~\cite{Foot:2012ai,CyrRacine:2012fz,Feng:2009mn}.  An ADM model that does not feature a dark radiation bath can still have suppression of structure below dwarf-galaxy scales provided there are interactions between the DM and SM states that cause the DM to kinetically decouple at an appropriate, sufficiently late time.  In this case, the suppression mechanism is solely acoustic oscillation damping.\footnote{For bounds on the protohalo minimum mass 
for the cases that the kinetic decoupling of DM is determined by its interactions with SM particles via effective operators, see Ref.~\refcite{Shoemaker:2013tda}. In the mass range $m\DM \sim (10-500) \MeV$, it is possible that the kinetic decoupling of DM from SM particles happens sufficiently late to affect the small-scale structure of the universe provided that DM couples preferentially to the left-handed neutrinos~\cite{Shoemaker:2013tda}. If the DM-neutrino interaction is not due to an effective operator, but is mediated by light vector bosons~\refcite{Aarssen:2012fx}, additional constraints apply~\cite{Laha:2013xua}. It should be noted that having a new coupling to left-handed neutrinos but not left-handed charged leptons violates weak-isospin gauge invariance, and must be considered a purely phenomenological prescription.}

Moreover, the direct coupling of DM to light degrees of freedom implies that DM may possess self-interactions sufficiently strong to affect the dynamics of galaxies. If DM interacts significantly inside haloes, through either short or long-range interactions, then the energy transfer among DM particles heats up the low-energy material and reduces the central densities~\cite{Spergel:1999mh}. However, this also tends to isotropize the DM haloes. Preserving the observed triaxiality of DM haloes yields in fact the most severe constraints on self-interacting DM~\cite{Feng:2009mn,Feng:2009hw}, stronger than the bounds arising from the Bullet Cluster~\cite{Markevitch:2003at} and from elliptical galaxy clusters~\cite{MiraldaEscude:2000qt}. Recent simulations show that for velocity-independent DM self-scattering cross-sections, DM self-interactions can affect the kinematics of haloes without spoiling their triaxiality, for a narrow range of values around
$\s_{_T} / m\DM \approx 0.6 \cm^2/ {\rm g}$~\cite{Rocha:2012jg,Peter:2012jh,Vogelsberger:2012sa,Vogelsberger:2012ku,Zavala:2012us},
where $\s_{_T} \equiv \int d\W \, (d\s_{\x\x}/d\W) \, (1-\cos \th) $ is the momentum transfer cross-section, with $d\s_{\x\x}/d\W$ being the differential DM self-scattering cross-section and $\th$ the scattering angle.
A much broader parameter space is available if the DM self-interaction is long-range.\footnote{Note that ``long range'' is not meant to imply astronomical distances.  For the case of a massless vector mediator -- a dark photon -- Debye screening due to the DM plasma makes the effective range of the interaction $\l_D \approx m\DM v\DM/\sqrt{4\p \aD \r\DM}$, i.e. typically $1 - 10^5 \cm$ depending on parameter choices.  A scalar mediator of any reasonable nonzero mass gives rise to a much shorter range, even if that range is large compared to typical particle physics distance scales.}
In this case, the DM self-interaction cross-section decreases with increasing velocity. As a result, DM self-scattering is more effective in smaller haloes with low velocity dispersion, such as the dwarf galaxies, while it is unimportant in larger galaxies and clusters, which have much higher velocity dispersions.
References~\cite{Vogelsberger:2012sa,Vogelsberger:2012ku} performed simulations for velocity-dependent cross-sections arising in Yukawa interactions via a light mediator~\cite{Feng:2009hw,Loeb:2010gj}. For some benchmark scenarios with roughly $\s_{_T} / m\DM \sim 10 \cm^2 / {\rm g}$ at $v \sim 20 \km / \snd$, they found that the ellipticity of the main halo was retained, while the subhalo inner profiles changed.  In particular, they found that no subhaloes are formed that are more concentrated than what is inferred from the kinematics of the dwarf spheroidals, thus addressing successfully the ``too big to fail" problem. Of course, to delineate the full range of possibilities, more simulations for a wider range of parameters and various types of velocity-dependence of the self-interaction cross-section are needed.\footnote{Another compelling possibility for DM self-interactions which can change the kinematics of haloes while preserving their ellipticity arises if DM is in the form of $Q$-balls. $Q$-balls can coalesce after collision, thus reducing the effective self-interaction rate to a negligible value after a few collisions per particle~\cite{Kusenko:2001vu}.}

As already stated, the ADM hypothesis motivates DM couplings to light degrees of freedom residing in the dark sector. In ADM models, DM particles are frequently assumed to couple to a light scalar, or a light or massless vector boson. 
The $t$-channel exchange of a light boson mediates DM self-scattering inside haloes. 
The same coupling of DM to a light bosonic species provides a $t$-channel DM annihilation mode, which may be responsible for setting the DM relic abundance.
We now describe some aspects of the phenomenology arising from direct couplings of ADM to a scalar or a vector boson.

\paragraph{Yukawa interactions.}
The DM coupling to a light scalar implies attractive DM self-interactions. 
For fermionic DM $\x$ that couples to a scalar $\f$ via
\beq 
{\cal \d L} = g_\x \f \bar{\x} \x  \ ,
\label{eq:phi chi chi}
\eeq
the various parametric regimes for the DM self-scattering have been delineated in Ref.~\refcite{Tulin:2013teo}.
In this setup, the effect of the DM self-interaction inside haloes depends on four parameters: the coupling $\a_\x \equiv g_\x^2/4\p$, the DM mass $m_\x$, the mass of the scalar mediator $m_\f$, and the velocity $v$ of DM in the haloes.
In the Born approximation, valid for $\a_\x m_\x/m_\f \ll 1$, the momentum-transfer cross-section can be computed perturbatively, and is~\cite{Feng:2009hw}
\beq
\frac{\a_\x m_\x}{m_\f} \ll 1 \ : \qquad
\s_{_T} = 
\frac{8 \p \a_\x^2}{m_\x^2 v^4}
\sqpare{
\ln \pare{1+\frac{m_\x^2 v^2}{m_\f^2}} 
- \frac{m_\x^2 v^2/m_\f^2}{1+m_\x^2 v^2/m_\f^2}
}
\  .
\label{eq:Born}
\eeq
Outside the Born approximation, non-perturbative effects become important. In the classical regime, $m_\x v / m_\f \gg 1$, the momentum-transfer cross-section is~\cite{Feng:2009hw}
\beq
\frac{\a_\x m_\x}{m_\f} \gtrsim 1 \, , \
\frac{m_\x v}{m_\f} \gg 1 \ : \quad
\s_{_T} = \left \{
\bal{5}
&\frac{4 \p \b^2}{m_\f^2} \ln \pare{1+ \b^{-1}}         \ ,   &\quad & \b \lesssim 10^{-1} & \\
&\frac{8 \p \b^2}{m_\f^2 \pare{1+ 1.5 \b^{1.65}} }     \ ,  &\quad & 10^{-1} \lesssim \b \lesssim 10^{3} & \\
&\frac{\p}{m_\f^2} \pare{\ln \b + 1 - \frac{1}{2} \ln^{-1} \b}  \ ,  &\quad & \b \gtrsim 10^{3} & \\
\eal
\right.
\label{eq:classical}
\eeq 
where $\b \equiv 2 \a_\x m_\f / m_\x v^2$. Because of the attractive nature of the interaction and the existence of bound states, outside the classical regime, for $v \ll m_\f/m_\x$, the DM self-scattering exhibits resonances. Reference~\cite{Tulin:2013teo} argued that in this limit the momentum-transfer cross-section can be approximated by
\beq
v \ll \frac{m_\f}{m_\x} \lesssim \a_\x  \ : \quad
\s_{_T} = \frac{16 \p}{m_\x^2 v^2} \sin^2 \d \  ,
\label{eq:reson}
\eeq
where
\beq
\d = \arg \sqpare{\frac{i \G\pare{\frac{i m_\x v}{\k m_\f}}}{\G (\l_+) \G(\l_-)}} 
\ , \quad 
\l_{\pm} = 1+ \frac{i m_\x v}{2 \k m_\f} \pm \sqrt{\frac{\a_\x m_\x}{\k m_\f} - \frac{m_\x^2 v^2}{4 \k^2 m_\f^2}} \ , \quad \k \approx 1.6 \ .
\label{eq:reson param}
\eeq

Based on \eqs{eq:Born} -- \eqref{eq:reson param}, the parametric analysis of Ref.~\refcite{Tulin:2013teo} showed that for $m_\x \gtrsim 300 \GeV$, the DM self-scattering cross-section can comfortably lie in the range required to affect the kinematics of dwarf spheroidal galaxies, and also exhibit strong velocity dependance which ensures negligible effect on Milky-Way and galaxy-cluster scales. This regime is described by the classical approximation of \eq{eq:classical}. A more limited parameter space for velocity-dependent self-interaction is available in the resonant regime, for $m_\x \gtrsim 60 \GeV$~\cite{Tulin:2013teo}. For lower DM masses, the Born approximation becomes applicable. Self-scattering cross-sections in the desirable range then imply $m_\x v /m_\f \ll 1$, and the DM self-scattering becomes effectively $v$-independent (c.f. \eq{eq:Born})~\cite{Tulin:2013teo}.

\paragraph{Atomic dark matter.} 
A distinct scenario arises if ADM interacts via a light or massless vector boson. In this case, DM is made up of (at least) two species of particles, so that the net gauge charge carried by one species (due to its asymmetric relic abundance) is compensated by an opposite gauge charge carried by the other species.\footnote{This obviously has to be the case if the gauge symmetry is unbroken.
Even if the gauge symmetry is broken, generating a net gauge charge, to be carried by the DM particles, becomes possible only after the phase transition of the universe to the broken phase. Presumably, this charge is protected at lower energies by an emergent global symmetry.
However, if the vector boson is to be significantly lighter than DM itself, as is the case of interest here, the phase transition occurs when DM is already non-relativistic and its population is strongly Boltzmann suppressed (and possibly already diminished below the observed DM relic density, due to strong annihilations in the absence, up to that point, of an asymmetry). It is then difficult to imagine how a sufficiently large asymmetry is generated in the underabundant population of the to-be DM particles, 
or how an asymmetry is transmitted from light abundant particles (participating in the asymmetry generation) to the much heavier and underabundant DM particles. Furthermore, the violation of the DM global number while DM is highly non-relativistic (necessary for the transmission/generation of the asymmetry), seems incompatible with the essential conservation of the DM global charge at low energies.}
This gives rise to the atomic DM scenario, already mentioned in Sec.~\ref{sec:DarkInter}.
Various aspects of the cosmology of atomic DM, for the case of dark atoms consisting of two fermionic species and bound by an unbroken U(1)$_D$, have been studied in Ref.~\refcite{CyrRacine:2012fz}. These include the process of dark recombination 
\beq
\pD + \eD \leftrightarrow  \HD + \gammaD
\ ,
\label{eq:recombination}
\eeq
the late-time ionization fraction of DM, the thermal decoupling of DM from the dark radiation, and the evolution of the DM density perturbations.
Even richer phenomenology emerges if the dark sector features also a strong force which binds dark particles into heavier states and gives rise to nuclear physics, as is the case, for example, in mirror DM models. 

The cosmology of the atomic DM scenario depends on four parameters: the dark fine-structure constant $\aD$, the mass of the dark bound state $\mD$, the dark proton-electron reduced mass $\muD$, and the ratio of temperatures of the dark and the visible sectors, $\ks~\equiv~\TD / \TV$,  at the time of dark recombination. The binding energy of the dark Hydrogen-like atom is $\ED = \aD^2 \muD/2$, and the mass of the bound state is related to the mass of the dark fermions by $\mD = m(\pD) + m(\eD) - \ED$. 
The residual ionization fraction is estimated to be~\refcite{CyrRacine:2012fz}
\beq
x_{\rm ion} \sim  10^{-6} \, \ks 
\pare{ \frac{ \mD \muD}{\GeV^2 } } 
\pare{\frac{0.1}{\a_{_D}}}^4
\  .
\label{eq:x_ion}
\eeq
This determines the self-interaction of the DM particles inside haloes, which involves atom-atom, atom-ion and ion-ion collisions. The corresponding momentum-transfer cross-sections are\footnote{For more exact expressions, see Ref.~\refcite{CyrRacine:2012fz}.}
\beq
\s_{_T} \pare{\HD \!-\! \HD }  \approx \frac{120 \p}{\aD^2 \muD^2 v^{1/4}} 
\ , \
\s_{_T} \pare{\HD \!- i }  \approx  \frac{240 \p} { \aD^2 \muD^2 } \pare{\frac{m_i}{\mD}}^\frac{1}{2}
, \
\s_{_T} \pare{i-j}  \approx  \frac{2 \p \aD^2} {  \m_{ij}^2 v^4} 
\  , 
\label{eq:sigma atomic DM}
\eeq
where $m_i = m(\pD) \text{ or } m(\eD)$ is the mass of the ion, and $\m_{ij}$ is the reduced mass of the $i-j$ ion pair.

Depending on the strength of the dark force, there are various regimes in the atomic DM scenario, which result in different phenomenology~\cite{CyrRacine:2012fz}:
\benu[(i)]
\item 
If the dark fine-structure constant is relatively large, $\aD \gtrsim 0.1$, the resulting cosmology resembles most closely the collisionless CDM scenario. Dark recombination happens in thermodynamic equilibrium and can be described by the Saha equation.  
Since the binding energy of the dark atoms is large, dark recombination and kinetic decoupling occur early, and the matter power spectrum differs from standard CDM only at very small (and unobservable) comoving scales.
The residual ionization fraction is small (c.f. \eq{eq:x_ion}). The DM self-interaction inside haloes is dominated by atom-atom collisions, whose cross-section is rather insensitive to $v$ (c.f. \eq{eq:sigma atomic DM}). Requiring $\s/m\DM \lesssim 1 \cm^2 / {\rm g}$ in order to preserve the observed ellipticity of haloes implies,
\beq
\aD \gtrsim  0.3 
\pare{\frac{10\GeV}{ \mD}}^{1/2} 
\pare{\frac{\GeV}{\muD}}
\  ,
\label{eq:alphaD strong}
\eeq
where parameter values which are close to satisfying the equality in \eq{eq:alphaD strong} could potentially  resolve the small-scale structure problems of collisionless CDM. 
Because of the large value of $\aD$, the atomic energy splittings are large and the collisions of dark atoms in the halo are not energetic enough to excite them. As a result, DM is not dissipative.

\item
For intermediate values of the fine structure constant, the recombination rate is comparable or faster than the Hubble expansion, and the recombination process depends on the details of the atomic transitions. 
Dark acoustic oscillations can imprint a new scale in the matter power spectrum, which determines the minimum DM protohalo mass. A significant residual ionization fraction may be present today, given roughly by \eq{eq:x_ion}. Rutherford scattering of the ionized component, with the characteristic velocity-dependence of the cross-section, $\s \propto v^{-4}$ (c.f. \eq{eq:sigma atomic DM}), can potentially alter the halo kinematics, resulting in subhaloes with reduced central density, without affecting the ellipticity of the main halo.

\item For very small $\aD$ and/or large dark proton and dark electron masses, the recombination rate is lower than the Hubble rate. The dark sector remains mostly ionized. In fact, dark atoms do not form if
\beq
\aD \lesssim  10^{-4} \, \ks
\pare{\frac{\mD}{\GeV}}^{1/4}  
\pare{\frac{\muD}{\keV}}^{1/4}  
\  .
\label{eq:no atoms}
\eeq

\eenu

\paragraph{Dissipative dark matter.}
The possibility of ionized dark atoms leads to another interesting aspect of ADM models with dark radiation: the DM may be dissipative. Dark atoms in the haloes today may be ionized either because dark recombination in the early universe was inefficient, as described above, or because (a portion of) the bound dark atoms formed in the early universe got reionized inside the haloes at late times. Reionization can occur via collisional excitation. However, requiring the DM self-interactions to be sufficiently rare to preserve the ellipticity of haloes, implies also that collisional excitation is not significant~\cite{CyrRacine:2012fz}.
Reionization can also occur for reasons similar to those thought to have caused the reionization of ordinary atoms in galaxies: if the dark baryons interact via a strong force which can bind them in nuclei, DM can form stars supported by nuclear burning and emitting dark radiation which can reionize the dark halo. This is indeed expected in the mirror DM scenario.
   
For ionized dark atoms, dissipation due to the bremsstrahlung of dark radiation will happen (there can be other dissipation mechanisms for other forms of ADM).  This is generically a problem, because the required spheroidal DM haloes around spiral galaxies will collapse into disks on relatively short time scales, unless a heating mechanism exists to counteract the dissipative cooling and maintain a pressure-supported spheroidal halo.  This problem has been studied in the context of mirror DM, where ordinary core-collapse supernovae (SN) have been argued to be energetically capable of supplying the right heating rate~\cite{Foot:2004wz}.  The gravitational binding energy released during the core collapse, which in the standard case goes almost entirely into neutrinos, is partially emitted through mirror photons, and subsequently mirror electrons and positrons, produced via kinetic mixing (see Eq.~\eqref{eq:kinetic mixing}).  For kinetic mixing strength $\epsilon \sim 10^{-9}$, about half of the binding energy is in fact released in the form of these mirror particles.  An $\epsilon$ of that magnitude is favored for explaining the DAMA, CoGeNT and CRESST direct detection signals~\cite{Foot:2013msa,Foot:2012cs}, and is consistent with cosmological requirements (see Sec.~\ref{sec:darkforce}). 

Recent work~\cite{Foot:2013vna,Foot:2013uxa} has argued that the SN heating mechanism for mirror DM, and similar dissipative DM candidates, may provide an explanation for some empirical scaling relations amongst galactic observables (see Refs.~\refcite{Foot:2013vna,Foot:2013uxa} for explicit forms of these relations and citations to the original literature).  From requiring that the galactic cooling rate due to mirror photon emission is balanced by the power injected from SN, the approximate scaling relation,
\begin{equation}
 \rho_0^2\, r_0^3 \propto R_{\rm SN} ,
\label{eq:scaling1}
\end{equation}
was derived, where $R_{\rm SN} \sim 0.03\ {\rm yr}^{-1}$ is the SN rate, $\rho_0$ is the galactic DM central density, and $r_0$ is the DM core radius.  The connection with DM halo properties arises because the cooling rate is proportional to the square of the mirror-electron number density, which in turn is related to the square of the DM density.  Equation~\eqref{eq:scaling1} is derived by using the Burkert cored DM density profile~\cite{Burkert:1995yz,Salucci:2000ps}, $\rho_{_{\rm DM}} = \rho_0 r_0^3/[(r^2 + r_0^2)(r+r_0)]$.  In addition, SN energy transport in the halo has been argued to lead to a second scaling relation,
\begin{equation}
r_0\, \rho_0 \sim {\rm const.}\ ,
\end{equation}
which has direct empirical support.  Combining the two scaling relations leads to $R_{\rm SN} \sim r_0$ which approximately agrees with data within observational uncertainties~\cite{Foot:2013vna,Foot:2013uxa}.

Another idea that involves dissipative DM is the Double-Disk DM proposal~\cite{Fan:2013yva,Fan:2013tia}.  The vision is that the DM haloes of spiral galaxies contain two components: collisionless CDM that forms the required spheroidal halo, plus a dissipative admixture that collapses into a rotationally-supported disk, by analogy with the VM in spiral galaxies.  The fraction of the DM that is dissipative is kept small enough so that the combination largely retains the phenomenological profile of collisionless CDM.  Nevertheless, the amount of mass in the dark disk could be comparable to that in the visible disk, and it could be detected via its gravitational effect on stellar motions.  The dissipative component is composed of partially asymmetric atomic DM (see Sec.~\ref{sec:freezeout}) to allow the possibility of indirect detection.

The existence of dark disks in addition to visible disks has also been invoked~\cite{Foot:2013nea} to explain the observation that a significant proportion of the dwarf satellite galaxies of the neighboring galaxy M31 (Andromeda) form a co-rotating disk~\cite{Ibata:2013rh}, matching a similar but more dilute structure seen for the Milky Way~\cite{Pawlowski:2012vz}.  It has not been established how such structures came about. However, it is known that similar configurations can be formed out of the tidal tails arising during violent galactic collision events (see Ref.~\refcite{Foot:2013nea} for additional discussion and references).  But if this were the case then these dwarf galaxies should be largely free of DM if it were non-dissipative, which is contrary to observations~\cite{Walker:2012td}.  It has been argued that if the tidal tail was, instead, ripped from a gravitationally-merged dark-plus-visible disk of dissipative DM and VM, then one naturally expects dwarf tidal galaxies to contain sufficient DM to match the observations~\cite{Foot:2013nea}.  If the dissipative DM is mirror matter, then subsequent heating by visible-sector SN~\cite{Foot:2004wz} could have expanded the mirror DM that was sufficiently metal rich, and in gaseous form, into the required spherical halo.

\medskip

Clearly, ADM presents a variety of possibilities, ranging from collisionless CDM to interacting and dissipative DM. The question then becomes, how self-interacting should  DM be in order for the anticipated halo dynamics to be in agreement with observations?
Identifying the parameter space for which this occurs, and deriving robust constraints, requires detailed simulations of the halo dynamics in the various DM scenarios, such as the Yukawa-interacting, the atomic and the dissipative DM cases.\footnote{Other solutions to the small-scale structure problems of collisionless CDM which rely on the properties of DM include warm DM~\cite{Bode:2000gq,Dalcanton:2000hn,Boyanovsky:2010pw} and late-decaying DM~\cite{Kaplinghat:2005sy,Abdelqader2008wa,Bell:2010fk,Bell:2010qt,Peter:2010au,Peter:2010jy}.}


\subsection{Direct detection}
\label{sec:directdetection}

Asymmetric DM may interact with nucleons via exchange of new gauge bosons or scalar particles, such as a $Z'_{_{B-L}}$, with $B-L$ being the generalized baryoleptonic symmetry under which both ordinary baryons and DM transform, or a dark photon which kinetically mixes with hypercharge, as per \eq{eq:kinetic mixing}, or a new scalar boson which mixes with the SM Higgs. In this section we shall denote the mediator of the DM-nucleon scattering by $\f$. Such an interaction between DM and the nuclei of the detector can be described by a Yukawa potential
\beq
V(r) = \pare{\frac{m_{_N} \l}{2 \p}}^{1/2} \frac{1}{r} \: e^{-m_\f \, r}
\ ,
\label{eq:YukawaDD}
\eeq
where $m_\f$ is the mass of the mediator and $m_{_N}$ is the mass of the nucleus participating in the collision. For interaction via dark photons, $\l = 2 \p \a_{_{\rm EM}} \aD \e^2 Z^2 Z'^2 / m_{_N}$, where  $\a_{_{\rm EM}}=1/137$ and $\aD$ are the fine structure constants of the ordinary and the dark electromagnetism respectively, $Z, \: Z'$ are the corresponding charges of the nucleus and the DM particle, and $\e$ is the kinetic mixing~\cite{Fornengo:2011sz}.\footnote{If the DM particle has internal structure, as is the case with mirror DM, then an appropriate  form factor that takes into account the finite size of the DM state should be included in  $\l$.} 

Equation~\eqref{eq:YukawaDD} results in the differential cross-section for DM-nucleus scattering~\cite{Fornengo:2011sz},
\beq
\frac{d \s (v, E_R)}{d q^2} = \frac{2 m_{_N} \l}{\pare{q^2 + m_\f^2}^2} \: \frac{1}{v^2} \: F^2(E_R)\ ,
\label{eq:sigmaDD differential}
\eeq
where $E_R$ is the recoil energy of the nucleus (which experiments aim to detect), $q^2 = 2 m_{_N} E_R$ is the momentum transfer in the DM-nucleus interaction, $v$ is the speed of the DM particle and $F(E_R)$ is the nuclear form factor. 
In \eq{eq:sigmaDD differential} we may discern two regimes with very different implications for DM direct detection and for the interpretation of the results of the various direct-detection experiments: 
in the limit $m_\f^2 \gg q^2$, the DM-nucleon interaction is contact-type; if   $m_\f^2 \ll q^2$ (including the case of a massless mediator),  then the DM-nucleon interaction is long-range. In these two regimes,
\beq
\frac{d\s(v, E_R)}{d E_R} =
\left \{
\bal{5}
&\frac{4\l \, m_{_N}^2}{m_\f^4} \, \frac{F^2(E_R)}{v^2}  \ , &
\qquad 
& m_\f^2 \gg 2 m_{_N} E_R : \text{ short-range}&
\\
&\frac{\l}{E_R^2} \, \frac{F^2(E_R)}{v^2}  \ , &
\qquad 
& m_\f^2 \ll 2 m_{_N} E_R : \text{ long-range} \ . &
\eal
\right.
\label{eq:sigmaDD regimes}
\eeq
We shall now discuss both cases.

In the short-range regime, the total scattering cross-section per nucleon is $\s_{n\x} =  Z^{-2} \pare{8 \m_{p\x}^2 m_{_N} \l/ m_\f^4}$, where $\m_{p\x}$ is the proton-DM reduced mass, and 
for simplicity we have considered DM scattering only on protons, as is the case for scattering via a dark photon. (For DM scattering on both neutrons and protons, the appropriate form factors have to be included.)
In this regime, the usual analysis and bounds presented for the WIMP DM scenario apply.  
For typical targets of mass $m_{_N} \sim 100 \GeV$ and nuclear recoil energies around $E_R \sim 10 \keV$, interactions manifest as short-range if $m_\f \gtrsim 50 \MeV$. 
Exchange of $Z'_{_{B-L}}$ yields the spin-independent scattering cross-section
\beq
\s_{_{B-L}}^{\rm SI}   \approx  \pare{10^{-46} \cm^2}  q\DM^2
\pare{ \frac{g_{_{B-L}}}{0.1} }^4 \!
\pare{ \frac{3 \TeV}{M_{_{B-L}}} }^4,
\label{eq:sigmaDD_B-L}
\ 
\eeq
where $q\DM \sim {\cal O}(1)$ is the charge of DM under the generalized $B-L$, $g_{_{B-L}}$ is the gauge coupling and $M_{_{B-L}}$ is the mass of $Z_{_{B-L}}$.
Exchange of a massive dark photon which mixes kinetically with hypercharge gives
\beq
\s_{_D}^{\rm SI}    \approx   \pare{10^{-40} \cm^2} 
\pare{ \frac{\e}{10^{-4}} }^2      
\pare{ \frac{g_{_D}}{0.1} }^2     
\pare{ \frac{1 \GeV}{M_{_D}} }^4 .
\label{eq:sigmaDD_dark}
\eeq
Equations~\eqref{eq:sigmaDD_B-L} and \eqref{eq:sigmaDD_dark} have been evaluated for $m\DM~=~5~\GeV$.
The cross-section of \eq{eq:sigmaDD_dark} can account for the low-mass-region signals favored by DAMA~\cite{Bernabei:2008yi,Bernabei:2010mq}, CoGeNT~\cite{Aalseth:2010vx,Aalseth:2011wp}, CRESST~\cite{Angloher:2011uu} and CDMS~\cite{Agnese:2013rvf}, but it can also vary by a few orders of magnitude and comfortably satisfy the current most stringent limits on DM with short-range interactions from XENON~\cite{Aprile:2011hi,Angle:2011th,Baudis:2012zs,Essig:2012yx}.  
It should be noted that within the interpretation of the direct-detection experiments in terms of short-range DM-nucleon interactions, the compatibility of the low-mass excesses reported by the various experiments is not optimal. Furthermore these signals appear to be in tension with bounds set by XENON.

In the long-range regime, which quite commonly appears in ADM models, a different picture emerges~\cite{Foot:2003iv,Foot:2010hu,Foot:2011pi,Foot:2013msa,Foot:2012cs,Fornengo:2011sz}. The $E_R^{-2}$ dependence of the differential cross-section of \eq{eq:sigmaDD regimes} implies that experiments with low energy thresholds, such as DAMA and CoGeNT, are more sensitive than experiments with higher energy thresholds, such as XENON100.  This has been shown to improve the compatibility of the various experiments, and in any case changes the interpretation of their results -- the signal regions and the bounds -- in terms of the fundamental particle-physics parameters involved~\cite{Foot:2003iv,Foot:2010hu,Foot:2011pi,Foot:2013msa,Foot:2012cs,Fornengo:2011sz}.  In fact, the value of the recoil energy above which the DM-nucleon interaction manifests itself as long-range depends on the mass of the target (if $m_\f \neq 0$), $E_{R, \, {\rm crit}} = m_\f^2/2 m_{_N}$.
Moreover, the interpretation of the direct-detection experiments, and in particular their mutual compatibility, depends on the velocity distribution of DM in the halo\footnote{The local DM density affects of course the overall normalization of the bounds and the reported signal strengths, but does not affect the mutual compatibility of the results from the various experiments.}~\cite{Foot:2003iv,Foot:2010hu,Foot:2011pi,Foot:2013msa,Foot:2012cs,
Fornengo:2011sz,Belli:2002yt,Friedland:2012fa}, which is of course uncertain.
The minimum velocity that can provide recoil energy $E_R$ in the detector is 
$v_{\rm min} (E_R) = (m_{_N} E_R / 2 \m_{_{N\x}}^2)^{1/2}$, where $\m_{_{N\x}}$ is the DM-nucleus reduced mass. 

The interpretation of the direct-detection experiments in terms of DM-nucleon long-range interactions, mediated by a massless dark photon, has been investigated within the mirror DM scenario~\cite{Foot:2003iv,Foot:2010hu,Foot:2011pi,Foot:2013msa}, and for general hidden-sector DM~\cite{Foot:2012cs}. Besides the long-range nature of the DM-nucleon interactions, the mirror DM scenario presents a natural framework in which DM has modified velocity dispersion: it features multi-component DM, with the different atomic species interacting via exchange of dark photons. 
The self-interactions thermalize the various species and result in mass-dependent velocity dispersions 
\beq 
v_i \simeq v_{\rm rot} \pare{\frac{\bar{m} }{ m_i}}^{1/2} 
\ , 
\label{eq:v of m}
\eeq 
where $v_i, \: m_i$ are the velocity dispersion and the mass of the $i$-th species, respectively, and $\bar{m} = \sum_j n_j m_j / \sum_j n_j$, with $n_j$ being the number density of the $j$-th species. 
Collisions of the heaviest species with the nuclei of the detector yield the largest recoil energies. The heaviest states are thus the most easily detectable by experiments, provided that they have significant abundances. These states are the mirror elements with atomic numbers around Fe. We shall collectively denote them by using the index $``h"$. 
Because the heaviest elements also have the smallest velocity dispersions, $v_h \ll v_{\rm rot}$, the tail of their distribution is shorter; this can partially explain why XENON100, having a higher energy threshold, does not see a signal, while the lower-threshold experiments do. This feature is absent in the collisionless CDM scenario, where $v = v_{\rm rot}$, even if DM is multicomponent. The interplay of mass-dependent velocity dispersion and long-range interaction has been shown to produce a good agreement of the DAMA~\cite{Bernabei:2008yi,Bernabei:2010mq}, CoGeNT~\cite{Aalseth:2010vx,Aalseth:2011wp} and CRESST-II~\cite{Angloher:2011uu} data for $v_{\rm rot} \sim 200 \km/\snd$ and $\e \cdot \ks_h^{1/2} \approx 2 \cdot 10^{-10}$, where $\ks_h \equiv m_h n_h/\sum_j m_j n_j$ is the halo mass fraction of species $h$. Other regions of the parameter space are also possible~\cite{Foot:2013msa}. The features leading to this result can appear in the context of a more general hidden sector; indeed, similar conclusions hold for a hidden sector composed of two or more stable species whose interactions reproduce a relation similar to \eq{eq:v of m}, and which interact with nucleons via a massless dark photon that kinetically mixes with the hypercharge gauge boson~\cite{Foot:2012cs}.

Reference~\cite{Fornengo:2011sz} generalized the above analysis (for single species DM) to massive mediators, considering  the transitional regime between the long-range and short-range limits. In this general case, 
$d\s / d E_R \propto (E_R + m_\f^2/2m_{_N})^{-2}$, as can been seen from \eq{eq:sigmaDD differential}. They found preferred values for the mass of the light mediator around $m_\f \sim 10 \MeV$, correlated with DM mass $m\DM \sim 10 \GeV$. They also found that large velocity dispersions $v_{\rm rot} \gtrsim 250 \km \snd^{-1}$ are disfavored.

It should be noted that even within the interpretation of the direct detection experiments involving  long-range DM-nucleon interactions, the excesses observed by the various experiments are in tension with bounds from XENON100~\cite{Aprile:2011hi}. However, the robustness of the XENON limits in this low-mass region is under active discussion~\cite{Bernabei:2008jm,Manzur:2009hp,Collar:2011kf,Collar:2011wq}, and it has been argued that consistency can be achieved when reasonable systematic uncertainties in the energy scale are considered\footnote{A different type of long-range interaction from those described above arises in the case of magnetic DM, and has been studied in the context of DM direct detection (for both symmetric and asymmetric DM) in Ref.~\refcite{DelNobile:2012tx}. Isospin-violating DM, which involves different DM couplings to protons and neutrons, has also been invoked to reconcile the results from direct-detection experiments~\cite{Feng:2011vu,Chang:2010yk,DelNobile:2011yb}. In the context of ADM, it has been studied in Refs.~\refcite{DelNobile:2011je,Okada:2013cba}.}~\cite{Foot:2013msa,Foot:2012cs}.

\bigskip

Self-interacting DM can produce another interesting signature: \emph{diurnal modulation} in DM detectors~\cite{Foot:2011fh}.   If DM interacts with nucleons, it can scatter off the nuclei in the Earth, lose energy and get captured in the Earth's core (see also Sec.~\ref{sec:captureinstars}). If DM self-interactions are stronger than DM-nucleon interactions,  the deceleration and capture of incoming DM particles due to scattering on DM already captured in the Earth dominates quickly over the capture due to scattering on nuclei. Self-scattering greatly enhances the DM capture rate. 
The captured DM sinks in the Earth's core, and because of the self-interactions, it can potentially shield a DM detector from incoming DM particles from the halo, if they originate from a direction which takes them through  (the core of) the Earth~\cite{Foot:2011fh}. 
Since the direction from which DM particles from the halo reach a detector has a daily modulation, the shielding of DM detectors due to DM self-interactions can potentially produce a diurnal modulation in the observed DM signal. Due to the relative orientation of the Earth's axis and the motion of the Earth through the DM halo,
this modulation is expected to be more pronounced for a DM detector located in the southern hemisphere. 
A detector similar to CoGeNT located in the southern hemisphere could potentially observe a statistically significant modulation signal within about 30 days~\cite{Foot:2011fh}.


\subsection{Indirect detection}
\label{sec:IndirectDet}

In the ADM scenario, the annihilation of DM in the present epoch is suppressed, if not completely absent, due to the small or negligible number of dark antiparticles left over from the early universe. For late-time fractional asymmetry $r_\infty$ (where $r$ is defined in \eq{eq:r}), the annihilation rate per unit volume is $\G_{\rm ann} /V = \ang{\s v}_{\rm ann} n(\x) n(\bar{\x}) = \ang{\s v}_{\rm ann} n\DM^2 r_\infty / (1+r_\infty)^2$, where the number density of DM includes both particles and antiparticles, $n\DM = n(\x) + n(\bar{\x})$. As always, $r_\infty = 1$ corresponds to the symmetric case.\footnote{Note, however, that in the case of Majorana fermion or real scalar DM, the annihilation rate is larger by a factor of 4 because particles are the same as antiparticles, and $n\DM = n(\x) = n(\bar{\x})$.}
The expected annihilation signals in the ADM case are suppressed with respect to the case of symmetric DM by~\cite{Graesser:2011wi}
\beq
\frac{\G_{\rm ADM}}{\G_{\rm SDM}} = 
\frac{\s_{_0} }{\s_{_\text{0,WIMP}}}  \frac{4 \, r_\infty}{(1+r_\infty)^2}
\quad \xrightarrow{r_\infty \ll 1} \quad
\frac{4 \s_{_0} }{\s_{_\text{0,WIMP}}} 
\exp \sqpare{ \frac{-2 \s_{_0}}{\s_{_\text{0,WIMP}}} }
\ ,
\label{eq:annihilation suppression}
\eeq
where in the last step we used \eq{eq:FractionalAsymmetry}. This is of course lower than 1 for any $\s_{_0} > \s_{_\text{0,WIMP}}/2$, and yields a rather severe (exponential) suppression whenever \eq{eq:sigma ann} is satisfied.\footnote{This also implies that ADM easily satisfies constraints on the annihilation of light DM into electromagnetically charged particles during CMB~\cite{Lin:2011gj}.}
However, in non-standard cosmological scenarios, the pre-BBN expansion of the universe is faster, freeze-out occurs earlier, and the late-time fractional asymmetry for a given annihilation cross-section, is larger than what is given by \eq{eq:FractionalAsymmetry}. The correct DM abundance is obtained for $\ang{\s v}_{\rm ann} \gg \ang{\s v}_{\rm ann}^{\rm WIMP}$. For $r_\infty > 0$ (including of course the completely symmetric case $r_\infty = 1$), such scenarios yield annihilation signals which can be even stronger than in the WIMP scenario, due to the larger annihilation cross-section~\cite{Gelmini:2013awa}.

Asymmetric DM can co-annihilate with ordinary matter. As described in detail in previous sections, relating the visible and the dark baryonic asymmetries relies on interactions which violate a non-trivial linear combination $X$ of $\BLV$ and $\BD$.  If perturbative, such interactions generate at low energies effective operators of the form (c.f. \eq{eq:sharing}) 
\beq
{\cal L}_{\not{X}, \, {\rm eff}} 
= {\cal O} (\text{SM},q_{\rm _V}) \, {\cal O} (\text{DS},q_{\rm _D}) \ . 
\label{eq:Leff Xviol}
\eeq
Here ${\cal O} (\text{SM},q_{\rm _V})$ is an operator involving SM fields that is invariant under the SM gauge group and carries charge $q_{\rm _V}$ under $\BLV$, and
${\cal O} (\text{DS},q_{\rm _D})$ is an operator involving dark-sector fields, among them the DM field itself, that is invariant under the dark-sector gauge group and carries charge $q_{\rm _D}$ under $\BD$. Specific examples of such operators have been given in Sec.~\ref{sec:models}. ${\cal L}_{\not{X}, \, {\rm eff}}$ can induce co-annihilations of DM with either SM baryons or leptons. If $q_{\rm _D} = q\DM$ and $q_{\rm _V} = +1(-1)$ for baryonic (leptonic) coupling, then one DM particle can co-annihilate with one SM baryon (lepton) into radiation. 

Co-annihilations of DM with ordinary matter can occur in galaxies, inside stars, and on the Earth. If the coupling of DM to VM in \eq{eq:Leff Xviol} is baryonic, their co-annihilation can be observed as \emph{induced nucleon decay}~\cite{Davoudiasl:2011fj} in terrestrial nucleon-decay experiments. 
Induced nucleon decay is expected to produce detectable mesons of higher energy than those expected in spontaneous proton decay. Because of the different kinematics, current limits from nucleon-decay experiments do not apply~\cite{Davoudiasl:2011fj}.
While DM co-annihilation with ordinary matter in the galaxy has not been extensively studied, the large viable range for the strength of ${\cal L}_{\not{X}, \, {\rm eff}}$ might potentially yield interesting observable signals and bounds.

If $q_{\rm _D} = q\DM$ in \eq{eq:Leff Xviol}, DM can decay in SM particles, provided that this is kinematically allowed. Decays of ADM resulting from this operator produce asymmetric amounts of SM particles and antiparticles. For example, if $q_{\rm _V} < 0$ and ${\cal O} (\text{SM},q_{\rm _V})$ is baryonic, or $q_{\rm _V} > 0$ and ${\cal O} (\text{SM},q_{\rm _V})$ is leptonic, ADM decay produces a larger number of SM fermions than anti-fermions. 
If the DM couplings to SM particles in \eq{eq:Leff Xviol} are flavor-dependent, the DM decay products will exhibit an energy-dependent charge asymmetry~\cite{Chang:2011xn,Masina:2011hu}. Overall charge neutrality is of course mandatory, but is ensured by oppositely-charged decay products produced with different energies, as for example in the decay chain $\x \to \ell_i^- \,  W^+  \to \ell_i^- \, \ell_j^+ \, \n$.~\footnote{This particular example may be in conflict with antiproton constraints from PAMELA~\cite{Chang:2011xn}. For more examples see Refs.~\refcite{Chang:2011xn,Masina:2011hu}.} Flavor-dependent couplings are necessary in order for the sum of the decay modes to produce a net charge asymmetry at a given energy~\cite{Chang:2011xn,Masina:2011hu}.
Energy-dependent charge asymmetry can of course be a powerful signature of ADM decay. It has been in fact invoked to explain the tension between the recent AMS-02 measurements of the positron-fraction spectrum and the Fermi-LAT measurements of the total electron+positron flux~\cite{Masina:2013yea,Feng:2013vva}.

Another exciting possibility for ADM indirect detection arises from bound-state formation in the galactic haloes today~\cite{Pearce:2013ola}. If DM possesses attractive self-interactions, two (or more) DM particles can form a bound state~\cite{Shepherd:2009sa}. This process is invariably accompanied by emission of radiation. 
Bound-state formation can occur in galaxies today, and the radiation emitted may be detectable. Reference~\refcite{Pearce:2013ola} considered the case of scalar ADM, interacting via a light gauge-singlet scalar field which mixes with the SM Higgs. The light singlet scalar mediates an attractive interaction between DM particles, and is responsible for the existence of DM bound states. The formation of a DM bound state is accompanied by emission of this mediator, which subsequently decays into SM particles via its mixing with the SM Higgs. This process may produce an observable gamma-ray signal or positron excess~\cite{Pearce:2013ola}.


\subsection{Capture in stars}
\label{sec:captureinstars}

If DM interacts with nucleons, it can scatter off the nuclei in stars, lose energy and get captured in their interiors. The accretion of DM in stars can have observable consequences. In the case of symmetric DM, the concentration and annihilation of DM inside the Sun can produce detectable neutrino signals (see e.g. Refs.~\refcite{Cirelli:2005gh,Blennow:2007tw,Bell:2011sn,Bell:2012dk}). Besides yielding observational signatures, the annihilation of DM regulates the DM density in the interior of stars, establishing a steady state between capture and annihilation.  
However, if DM is asymmetric, DM particle-antiparticle annihilations are suppressed, and the DM density in the stars keeps increasing over time. The denser accumulation of ADM in stars can change their thermal evolution, or affect them in other ways. Considering these effects can yield potentially observable signatures and bounds on the ADM scenario~\cite{Gelmini:1986zz,Bertone:2007ae,Kouvaris:2007ay,Frandsen:2010yj,Cumberbatch:2010hh,
Taoso:2010tg,Casanellas:2010he,Iocco:2012wk,Zentner:2011wx,Lopes:2012af,Casanellas:2012jp,
Kouvaris:2010jy,Kouvaris:2011gb,Kouvaris:2011fi,McDermott:2011jp,Guver:2012ba,
Kouvaris:2012dz,Bramante:2013hn,Bell:2013xk} (for the specific case of mirror DM, see Ref.~\refcite{Sandin:2008db}).

The accumulation of ADM in stars is described by the equation
\beq
\frac{dN\DM}{dt} = C_{n \x} 
+ \pare{C_{\x\x} - C_{\rm evap} - C_{\rm coann} } N\DM 
- C_{\rm self-ann} \, N\DM^2
\ ,
\label{eq:capture}
\eeq
where $N\DM$ is the number of DM particles in the star.\footnote{We assume that $r_\infty \ll 1$, so that we can ignore DM antiparticles. For $r_\infty$ a substantial fraction of 1, the accumulation of DM is described by a pair of (coupled) differential equations, one for DM particles and one for DM antiparticles.}
$C_{n \x}$ is the capture rate due to DM scattering on nucleons, and
$C_{\x\x} N\DM$ is the capture rate due to DM scattering on the already captured DM particles (self-capture).  
After DM is captured, it thermalizes via collisions with nuclei. 
$C_{\rm evap} N\DM$ stands for the evaporation rate due to the thermal velocities of the DM particles in the interior of the star, but it can be neglected if DM is sufficiently heavy -- how heavy depends on the temperature of the star under consideration. 
$C_{\rm coann} N\DM$ takes into account the possibility of DM co-annihilation with nucleons or leptons, arising from an operator of the form of \eq{eq:Leff Xviol}. $C_{\rm self-ann} N\DM^2$ incorporates the possibility of DM self-annihilation, which may arise if $\BD$ is broken into a $\mathbb{Z}_2$ symmetry by some operator which, however, has to be very suppressed to ensure that the dark baryonic asymmetry of the universe is not washed-out.

While the general solution of \eq{eq:capture} can be easily obtained, it is more illustrative to examine specific cases. In fact, \eq{eq:capture} describes also the capture of self-conjugate DM in stars, such as that appearing in most WIMP-miracle models. We shall use this fact to draw a comparison with ADM.    
For self-conjugate DM with sizable (weak-scale) self-annihilation cross-section, \eq{eq:capture} gives $ N\DM (t) = \sqrt{C_{n\x}/C_{\rm self-ann}} \: \tanh \pare{\sqrt{C_{n\x} C_{\rm self-ann}} \: t}$, where we ignored self-capture and evaporation, which are typically negligible, and DM-nucleon and DM-lepton coannihilations, which are absent. Thus, in this case, the DM concentration inside the star reaches its steady state value, $N\DM (\infty)= \sqrt{C_{n\x}/C_{\rm self-ann}}$, within time $\t \sim 1/\sqrt{C_{n\x} C_{\rm self-ann}}$, which is typically much lower than the lifetime of the star.
On the other hand, for asymmetric DM which does not co-annihilate with ordinary matter and has negligible self-annihilation rate,
\beq
N\DM(t) = \frac{C_{n \x}}{C_{\x\x}} \pare{e^{C_{\x\x} \, t} - 1} \ .
\label{eq:N ADM}
\eeq 
That is, the concentration of DM in the star grows linearly until $t \sim C_{\x\x}^{-1}$; after that point it grows exponentially, until the geometric limit for DM self-capture is reached (see below), thence the DM concentration resumes its linear growth.
Obviously, the density-regulating effect of self-annihilations -- if at all present -- will become important at  much longer time-scales than for self-conjugate DM. 
Co-annihilations of DM with nucleons or leptons, which appear in ADM scenarios, can also cap the DM density in the interior of stars, at the steady state value
$N\DM(\infty) = C_{n\x} / \pare{C_{\rm coann} - C_{\x\x}}$, obtained within time $\t \sim \pare{C_{\rm coann}-C_{\x\x}}^{-1}$.
Below, we describe some effects that arise from the capture of ADM in stars.
For completeness, we first summarize the estimates for the various rates which appear in \eq{eq:capture}.

The DM capture rate due to scattering on nucleons is~\cite{Press:1985ug,Gould:1987ir,Kouvaris:2007ay}
\beq
C_{n\x}  = \sqrt{6 \p} \pare{\frac{\r\DM}{m\DM} } \frac{1}{\bar{v}\DM} 
\: \pare{ \frac{2 G M_\star R_\star}{1-\frac{2G M_\star}{R_\star}}} 
\, f \ ,
\label{eq:C nx}
\eeq
where $\r\DM, \bar{v}\DM$ are the DM density and average velocity in the vicinity of the star, and $m\DM$ is the DM mass. $M_\star$ and $R_\star$ are the mass and radius of the star, respectively. The efficiency factor $f$ takes into account the saturation (geometric limit) of the capture rate at sufficiently large cross-sections, when all incident DM is captured:
\beq
f = \min \sqpare{1, \ \frac{\s_{n\x}}{\s_\text{sat}}} \: , \quad
\s_\text{sat} \equiv  \frac{R_\star^2}{0.45 N_n \, \z}
\eeq
with $N_n$ the number of nucleons in the stellar object. The factor $\z$ takes into account the possible  suppression in the capture rate due to nucleon degeneracy, as is the case in neutron stars, where $\z \approx \min \sqpare{1, \, m\DM / 0.2 \GeV}$.

The capture rate of incident DM particles scattering on DM particles already captured in a star can be estimated according to Ref.~\refcite{Zentner:2009is} to be
\beq
C_{\x\x} \, N\DM = 
 \sqrt{\frac{3}{2}} \pare{\frac{\r\DM}{m\DM} } \frac{\f\DM}{\bar{v}\DM} 
 \pare{ \frac{2 G M_\star}{R_\star}}
 \pare{\frac{\rm{erf} \, \h }{\h}}
 \min \sqpare{ \s_{\x\x} \, N\DM, \p r\DM^2}\, ,
\label{eq:C xx}
\eeq
where $\f\DM = \ang{ v_\text{esc}^2(r) / v_\text{esc}^2(R_\star)}$. The average here 
is over the DM distribution in the star, and the numerical value arises assuming a homogeneous star with the DM distributed within radius $r\DM \ll R_\star$ in the center of the star.
The parameter $\h = 3v_\star^2/2\bar{v}\DM^2$ takes into account the motion of the star in the galaxy, with $v_\star$ being the velocity of the star.
The last factor in \eq{eq:C xx} shows that the DM self-capture saturates when all DM particles incident to the region of the star where the captured DM is concentrated ($r \leqslant r\DM$) are captured. If this geometric limit is reached, the capture rate due to DM self-scattering becomes constant, and drives only linear, rather than exponential, growth of the DM population in the star.

The DM particles captured in the star thermalize via their collisions with nuclei. The thermalized DM is concentrated within radius $r\DM$, which can be estimated from the virial theorem in the harmonic gravitational potential in the interior of the star~\cite{Griest:1986yu}, $GM_\star m\DM r\DM^2 / 2 R_\star^3 = m\DM v\DM^2 / 2 = 3T_\star/2$, or
\beq
r\DM = \pare{\frac{9 T_\star}{4 \pi G \r_\star m\DM}}^{1/2}  
\: ,
\label{eq:rDM}
\eeq
where $T_\star$ is the temperature in the core of the star.
If $m\DM$ is very small, $r\DM$ may exceed the radius of the star, and DM evaporation becomes important. 
Detailed studies show that the evaporation rate, $C_{\rm evap}$, decreases exponentially with the particle mass, hence there is a mass threshold above which evaporation can be safely ignored. This depends on the scattering cross-section: while scatterings can impart large velocities to the DM particles which will allow them to escape, a very large scattering cross-section also means that the multiple scattering events cause the DM particles to lose their velocities quickly~\cite{Gould:1987ju,Gould:1990}.

The DM  self-annihilation coefficient is
\beq
C_{\rm self-ann} =  \frac{\ang{\s v}_{\rm self-ann} }{ \cal V}  \  ,
\label{eq:Cann}
\eeq
where $\ang{\s v}_{\rm self-ann}$ is the self-annihilation cross-section, and ${\cal V} \approx 4 \p r\DM^3/3$ is the volume that DM occupies in the star. Note that for a successful ADM scenario, it is necessary that 
$\ang{\s v}_{\rm self-ann} \lll \ang{\s v}_{\rm WIMP}$, to ensure that the dark baryonic asymmetry is not washed-out.
Even so, self-annihilations may have a sizable effect inside stars, when the DM concentration is very high.

If DM can coannihilate with nucleons or leptons (see discussion in Sec.~\ref{sec:IndirectDet}), the co-annihilation rate per DM particle in the star is
\beq
C_{\rm coann}   = 
\ang{\s v}_{\rm coann}  \: n_{n,\ell}  
\label{eq:C coann} 
\ ,
\eeq
where $\ang{\s v}_{\rm coann}$ is the DM-nucleon or DM-lepton co-annihilation cross-section times relative velocity, averaged over the (thermal) distribution of DM and nucleons or leptons in the neutron star. 
Here $n_{n,\ell}$ are the ordinary baryon and lepton number densities in the interior of the star. 

\subsubsection{Effect on the Sun and other main-sequence stars}

The accretion of ADM in the Sun can potentially alter helioseismology and the low-energy neutrino fluxes.
Although ADM does not annihilate, the energy transport due to the scattering of DM particles off nuclei can affect the thermal conductivity, the sound speed, the depth and the helium abundance of the  convection zone, and the oscillation modes of the Sun. Thermalized DM is localized roughly within radius $r\DM$, given in \eq{eq:rDM}. However, the mean free path of the DM particles can be larger, $l\DM > r\DM$, thus resulting in  non-local energy transport from the innermost part of the Sun to its outer regions~\cite{Gelmini:1986zz,Frandsen:2010yj,Cumberbatch:2010hh,Taoso:2010tg,Lopes:2012af}.

The non-local energy transport caused by ADM in the Sun may alter the helioseismological predictions. In fact, the recent revision of the solar composition~\cite{Asplund:2004eu,Asplund:2009fu} implies that the standard solar model is not in agreement with helioseismology data~\cite{PenaGaray:2008qe}. 
It was proposed that the modified solar properties due to the presence of ADM in the Sun
can reconcile the two~\refcite{Frandsen:2010yj}.
However, detailed simulations have shown that the presence of ADM in the Sun does not modify the solar properties in a way that can bring them in agreement with helioseismological data~\cite{Taoso:2010tg}. This is true for both spin-independent and spin-dependent DM-nucleon scattering, with the latter being less constrained by direct-detection experiments and allowed to occur at larger cross-sections~\cite{Cumberbatch:2010hh,Taoso:2010tg}.

A more promising probe of the DM effect on the Sun is the solar neutrino fluxes which are very sensitive to the variation of the temperature and the density profile of the inner regions of the Sun. Reference~\refcite{Taoso:2010tg} found that fermionic ADM in the mass range $m\DM \sim (5 - 20) \GeV$, having short-range spin-dependent interactions with nucleons at cross-sections $\s_{n\x}^{\rm SD} \sim (10^{-36} - 10^{-33})\, \cm^2$, alters the $^8 {\rm B}$ neutrino flux by more than $25\% $, and is thus in tension with present data.\footnote{Particles with similar properties had been, in fact, proposed in the past as a possible solution to the solar neutrino problem~\cite{Gelmini:1986zz}, which is now known to be due to neutrino oscillations.}$^,$\footnote{Particles with mass lower than 5~GeV would evaporate from the Sun, while particles heavier than 20~GeV would get concentrated in a very small region in the center of the Sun (c.f. \eq{eq:rDM}) and would not affect the  energy transport. $\s_{n\x} \lesssim 10^{-36} \cm^2$ implies infrequent scatterings and inefficient heat transport. High cross-sections, $\s_{n\x} \gtrsim 10^{-33} \cm^2$, imply frequent DM-nucleon scatterings which result in local heat transport that does not affect the neutrino fluxes.}
Reference~\refcite{Lopes:2012af} examined (partially) asymmetric DM with short-range spin-independent DM-nucleon interaction at scattering cross-section $\s_{n\x}^{\rm SI} \sim 10^{-36} \cm^2$. It allowed for a range of dark-baryonic asymmetries, assuming always a DM annihilation cross-section such that the combination of DM asymmetry and mass yields the correct DM abundance (see Refs.~\refcite{Graesser:2011wi,Iminniyaz:2011yp}). It found that the range $m\DM \sim (5-15)\, \GeV$ is in disagreement with current neutrino flux measurements, for dark-baryonic charge-to-entropy ratio in the range $\h\DM \sim 10^{-12}-10^{-10}$.

\medskip

The ADM-driven energy-transport effects are expected to be more pronounced in solar-mass stars located in regions of higher DM density than the Sun~\cite{Iocco:2012wk}, and in lower-mass stars~\cite{Zentner:2011wx}. Reference~\cite{Iocco:2012wk} found that solar-mass stars in DM densities $\r\DM \gtrsim 10^2 \GeV/\cm^3$ are sensitive to short-range spin-dependent DM-nucleon interactions with cross-section $\s_{n\x}^{\rm SD} \gtrsim 10^{-37} \cm^2$, for DM masses as low as $m\DM \sim 5 \GeV$. Even smaller DM-nucleon cross-sections can affect stars at DM-denser environments. The expected observable implication is deviations from the standard path in the temperature-luminosity plane of the Hertzsprung-Russell diagram~\cite{Iocco:2012wk}.

Very low mass stars are quite sensitive to changes in the energy transport in their interior, due to their low luminosity (which scales as $L_\star \propto M_\star^3$) and their low core temperature. A relatively small energy transport is thus sufficient to alter the stellar evolution~\cite{Zentner:2011wx}. In particular, increased cooling of the stellar core implies that a collapsing low-mass gas cloud may not achieve sufficiently high core temperature to ignite Hydrogen burning at the level that can halt gravitational collapse. This suggests an increased minimum stellar mass for hydrogen burning, and consequently a deficit of very low mass stars with respect to the standard picture without ADM. In fact,  ADM with mass $4\GeV \lesssim m\DM \lesssim 10 \GeV$ and $\s_{n\x}^{\rm SD} \sim 10^{-37} \cm^2$ or $\s_{n\x}^{\rm SI} \sim 10^{-40} \cm^2$ increases the minimum stellar mass for main-sequence hydrogen burning significanlty~\cite{Zentner:2011wx}. While current observations of low-mass stars are insufficient for the purpose of constraining ADM models based on the above considerations, future observations could allow for this possibility~\cite{Zentner:2011wx}.

Reference~\cite{Casanellas:2012jp} studied the effect of ADM capture in main-sequence stars at galactic regions with local DM density similar to that of the solar neighborhood. It was found that the non-local energy transport caused by ADM particles can change the stars' central temperatures. The reduced temperature gradients, in turn, suppressed the stars' convective cores. 
Such changes can modify the oscillation frequencies and frequency seperations of stars, which are constrained from precision asteroseismology~\cite{Casanellas:2010he}. 
It was found that ADM with mass in the interval $4 \GeV \lesssim m\DM \lesssim 14 \GeV$ and with short-range spin-dependent DM-nucleon scattering cross-sections in the range $3 \cdot 10^{-36}\cm^2 \lesssim \s_{n\x} \lesssim 10^{-33} \cm^2$ (with the exact range depending on the DM mass) is in tension with the asteroseismic analysis of the star $\a$~Cen~B~\cite{Casanellas:2012jp}.

\medskip

Note that the bounds discussed in this section are not necessarily applicable to ADM with long-range interactions, whose strength increases with decreasing velocity. In this case, the mean free path of the captured DM particles decreases as they lose more of their energy. It is then reasonable to anticipate that the energy transport within the star due to DM scatterings is local and does not alter the temperature profile of the star. Moreover, DM self-interaction, if sufficiently strong, can potentially reduce the mean free path of the DM particles in the star, and result in local energy transport. In this case, the constraints are also relaxed.

\subsubsection{Fermionic ADM in compact stars}
\label{sec:fermionicADMinStars}

If asymmetric DM is captured efficiently in compact objects, such as white dwarfs and neutron stars, it can accumulate over time, condense collectively, and eventually reach the critical density for gravitational collapse. A mini black hole formed in the center of the star can potentially consume it. The observation of old neutron stars can thus constrain the properies of ADM~\cite{Goldman:1989nd}.

Fermionic matter in its ground state is supported against gravitational collapse by the Fermi degeneracy pressure. Collapse occurs if the gravitational attraction dominates, i.e.\ when
\beq
\frac{G N m^2}{r}  > k_F = \pare{\frac{3 \p^2 N}{\cal V}}^{1/3} = \pare{\frac{9 \p}{4}}^{1/3} \frac{N^{1/3}}{r}
\ , 
\eeq
where $N$ is the total number of particles, ${\cal V}$ is the volume they occupy, $m$ is the particle mass, and $k_F$ is the Fermi momentum. This yields the Chandrashekhar limit for fermions,
\beq
N_{\rm Cha}^f \approx \pare{\frac{\mp}{m}}^3 \ .
\label{eq:N Cha F}
\eeq
If the number of DM particles in the star exceeds the above limit,  $N\DM \gtrsim N_{\rm Cha}^f$, then gravitational collapse follows, and a black hole forms.\footnote{A black hole forms unless some other stabilising mechanism takes over. For example if the DM particles are not fundamental, then the degeneracy pressure or zero-point energy of the constituent (lighter) particles can withhold gravitational collapse until the new Chandrashekhar limit is reached.} Equation~\eqref{eq:N Cha F} assumes no self-interactions; if DM possesses self-interactions,\footnote{Note that DM self-interactions are expected to arise due to the DM-nucleon interaction -- necessary for the DM capture in the star -- at least at one extra loop order~\cite{Bell:2013xk}. For more details, see Sec.~\ref{sec:bosonicADMinStars}. }  the above limit is modified. Repulsive (attractive) self-interactions would tend to increase (decrease) $N_{\rm Cha}$.

Reference~\refcite{Kouvaris:2010jy} considered fermionic ADM with short-range spin-dependent interactions with nucleons and no (repulsive) self-interactions. In this case, the DM captured in a star is simply $N\DM \approx C_{n\x} \: \t_\star$,
where $C_{n\x}$ is given in \eq{eq:C nx} and $\t_\star$ is the lifetime of the star. Using the above, Ref.~\refcite{Kouvaris:2010jy} placed constraints on $\s_{n\x}^{\rm SD}$ from observations of old neutron stars in the globular clusters, assuming local DM density $\r\DM \sim(10^3 - 10^4) \GeV \cm^{-3}$.
Both the DM capture in the neutron-star progenitor and the neutron star itself were taken into account.
While the progenitor lives for a much shorter time than the neutron star, its geometrical cross-section is much larger, and in the case of spin-dependent DM-nucleon interactions it accretes more DM than the neutron star.
For DM masses $m\DM \gtrsim \TeV$, the constraints of Ref.~\refcite{Kouvaris:2010jy} are competitive with the ones from direct-detection experiments.

Reference~\refcite{Kouvaris:2011gb} considered fermionic ADM with attractive self-interactions described by the Yukawa potential 
\beq
V(r) = - \frac{\a}{r} \exp(-\m r)  \  .
\label{eq:Yukawa}
\eeq
The attractive self-interactions lower the number of particles necessary for gravitational collapse. By considering the nearby old pulsars J0437-4715 and J2124-3358, with estimated ages $\t_\star \approx 6.7 \Gyr$ and $7.8 \Gyr$ respectively, and assuming DM-nucleon cross-section $\s_{n\x} \gtrsim 10^{-48} \cm^2$, Ref.~\refcite{Kouvaris:2011gb} derived constraints on the parameters $\a, \m$. 
For example, for $\a = 0.01$, the excluded range of values for $\m$ is 
\beq
15 \MeV \pare{\frac{m\DM}{100 \GeV}}^{2/5} \lesssim \ \m \ \lesssim 170 \MeV \pare{\frac{m\DM}{100 \GeV}}^{2/5}\ ,
\eeq 
with this range expanding for larger $\a$ and shrinking for smaller $\a$.
These constraints extend beyond those from the Bullet Cluster.

\subsubsection{Bosonic ADM in neutron stars}
\label{sec:bosonicADMinStars}

The gravitational equilibria of bosonic matter differ significantly from those of fermionic matter, thus yielding different bounds on bosonic ADM captured in neutron stars.
Bosonic matter in its ground state is supported against gravitational collapse by the zero-point energy. 
Heisenberg's uncertainty principle ensures non-zero ground-state momentum, $p \sim 1/r$, where $r$ is the size of the region in which the particles are confined. When $r> 1/m$, particles are non-relativistic, and their kinetic energy, $E_k \simeq p^2/2m \sim 1/2m r^2$, can always balance the gravitational attraction, $E_g = -GN m^2/r$, by $r$ becoming sufficiently small. As $r$ becomes lower than $1/m$ though, particles become relativistic and their kinetic energy $E_k \simeq p \sim 1/r$ cannot withhold gravitational collapse 
if $N \gtrsim N_{\rm Cha}^b \approx \pare{\mp / m}^2 $~\cite{Ruffini:1969qy}. Evidently, for masses much lower than $\mp$ (which is the case of interest here), the Chandrashekhar limit for non-interacting bosons is much lower than the corresponding limit for fermions. Gravitational collapse can thus occur even if a much smaller amount of bosonic ADM accumulates in a neutron star, thus yielding more stringent bounds~\cite{Kouvaris:2011fi}.
However, $N_{\rm Cha}^b$ is very sensitive to self-interactions. For a repulsive contact-type self-interaction of a scalar field $\x$, described by the potential  $V_{\rm self} = \l_n |\x|^n /n$, with $n \geqslant 4$ and even, and $\l_n \geqslant 0$, 
the maximum number of particles that can exist in gravitational equilibrium is~\cite{Colpi:1986ye,Ho:1999hs} 
\beq 
N_{\rm Cha}^b \approx \frac{2 \mp^2}{\p m^2} \pare{1 +\frac{\l_n}{8 \p n} \frac{\mp^{n-2}}{m^2}}^\frac{1}{n-2}  
\  .
\label{eq:N Cha B inter}
\eeq 
Note that for a renormalizable coupling $\l_4 \sim {\cal O}(1)$, $N_{\rm Cha}^b \!\sim \! N_{\rm Cha}^f$.

In the absence of annihilations and co-annihilations, the number of DM particles captured in the neutron star due to DM-nucleon scattering is
\bea
N_\text{capt} &\approx& 7 \cdot 10^{45} \pare{\frac{\GeV}{m\DM}}
\pare{\frac{\r\DM}{100 \GeV  \cm^{-3} }}
\pare{\frac{100 \, \km/\snd }{\bar{v}\DM}} 
\pare{\frac{\t_\ns}{10\Gyr}}
\times
\nn
\\
&& \min \sqpare{1 \, , \, \frac{\s_{n\x}}{10^{-45} \cm^2} \, , \, \frac{\s_{n\x}}{10^{-45} \cm^2} \, \frac{m\DM}{0.2 \GeV}}
\  .
\label{eq:Ncapt}
\eea
For some range of parameters, this exceeds the critical number for Bose-Einstein condensation, 
\beq
N_{_{\rm BEC}} \simeq 3 \cdot 10^{42} 
\pare{\frac{T_\ns}{10^7 \, {\rm K}}}^3
\  .
\label{eq:N BEC}
\eeq
All DM particles captured in excess of this amount will go to the ground state, forming a condensate with $N_{\rm cond} = N_{\rm capt} - N_{_{\rm BEC}}$. If $N_{\rm cond} > N_{\rm Cha}^b$, gravitational collapse occurs.\footnote{Note that the formation of the condensate is necessary for gravitational collapse to take place~\cite{Kouvaris:2012dz}. If matter has not reached its ground state, the pressure associated with the energy of the excited states counteracts the gravitational attraction. In the presence of stabilizing pressure other than that in the ground state, the Chandrashekhar limit is evidently not relevant.}

The fate of the mini black hole formed is determined by its initial mass. The rate of accretion of surrounding matter onto the black hole increases with its mass. In the hydrodynamic spherical approximation (Bondi regime), the rate of accretion of neutron star matter is $\pare{dM_{\rm BH}/dt}_\ns = \p \r_\ns G^2 M_{\rm BH}^2/c_s^3$~\cite{ShapiroTeukolsky}, where $M_{\rm BH}$ is the mass of the black hole, $\r_\ns \approx 5 \cdot 10^{38} \GeV \cm^{-3}$ is the neutron-star core density and $c_s \approx 0.17$ the speed of sound in its interior. On the other hand, the black hole will emit Hawking radiation at a rate which decreases with its mass: $\pare{dM_{\rm BH}/dt}_{\rm Haw} = - \pare{15360 \p G^2 M_{\rm BH}^2}^{-1}$. Thus, the black hole will grow and consume the star if $\pare{dM_{\rm BH}/dt}_\ns > \pare{dM_{\rm BH}/dt}_{\rm Haw}$, that is if $M_{\rm BH} > M_{\rm Haw} \approx 5 \cdot 10^{36}\GeV$ at its formation.

Collecting everything together, regions of the parameter space for which 
\beq
m\DM \pare{N_{\rm capt} - N_{_{\rm BEC}}} \ > \ m\DM N_{\rm Cha}^b  \ > \  M_{\rm Haw} \ ,
\label{eq:BH forms}
\eeq
are disfavored. Considering observations of old neutron stars in the Milky Way, such as PSR J0437-4715 with mass and lifetime $M_\ns \approx 1.76 M_\odot$ and $\t_\ns \approx 6.7 \Gyr$, estimated core temperature $T_{_{\rm NS, \, core}} \sim 3 \cdot 10^6 \: {\rm K}$, located in the solar neighborhood with DM density and average velocity $\r\DM \simeq 0.3 \GeV \cm^{-3}$ and $v\DM \simeq 220 \km/{\rm s}$, 
the above considerations exclude ADM masses in the range $m\DM = 1\MeV - 16 \GeV$ and $\s_{n\x} \gtrsim 10^{-43}\cm^2$ (with a more narrow mass range excluded at smaller $\s_{n\x}$), assuming vanishing self-interactions~\cite{Kouvaris:2011fi,McDermott:2011jp,Bramante:2013hn,Bell:2013xk}. 

However, the presence of DM self-interactions changes the limits dramatically~\cite{Kouvaris:2011fi,Bramante:2013hn,Bell:2013xk}. 
Importantly, the existence of DM self-interactions is nearly inevitable if DM possesses couplings with other particles which allow it to scatter off nucleons and get captured in stars~\cite{Bell:2013xk}. 
For most DM-nucleon effective operators and their possible ultraviolet completions, DM self-scattering arises in fact at the same loop-order as DM-nucleon scattering, and in all cases DM self-scattering arises at most at one extra loop order. This implies $\s_{\x\x} \sim \s_{n\x}$. At a more fundamental level, for bosonic DM, the renormalizable $|\x|^4$ operator cannot be forbidden by any unitary symmetry transformation of the DM field~\cite{Bramante:2013hn,Bell:2013xk}. In some cases only, it may be absent due to a space-time symmetry such as supersymmetry~\cite{Bell:2013xk}.  

If not disallowed by any symmetry, the $\l_4$ coupling receives contributions both from the bare Lagrangian and the loop corrections involving DM interactions with other fields. Being renormalizable, the $\l_4$ coupling is of course a free parameter, and can be set to vanish. However, this would  amount to detailed cancellation of the various contributions, which would be spoiled by the running of the coupling. The running of the coupling with the field value $\x$ is directly relevant to the DM condensate inside the neutron star, where the field expectation value  is the order parameter of the condensation, and it increases as condensation proceeds~\cite{Bell:2013xk}. Equivalently, loop corrections generate higher-dimensional DM self-interaction operators, $\d V \supset \l_n |\x|^n / n$ with $n > 4$, which contribute to the Chandrashekhar limit (c.f. \eq{eq:N Cha B inter}), and thus affect the bounds on bosonic ADM from black hole formation in neutron stars. 
The DM self-coupling generated due to a DM-nucleon interaction with cross-section $\s_{n\x} \gtrsim 10^{-45} \cm^2$ is expected to be $|\l_{\x\x}|  \equiv 32\p \pare{|\l_n| m\DM^{n-4} / 8 \p n}^{2/(n-2)} \gtrsim 10^{-6}$~\cite{Bell:2013xk}. 
On the other hand, supersymmetric theories typically possess many flat directions in the scalar potential, along which the quartic and possibly higher-order couplings vanish~\cite{Gherghetta:1995dv}. However, even if DM corresponds to such a flat direction,  SUSY-breaking induces DM self-interactions. In the gravity-mediated and gauge-mediated SUSY-breaking scenarios, typical coupling strengths are $|\l_4| \sim m_s^2/\mp^2 \sim 10^{-32} (m_s/\TeV)^2$ and $|\l_4| \sim m_s^2/M_m^2 \sim 10^{-2} \sqpare{(m_s/M_m)/0.1}^2$, respectively, where $m_s$ is the soft mass and $M_m$ the messenger scale~\cite{deGouvea:1997tn,Enqvist:1997si}.

Repulsive self-interactions ($\l_n >0$) increase the critical particle number necessary for gravitational collapse, as per \eq{eq:N Cha B inter}. The critical number of particles can then be accreted only for larger DM masses. The resulting black hole, being larger, evaporates more slowly. The overall effect is that repulsive self-interactions shift the range of excluded masses to higher values.  However, for sufficiently strong repulsive self-interaction, the observed neutron stars cannot have accreted the critical DM mass for gravitational collapse to the present day, and no limits apply. 
In fact, there are no constraints on $\s_{n\x}$ at any mass range, if the self-coupling $\l_{\x\x} \equiv 32\p \sqpare{\l_n m\DM^{n-4} / 8 \p n}^{2/(n-2)}$  is $\l_{\x\x} \gtrsim 10^{-18}$~\cite{Bell:2013xk}. (Even smaller self-couplings are sufficient to guarantee the viability of bosonic ADM at masses lower than $\sim 200 \GeV$.) 
This means that, while in the particular case of supersymmetric DM corresponding to a flat-direction field in a scenario of high-scale SUSY-breaking mediation, the neutron-star constraints may exclude some of the parameter space, in the rest of the cases bosonic ADM remains rather unconstrained from neutron star observations.

Attractive self-interactions ($\l_n < 0$) imply that DM bound states may exist; if so, the ground state of the accreted DM in the neutron star involves the formation of bound states among DM particles. This changes the dynamics of collapse, and has to be taken into account. The bounds derived for non-interacting bosonic ADM are not applicable in this case either.

It should be noted that the presence of self-interactions can enhance the capture rate of DM in the neutron star  (c.f. \eqs{eq:capture} and \eqref{eq:N ADM}), and strengthen the bounds~\cite{Guver:2012ba}. However, in all cases, the total amount of DM captured cannot exceed the estimate for the case $\s_{n\x} \gtrsim \s_{\rm sat}$, when the geometric limit for capture via DM-nucleon scatterings has been reached. Thus the limits cannot be further strengthened. Moreover, self-interaction strengths which enhance the DM capture rate significantly, also increase the Chandrashekhar limit dramatically (c.f. \eq{eq:N Cha B inter}), and lie far beyond what can be excluded by observations of neutron stars.

The constraints on bosonic ADM are relaxed if the dark baryonic symmetry $\BD$ is broken at high energies into a $\mathbb{Z}_2$ symmetry by an operator which allows for DM self-annihilations~\cite{Bramante:2013hn}, or if DM can co-annihilate with nucleons or leptons~\cite{Bell:2013xk}, due to an operator of the form of \eq{eq:Leff Xviol}.

\subsubsection{Admixed stars}

The possibility of stars made of comparable admixtures of dark and ordinary matter has been recently considered~\cite{Goldman:2011aa,Goldman:2013,Li:2012ii,Leung:2013pra}, as a possible consequence of non-annihilating (asymmetric) DM.
If such stars can form, their gravitational equilibria are expected in general to be different from the gravitational equilibria of stars made purely of ordinary matter~\cite{Goldman:2011aa,Goldman:2013,Li:2012ii,Leung:2013pra}.
For example, non-annihilating fermionic DM with mass $m\DM < {\rm GeV}$ can increase the Chandrashekhar limit (c.f. \eq{eq:N Cha F}), and allow for neutron stars heavier than  those predicted by SM physics~\cite{Goldman:2011aa,Goldman:2013,Li:2012ii}. This could potentially explain recent observations of neutron stars with masses as large as $\sim 2 M_\odot$~\cite{Demorest:2010bx}. The existence of so massive neutron stars can be explained by ordinary matter only if its equation of state in the neutron star is very stiff.

However, how stars with significant admixture of dark and ordinary matter can form is very unclear. The accretion of DM from the halo onto stars of ordinary matter results in DM concentrations which are many orders of magnitude below what is needed to affect the gravitational equilibrium of the star, even when the star is at a very DM-dense environment and the DM capture rate is saturated to its geometric limit (see e.g. \eq{eq:Ncapt}). 
The formation of stars by DM itself requires DM to be dissipative, which appears in mirror DM models with unbroken mirror symmetry. However this does not explain the joint clustering of dark and ordinary matter. 
One speculation for the joint DM-VM clustering is that ordinary matter falls into potential wells of non-dissipative but self-interacting DM, interacts with it gravitationally and draws its energy, which it subsequently dissipates~\cite{Goldman:2013}.


\subsection{Collider signatures}
\label{sec:colliders}

\subsubsection{$Z'$ decays to the dark sector}
\label{sec:Zprime}

Asymmetric DM models can feature a U(1) gauge interaction that couples to both the visible and dark sectors.  
The primary example takes the generator of this Abelian gauge symmetry to be the conserved $\BLV - \BD$ charge of baryon-symmetric models.  
The gauged U(1) is spontaneously broken, resulting in a massive $Z'$ boson that has decay channels to both VM and dark-sector particles.\footnote{Since we expect the $Z'$ mass to be at least 100s of GeV and the DM mass to be much lower, the decay of the $Z'$ to DM and anti-DM should be kinematically allowed.} Experimentally, this manifests as a $Z'$ resonance with an invisible width that cannot be accounted for by standard neutrinos.

There are two important questions:  Can the invisible width of the $Z'$ be measured with sufficient accuracy?  And, can experiment distinguish between an invisible width produced solely from decays to neutrinos from one that also includes dark-sector final states?    References \refcite{Petriello:2008pu} and \refcite{Gershtein:2008bf} argue that such measurements are indeed possible at the LHC, based on the processes $pp \to Z Z' \to \ell^+ \ell^- + {\rm missing}\ E_T$ and $pp \to Z Z' \to \gamma + {\rm missing}\ E_T$, respectively.  The ability to discriminate between neutrino-only invisible final states and those that also contain dark matter rests on the determination of the $Z'$-neutrino coupling from Drell-Yan $Z'$ production~\cite{Petriello:2008zr}.  This in turn depends on the (mild) assumption that the $Z'$ couplings obey weak-isospin invariance, so that the $\nu_L$ coupling is the same as the $e_L$ coupling. Their analysis indicates that a $Z'$ with a 1 TeV mass and O(1) gauge couplings can be probed at a 14 TeV LHC with a few $\times$ 10 fb$^{-1}$ of data.

\subsubsection{Monojet signatures}
\label{sec:monojets}

The existence of asymmetry transfer operators of the form of Eq.~\eqref{eq:sharing} implies that dark-sector particles can be searched for at colliders through missing-energy signals, provided that the SM states in the transfer term are appropriate.  For the LHC, the highest sensitivity will be to operators containing up and down quarks (for obvious reasons, gluons are not typically present in these operators).

To be specific, Ref.~\refcite{Davoudiasl:2011fj} considered a neutron-portal effective operator of the form
\begin{equation}
\frac{1}{\Lambda^3}\, \overline{(u_R)^c}\, d_R\, \overline{(d_R)^c}\, \Psi_R\, \Phi + H.c.\ ,
\label{eq:hyloeffop}
\end{equation}
where $\Psi$ and $\Phi$ are a DM fermion and scalar in a certain scheme the authors term ``hylogenesis''.  The connection between VM and the dark sector is accomplished through multiple copies of Dirac fermions $X$; when integrated out, they produce Eq.~\eqref{eq:hyloeffop}.  The interactions are
\begin{equation}
\frac{\lambda}{M^2} \bar{X}_L\, s_R\, \overline{(u_R)^c}\, d_R + \zeta\, \bar{X}^c\, \Psi\, \Phi + H.c.\ ,
\end{equation}
where a certain flavor structure has been specified for the purposes of the collider phenomenology, and $M$ is the scale above which a UV complete theory should be specified.  Through real or virtual $X$ exchange, these interactions produce the process
\begin{equation}
q q' \to \bar{q} \bar{\Psi} \Phi^*\ ,
\end{equation}
leading to a monojet-plus-missing-energy signature at hadron colliders. Their conclusion is that a monojet cross-section sensitivity down to about 7 fb is possible with 100 fb$^{-1}$ of data at a 14 TeV LHC, allowing scales in the range $M = 1-4$ TeV to be explored.

\subsection{Dark force constraints}
\label{sec:darkforce}

As reviewed in Sec.~\ref{sec:DarkInter}, an important way that VM and the dark sector may interact is via kinetic mixing between the hypercharge gauge boson and a hypothetical dark photon, as per Eq.~\eqref{eq:kinetic mixing} with $\epsilon$ the free parameter governing the strength of the mixing.\footnote{In models with no U(1) factors in the gauge group, $\epsilon$ can be induced radiatively and it then becomes a function of other parameters in the theory \cite{Holdom:1985ag}.}  At low energies, this kinetic mixing may be considered as effectively a mixing between the ordinary photon and the dark photon.

Two cases have been considered at length in the literature: (i) photon--mirror-photon mixing in the context of mirror matter, where the mirror photon is massless and its couplings to mirror states are identical in form and strength to the coupling of ordinary photons to ordinary charged particles~\cite{Carlson:1987si,Foot:2000vy,Ciarcelluti:2008qk,Foot:2012ai}, and (ii) a more generic setup where the dark photon has a mass $M_{_D}$, and phenomenological constraints and detection prospects are analysed in terms the two-dimensional parameter space $(\epsilon, M_{_D})$ ~\cite{Williams:1971ms,Bartlett:1988yy,Jaeckel:2008fi,Mirizzi:2009iz,Ruoso:1992nx,Cameron:1993mr,Robilliard:2007bq,Ahlers:2007rd,Chou:2007zzc,Ahlers:2007qf,Afanasev:2008jt,Fouche:2008jk,Afanasev:2008fv,Ehret:2009sq,Andriamonje:2007ew,Redondo:2008aa,Andreas:2012mt,Jaeckel:2012yz,Gninenko:2013sr,An:2013yfc,Jaeckel:2007ch,Gninenko:2008pz,Jaeckel:2008sz,Caspers:2009cj,Bjorken:2009mm,Essig:2010xa,Dent:2012mx} (for a recent compilation of bounds, see Ref.~\refcite{Jaeckel:2013ija}).  We briefly review these in turn.

\paragraph{Mirror photons.}
The most important constraint comes from BBN.  If $\epsilon$ is too large, then kinetic-mixing-induced processes such as $e^+ e^- \to e'^+ e'^-$, where $e'$ is the mirror electron, will bring the mirror sector into thermal equilibrium with the visible sector and spoil BBN~\cite{Carlson:1987si,Ciarcelluti:2008qk}. The critical parameter is the ratio of the mirror- and visible-sector temperatures, $T'/T$, and one may track the thermal production of mirror particles from the visible-sector bath as a function of $\epsilon$.  Note that once enough mirror matter has been produced, the mirror plasma will be in thermal equilibrium with itself through mirror electroweak and other interactions, so will have some temperature $T'$.  The result of the calculation is~\cite{Ciarcelluti:2008qk},
\begin{equation}
\frac{T'}{T} \simeq 0.31 \sqrt{\frac{\epsilon}{10^{-9}}} .
\end{equation}
The BBN constraint is met for $T'/T \lesssim 0.5$, with structure formation favoring $T'/T \lesssim 0.3$.  The upshot is that an $\epsilon$ of about $3 \times 10^{-9}$ or less is required.  An $\epsilon$ in the range $(1-3) \times 10^{-9}$ also leaves an imprint on the CMB temperature-anisotropy spectrum by suppressing the third and higher odd peaks~\cite{Foot:2012ai} (see Ref.~\refcite{Berezhiani:2003wj} for another analysis of mirror DM implications for the CMB).

Terrestrially, photon--mirror-photon kinetic mixing can be probed through orthopositronium lifetime measurements~\cite{Badertscher:2003rk}.

\paragraph{Massive dark photons.}
The constraints arise from different considerations as one scans over the $(\epsilon, M_{_D})$ plane.  In terms of mass range there are constraints at some level from as low as about $M_{_D} \sim 10^{-15}$ eV to as high as the TeV energies being explored by the LHC.  The constraints on $\epsilon$ are weakest at the lowest and highest ends of this mass range, and they are most severe in the 100 eV to 1 MeV interval where upper limits of $10^{-13,-14}$ have been derived.

A very abbreviated review of how the constraints are derived is as follows (see, for example, Ref.~\refcite{Jaeckel:2013ija} for a longer summary):  For extremely low $M_{_D}$, the dark force manifests as a very long-ranged force between ordinary matter with strength suppressed by $\epsilon^2$.  The lowest-mass bounds are obtained by matching the precisely mapped magnetic fields of the Earth and Jupiter to the expectations from pure electromagnetism.  Relatively weak upper bounds of order 0.1 for $\epsilon$ are obtained in the $10^{-15} - 10^{-12}$ eV mass range.  Moving up in mass to the $10^{-12} - 10^{-3}$ eV regime, the most severe constraint comes from testing Coulomb's law through a Cavendish kind of experiment.  At $M_{_D} \sim 10^{-6}$ eV, one has $\epsilon \lesssim 10^{-8}$.  From about $10^{-3}$ eV to 100 keV, the most stringent bounds arise from the anomalous cooling that would take place in the sun and other stellar systems through dark photon emission.  It is these processes that give the $\epsilon \lesssim 10^{-14}, 10^{-13}$ restrictions mentioned above.  Cosmic microwave background measurements are a strong constraint near $M_{_D}$ of 1 MeV.  The limits then become rapidly less severe as the mass is pushed higher.


\section{Conclusion}
\label{sec:conclusion}

The observational fact that $\Omega_{\rm DM}$ and $\Omega_{\rm VM}$ differ by only a factor of five, rather than the default expectation of many orders of magnitude, may be an extremely important clue to the nature of DM, one of the few clues we have.  It suggests that DM and VM are closely related in both microphysics and cosmological history.  The idea of asymmetric DM is motivated by these facts and suppositions.

The simplest possible dark sector suitable for ADM consists of a stable particle and its antiparticle, with one of these constituting the DM today by virtue of an asymmetry in their number densities that developed in the early universe, just as the VM density today is known to be due to the baryon asymmetry.  But just as the proton and antiproton are not the only particles in the visible sector, not even the only stable particles, so might the dark sector consist of a quite complicated set of states and interactions, perhaps including gauge bosons as well as fermions and/or scalars.  The dark sector may share the visible-sector feature of having more than one stable species, and some of these species may be massless or very light and thus constitute dark radiation.  Perhaps the symmetric part of the dark plasma annihilated into dark radiation that will eventually be discovered through BBN and CMB observations, or perhaps it annihilated into SM particles instead, either directly or, more likely, through intermediate states.  At some early epoch in the evolution of the universe, the visible and dark sectors may have been in chemical and thermal contact, which is when the baryon and DM asymmetries became related.  Although not mandatory, most ADM scenarios have a DM mass in the few GeV regime, or a little higher, which strongly suggests that the microphysics behind the origin of mass in the dark sector is related to SM scales such as the QCD or electroweak scales.  A striking example of this is the mirror matter model, where the DM mass scale is exactly the same QCD scale that sets the proton mass.

The (potential) richness of the dark-sector physics suggests also a rich phenomenology: in cosmology, astrophysics, and terrestrial contexts such as direct-detection and accelerator experiments.  In this review, we have surveyed the ways in which DM asymmetries can be produced, how dark sectors can be constructed, and how chemical exchange can relate the VM and DM asymmetries.  A general symmetry framework based on ordinary baryon number and a dark analog was used to systematically think through the possible structures of ADM models.  The phenomenological consequences of these diverse scenarios were examined, both to understand the constraints that must already be applied to ADM model building, but also to explore how ADM might solve some of the persistent problems in DM astrophysics and physics: the difficulties faced by standard collisionless cold DM in accounting for galactic structure on small scales~\cite{Gilmore:2006iy,Gilmore:2007fy,Gilmore:2008yp,deBlok:2009sp,Walker:2011zu,deNaray:2011hy,Klypin:1999uc,Moore:1999nt,Kravtsov:2009gi,BoylanKolchin:2011de}, and the puzzling results from some direct-detection experiments~\cite{Bernabei:2008yi,Bernabei:2010mq,Aalseth:2010vx,Aalseth:2011wp,Angloher:2011uu,Agnese:2013rvf} that hint at the few-GeV DM mass scale, but whose mutual consistency, and consistency with null experiments, is unclear.  

The small-scale structure problem encourages us to ask:  How strongly self-interacting should DM be, and what types of ADM models are then the best motivated?  We can also ask:  Was structure on small scales washed out because of the late kinetic decoupling of ADM from a bath of dark radiation (the same radiation into which the symmetric part annihilated)?  Could both effects be important?  The hints for a positive DM signal from some of the direct-detection experiments suggest other DM properties that the asymmetric paradigm can provide:  DM can be multi-component, with the resulting non-standard velocity dispersion characteristics of galactic DM haloes argued to be an important factor in reconciling the different positive results~\cite{Foot:2003iv,Foot:2012cs,Foot:2013msa}.  Perhaps DM-nucleon scattering is not due to a contact interaction but rather a long-range force associated with a light or massless mediator~\cite{Foot:2003iv,Foot:2012cs,Foot:2013msa,Fornengo:2011sz}?  Precisely this occurs through the kinetic mixing of ordinary photons with dark photons in some ADM models.  More observational and experimental data, together with detailed studies of low-energy phenomenology and astrophysical effects, are needed to identify and quantify the properties of DM that reproduce observations.

Asymmetric DM may be placed along a quasi-continuum of models based on how strongly DM interacts with either VM or itself, taking as a reference point the canonical WIMP self-annihilation cross-section, $\sigma_{_{\rm WIMP}}$, that leads to the correct DM relic density from standard thermal freeze-out.  The ADM world lives at $\sigma > \sigma_{_{\rm WIMP}}$, because without an asymmetry the relic density after freeze-out is much too small.  The opposite situation, $\sigma < \sigma_{_{\rm WIMP}}$, takes us into another broad class of models.  In these cases the DM should not be a relic from a thermal freeze-out process, because the resulting density would then be much too high.  Instead, it must never have been in thermal equilibrium with VM and should have been produced by a process such as freeze-in.  From this perspective, ADM is a natural product of a large class of theories.

Standard collisionless cold DM -- the reference point along the continuum -- remains viable, but it is becoming increasing constrained from lack of direct production in accelerators, from the absence of indirect signatures from astrophysical sources, and from the fact that the direct-detection experiments reporting null results were designed to probe the favored WIMP parameter region~\cite{Bai:2010hh,Buckley:2011kk,Fox:2012ee}.  If the experiments indicating a few-GeV DM mass are broadly correct, then this fact does not sit well with the standard WIMP paradigm (claiming anything stronger is premature).  It also faces the small-scale structure challenge, though that may yet be resolved through standard astrophysics.  But most of all, the standard thermal-WIMP scenario has to accept $\Omega_{\rm DM} \simeq 5 \Omega_{\rm VM}$ as a coincidence.  These remarks serve to justify serious consideration being given to alternatives.  Asymmetric DM could emerge as the new paradigm, to be studied in parallel with other interesting frameworks such as sterile neutrinos, axions and Q-balls.

\section*{Acknowledgments}

We thank I. Baldes, N. Bell and R. Foot for useful comments on an early draft of this review. RV also thanks R. Foot for many enlightening discussions.  KP thanks A. Kusenko, M. Postma, I. Shoemaker and M. Wiechers for useful discussions.  KP was supported by the Netherlands Foundation for Fundamental Research of Matter (FOM) and the Netherlands Organisation for Scientific Research (NWO). RRV was supported in part by the Australian Research Council.

\def\bibfont{\small}
\setlength{\bibsep}{0.5pt}
\bibliography{ADMreview_arXiv_v3.bbl}

\begin{thebibliography}{370}
\expandafter\ifx\csname natexlab\endcsname\relax\def\natexlab#1{#1}\fi
\expandafter\ifx\csname bibnamefont\endcsname\relax
  \def\bibnamefont#1{#1}\fi
\expandafter\ifx\csname bibfnamefont\endcsname\relax
  \def\bibfnamefont#1{#1}\fi
\expandafter\ifx\csname citenamefont\endcsname\relax
  \def\citenamefont#1{#1}\fi
\expandafter\ifx\csname url\endcsname\relax
  \def\url#1{\texttt{#1}}\fi
\expandafter\ifx\csname urlprefix\endcsname\relax\def\urlprefix{URL }\fi
\providecommand{\bibinfo}[2]{#2}
\providecommand{\eprint}[2][]{\url{#2}}

\bibitem[{\citenamefont{Bertone et~al.}(2005)\citenamefont{Bertone, Hooper, and
  Silk}}]{Bertone:2004pz}
\bibinfo{author}{\bibfnamefont{G.}~\bibnamefont{Bertone}},
  \bibinfo{author}{\bibfnamefont{D.}~\bibnamefont{Hooper}}, \bibnamefont{and}
  \bibinfo{author}{\bibfnamefont{J.}~\bibnamefont{Silk}},
  \bibinfo{journal}{Phys.Rept.} \textbf{\bibinfo{volume}{405}},
  \bibinfo{pages}{279} (\bibinfo{year}{2005}), \eprint{hep-ph/0404175}.

\bibitem[{\citenamefont{Davoudiasl and Mohapatra}(2012)}]{Davoudiasl:2012uw}
\bibinfo{author}{\bibfnamefont{H.}~\bibnamefont{Davoudiasl}} \bibnamefont{and}
  \bibinfo{author}{\bibfnamefont{R.~N.} \bibnamefont{Mohapatra}},
  \bibinfo{journal}{New J.Phys.} \textbf{\bibinfo{volume}{14}},
  \bibinfo{pages}{095011} (\bibinfo{year}{2012}), \eprint{1203.1247}.

\bibitem[{\citenamefont{Hinshaw et~al.}(2012)}]{Hinshaw:2012aka}
\bibinfo{author}{\bibfnamefont{G.}~\bibnamefont{Hinshaw}} \bibnamefont{et~al.}
  (\bibinfo{collaboration}{WMAP Collaboration}) (\bibinfo{year}{2012}),
  \eprint{1212.5226}.

\bibitem[{\citenamefont{Ade et~al.}(2013)}]{Ade:2013zuv}
\bibinfo{author}{\bibfnamefont{P.}~\bibnamefont{Ade}} \bibnamefont{et~al.}
  (\bibinfo{collaboration}{Planck Collaboration}) (\bibinfo{year}{2013}),
  \eprint{1303.5076}.

\bibitem[{\citenamefont{Feng}(2010)}]{Feng:2010gw}
\bibinfo{author}{\bibfnamefont{J.~L.} \bibnamefont{Feng}},
  \bibinfo{journal}{Ann.Rev.Astron.Astrophys.} \textbf{\bibinfo{volume}{48}},
  \bibinfo{pages}{495} (\bibinfo{year}{2010}), \eprint{1003.0904}.

\bibitem[{\citenamefont{McDonald}(2011{\natexlab{a}})}]{McDonald:2011zz}
\bibinfo{author}{\bibfnamefont{J.}~\bibnamefont{McDonald}},
  \bibinfo{journal}{Phys.Rev.} \textbf{\bibinfo{volume}{D83}},
  \bibinfo{pages}{083509} (\bibinfo{year}{2011}{\natexlab{a}}),
  \eprint{1009.3227}.

\bibitem[{\citenamefont{McDonald}(2011{\natexlab{b}})}]{McDonald:2011sv}
\bibinfo{author}{\bibfnamefont{J.}~\bibnamefont{McDonald}},
  \bibinfo{journal}{Phys.Rev.} \textbf{\bibinfo{volume}{D84}},
  \bibinfo{pages}{103514} (\bibinfo{year}{2011}{\natexlab{b}}),
  \eprint{1108.4653}.

\bibitem[{\citenamefont{Cui et~al.}(2012{\natexlab{a}})\citenamefont{Cui,
  Randall, and Shuve}}]{Cui:2011ab}
\bibinfo{author}{\bibfnamefont{Y.}~\bibnamefont{Cui}},
  \bibinfo{author}{\bibfnamefont{L.}~\bibnamefont{Randall}}, \bibnamefont{and}
  \bibinfo{author}{\bibfnamefont{B.}~\bibnamefont{Shuve}},
  \bibinfo{journal}{JHEP} \textbf{\bibinfo{volume}{1204}}, \bibinfo{pages}{075}
  (\bibinfo{year}{2012}{\natexlab{a}}), \eprint{1112.2704}.

\bibitem[{\citenamefont{Bai et~al.}(2010)\citenamefont{Bai, Fox, and
  Harnik}}]{Bai:2010hh}
\bibinfo{author}{\bibfnamefont{Y.}~\bibnamefont{Bai}},
  \bibinfo{author}{\bibfnamefont{P.~J.} \bibnamefont{Fox}}, \bibnamefont{and}
  \bibinfo{author}{\bibfnamefont{R.}~\bibnamefont{Harnik}},
  \bibinfo{journal}{JHEP} \textbf{\bibinfo{volume}{1012}}, \bibinfo{pages}{048}
  (\bibinfo{year}{2010}), \eprint{1005.3797}.

\bibitem[{\citenamefont{Buckley}(2011)}]{Buckley:2011kk}
\bibinfo{author}{\bibfnamefont{M.~R.} \bibnamefont{Buckley}},
  \bibinfo{journal}{Phys.Rev.} \textbf{\bibinfo{volume}{D84}},
  \bibinfo{pages}{043510} (\bibinfo{year}{2011}), \eprint{1104.1429}.

\bibitem[{\citenamefont{Fox et~al.}(2012)\citenamefont{Fox, Harnik, Primulando,
  and Yu}}]{Fox:2012ee}
\bibinfo{author}{\bibfnamefont{P.~J.} \bibnamefont{Fox}},
  \bibinfo{author}{\bibfnamefont{R.}~\bibnamefont{Harnik}},
  \bibinfo{author}{\bibfnamefont{R.}~\bibnamefont{Primulando}},
  \bibnamefont{and} \bibinfo{author}{\bibfnamefont{C.-T.} \bibnamefont{Yu}},
  \bibinfo{journal}{Phys.Rev.} \textbf{\bibinfo{volume}{D86}},
  \bibinfo{pages}{015010} (\bibinfo{year}{2012}), \eprint{1203.1662}.

\bibitem[{\citenamefont{Dodelson and Widrow}(1994)}]{Dodelson:1993je}
\bibinfo{author}{\bibfnamefont{S.}~\bibnamefont{Dodelson}} \bibnamefont{and}
  \bibinfo{author}{\bibfnamefont{L.~M.} \bibnamefont{Widrow}},
  \bibinfo{journal}{Phys.Rev.Lett.} \textbf{\bibinfo{volume}{72}},
  \bibinfo{pages}{17} (\bibinfo{year}{1994}), \eprint{hep-ph/9303287}.

\bibitem[{\citenamefont{Shi and Fuller}(1999)}]{Shi:1998km}
\bibinfo{author}{\bibfnamefont{X.-D.} \bibnamefont{Shi}} \bibnamefont{and}
  \bibinfo{author}{\bibfnamefont{G.~M.} \bibnamefont{Fuller}},
  \bibinfo{journal}{Phys.Rev.Lett.} \textbf{\bibinfo{volume}{82}},
  \bibinfo{pages}{2832} (\bibinfo{year}{1999}), \eprint{astro-ph/9810076}.

\bibitem[{\citenamefont{Kusenko}(2006)}]{Kusenko:2006rh}
\bibinfo{author}{\bibfnamefont{A.}~\bibnamefont{Kusenko}},
  \bibinfo{journal}{Phys.Rev.Lett.} \textbf{\bibinfo{volume}{97}},
  \bibinfo{pages}{241301} (\bibinfo{year}{2006}), \eprint{hep-ph/0609081}.

\bibitem[{\citenamefont{Petraki and Kusenko}(2008)}]{Petraki:2007gq}
\bibinfo{author}{\bibfnamefont{K.}~\bibnamefont{Petraki}} \bibnamefont{and}
  \bibinfo{author}{\bibfnamefont{A.}~\bibnamefont{Kusenko}},
  \bibinfo{journal}{Phys.Rev.} \textbf{\bibinfo{volume}{D77}},
  \bibinfo{pages}{065014} (\bibinfo{year}{2008}), \eprint{0711.4646}.

\bibitem[{\citenamefont{Petraki}(2008)}]{Petraki:2008ef}
\bibinfo{author}{\bibfnamefont{K.}~\bibnamefont{Petraki}},
  \bibinfo{journal}{Phys.Rev.} \textbf{\bibinfo{volume}{D77}},
  \bibinfo{pages}{105004} (\bibinfo{year}{2008}), \eprint{0801.3470}.

\bibitem[{\citenamefont{Kusenko}(2009)}]{Kusenko:2009up}
\bibinfo{author}{\bibfnamefont{A.}~\bibnamefont{Kusenko}},
  \bibinfo{journal}{Phys.Rept.} \textbf{\bibinfo{volume}{481}},
  \bibinfo{pages}{1} (\bibinfo{year}{2009}), \eprint{0906.2968}.

\bibitem[{\citenamefont{Wu et~al.}(2009)\citenamefont{Wu, Ho, and
  Boyanovsky}}]{Wu:2009yr}
\bibinfo{author}{\bibfnamefont{J.}~\bibnamefont{Wu}},
  \bibinfo{author}{\bibfnamefont{C.-M.} \bibnamefont{Ho}}, \bibnamefont{and}
  \bibinfo{author}{\bibfnamefont{D.}~\bibnamefont{Boyanovsky}},
  \bibinfo{journal}{Phys.Rev.} \textbf{\bibinfo{volume}{D80}},
  \bibinfo{pages}{103511} (\bibinfo{year}{2009}), \eprint{0902.4278}.

\bibitem[{\citenamefont{Shoemaker et~al.}(2010)\citenamefont{Shoemaker,
  Petraki, and Kusenko}}]{Shoemaker:2010fg}
\bibinfo{author}{\bibfnamefont{I.~M.} \bibnamefont{Shoemaker}},
  \bibinfo{author}{\bibfnamefont{K.}~\bibnamefont{Petraki}}, \bibnamefont{and}
  \bibinfo{author}{\bibfnamefont{A.}~\bibnamefont{Kusenko}},
  \bibinfo{journal}{JHEP} \textbf{\bibinfo{volume}{1009}}, \bibinfo{pages}{060}
  (\bibinfo{year}{2010}), \eprint{1006.5458}.

\bibitem[{\citenamefont{Kusenko et~al.}(2010)\citenamefont{Kusenko, Takahashi,
  and Yanagida}}]{Kusenko:2010ik}
\bibinfo{author}{\bibfnamefont{A.}~\bibnamefont{Kusenko}},
  \bibinfo{author}{\bibfnamefont{F.}~\bibnamefont{Takahashi}},
  \bibnamefont{and} \bibinfo{author}{\bibfnamefont{T.~T.}
  \bibnamefont{Yanagida}}, \bibinfo{journal}{Phys.Lett.}
  \textbf{\bibinfo{volume}{B693}}, \bibinfo{pages}{144} (\bibinfo{year}{2010}),
  \eprint{1006.1731}.

\bibitem[{\citenamefont{Canetti et~al.}(2013)\citenamefont{Canetti, Drewes, and
  Shaposhnikov}}]{Canetti:2012vf}
\bibinfo{author}{\bibfnamefont{L.}~\bibnamefont{Canetti}},
  \bibinfo{author}{\bibfnamefont{M.}~\bibnamefont{Drewes}}, \bibnamefont{and}
  \bibinfo{author}{\bibfnamefont{M.}~\bibnamefont{Shaposhnikov}},
  \bibinfo{journal}{Phys.Rev.Lett.} \textbf{\bibinfo{volume}{110}},
  \bibinfo{pages}{061801} (\bibinfo{year}{2013}), \eprint{1204.3902}.

\bibitem[{\citenamefont{Canetti et~al.}(2012)\citenamefont{Canetti, Drewes,
  Frossard, and Shaposhnikov}}]{Canetti:2012kh}
\bibinfo{author}{\bibfnamefont{L.}~\bibnamefont{Canetti}},
  \bibinfo{author}{\bibfnamefont{M.}~\bibnamefont{Drewes}},
  \bibinfo{author}{\bibfnamefont{T.}~\bibnamefont{Frossard}}, \bibnamefont{and}
  \bibinfo{author}{\bibfnamefont{M.}~\bibnamefont{Shaposhnikov}}
  (\bibinfo{year}{2012}), \eprint{1208.4607}.

\bibitem[{\citenamefont{Merle}(2013)}]{Merle:2013gea}
\bibinfo{author}{\bibfnamefont{A.}~\bibnamefont{Merle}} (\bibinfo{year}{2013}),
  \eprint{1302.2625}.

\bibitem[{\citenamefont{Drewes}(2013)}]{Drewes:2013gca}
\bibinfo{author}{\bibfnamefont{M.}~\bibnamefont{Drewes}}
  (\bibinfo{year}{2013}), \eprint{1303.6912}.

\bibitem[{\citenamefont{Peccei and Quinn}(1977{\natexlab{a}})}]{Peccei:1977hh}
\bibinfo{author}{\bibfnamefont{R.}~\bibnamefont{Peccei}} \bibnamefont{and}
  \bibinfo{author}{\bibfnamefont{H.~R.} \bibnamefont{Quinn}},
  \bibinfo{journal}{Phys.Rev.Lett.} \textbf{\bibinfo{volume}{38}},
  \bibinfo{pages}{1440} (\bibinfo{year}{1977}{\natexlab{a}}).

\bibitem[{\citenamefont{Peccei and Quinn}(1977{\natexlab{b}})}]{Peccei:1977ur}
\bibinfo{author}{\bibfnamefont{R.}~\bibnamefont{Peccei}} \bibnamefont{and}
  \bibinfo{author}{\bibfnamefont{H.~R.} \bibnamefont{Quinn}},
  \bibinfo{journal}{Phys.Rev.} \textbf{\bibinfo{volume}{D16}},
  \bibinfo{pages}{1791} (\bibinfo{year}{1977}{\natexlab{b}}).

\bibitem[{\citenamefont{Kim}(1979)}]{Kim:1979if}
\bibinfo{author}{\bibfnamefont{J.~E.} \bibnamefont{Kim}},
  \bibinfo{journal}{Phys.Rev.Lett.} \textbf{\bibinfo{volume}{43}},
  \bibinfo{pages}{103} (\bibinfo{year}{1979}).

\bibitem[{\citenamefont{Zhitnitsky}(1980)}]{Zhitnitsky:1980tq}
\bibinfo{author}{\bibfnamefont{A.}~\bibnamefont{Zhitnitsky}},
  \bibinfo{journal}{Sov.J.Nucl.Phys.} \textbf{\bibinfo{volume}{31}},
  \bibinfo{pages}{260} (\bibinfo{year}{1980}).

\bibitem[{\citenamefont{Dine et~al.}(1981)\citenamefont{Dine, Fischler, and
  Srednicki}}]{Dine:1981rt}
\bibinfo{author}{\bibfnamefont{M.}~\bibnamefont{Dine}},
  \bibinfo{author}{\bibfnamefont{W.}~\bibnamefont{Fischler}}, \bibnamefont{and}
  \bibinfo{author}{\bibfnamefont{M.}~\bibnamefont{Srednicki}},
  \bibinfo{journal}{Phys.Lett.} \textbf{\bibinfo{volume}{B104}},
  \bibinfo{pages}{199} (\bibinfo{year}{1981}).

\bibitem[{\citenamefont{Asztalos et~al.}(2010)}]{Asztalos:2009yp}
\bibinfo{author}{\bibfnamefont{S.}~\bibnamefont{Asztalos}} \bibnamefont{et~al.}
  (\bibinfo{collaboration}{ADMX Collaboration}),
  \bibinfo{journal}{Phys.Rev.Lett.} \textbf{\bibinfo{volume}{104}},
  \bibinfo{pages}{041301} (\bibinfo{year}{2010}), \eprint{0910.5914}.

\bibitem[{\citenamefont{Sikivie}(2011)}]{Sikivie:2010bq}
\bibinfo{author}{\bibfnamefont{P.}~\bibnamefont{Sikivie}},
  \bibinfo{journal}{Phys.Lett.} \textbf{\bibinfo{volume}{B695}},
  \bibinfo{pages}{22} (\bibinfo{year}{2011}), \eprint{1003.2426}.

\bibitem[{\citenamefont{Sikivie and Yang}(2009)}]{Sikivie:2009qn}
\bibinfo{author}{\bibfnamefont{P.}~\bibnamefont{Sikivie}} \bibnamefont{and}
  \bibinfo{author}{\bibfnamefont{Q.}~\bibnamefont{Yang}},
  \bibinfo{journal}{Phys.Rev.Lett.} \textbf{\bibinfo{volume}{103}},
  \bibinfo{pages}{111301} (\bibinfo{year}{2009}), \eprint{0901.1106}.

\bibitem[{\citenamefont{Aune et~al.}(2011)}]{Arik:2011rx}
\bibinfo{author}{\bibfnamefont{S.}~\bibnamefont{Aune}} \bibnamefont{et~al.}
  (\bibinfo{collaboration}{CAST Collaboration}),
  \bibinfo{journal}{Phys.Rev.Lett.} \textbf{\bibinfo{volume}{107}},
  \bibinfo{pages}{261302} (\bibinfo{year}{2011}), \eprint{1106.3919}.

\bibitem[{\citenamefont{Kusenko and Shaposhnikov}(1998)}]{Kusenko:1997si}
\bibinfo{author}{\bibfnamefont{A.}~\bibnamefont{Kusenko}} \bibnamefont{and}
  \bibinfo{author}{\bibfnamefont{M.~E.} \bibnamefont{Shaposhnikov}},
  \bibinfo{journal}{Phys.Lett.} \textbf{\bibinfo{volume}{B418}},
  \bibinfo{pages}{46} (\bibinfo{year}{1998}), \eprint{hep-ph/9709492}.

\bibitem[{\citenamefont{Kasuya and Kawasaki}(2000)}]{Kasuya:1999wu}
\bibinfo{author}{\bibfnamefont{S.}~\bibnamefont{Kasuya}} \bibnamefont{and}
  \bibinfo{author}{\bibfnamefont{M.}~\bibnamefont{Kawasaki}},
  \bibinfo{journal}{Phys.Rev.} \textbf{\bibinfo{volume}{D61}},
  \bibinfo{pages}{041301} (\bibinfo{year}{2000}), \eprint{hep-ph/9909509}.

\bibitem[{\citenamefont{Demir}(2000)}]{Demir:2000gj}
\bibinfo{author}{\bibfnamefont{D.~A.} \bibnamefont{Demir}},
  \bibinfo{journal}{Phys.Lett.} \textbf{\bibinfo{volume}{B495}},
  \bibinfo{pages}{357} (\bibinfo{year}{2000}), \eprint{hep-ph/0006344}.

\bibitem[{\citenamefont{Kusenko and Steinhardt}(2001)}]{Kusenko:2001vu}
\bibinfo{author}{\bibfnamefont{A.}~\bibnamefont{Kusenko}} \bibnamefont{and}
  \bibinfo{author}{\bibfnamefont{P.~J.} \bibnamefont{Steinhardt}},
  \bibinfo{journal}{Phys.Rev.Lett.} \textbf{\bibinfo{volume}{87}},
  \bibinfo{pages}{141301} (\bibinfo{year}{2001}), \eprint{astro-ph/0106008}.

\bibitem[{\citenamefont{Kusenko et~al.}(2005)\citenamefont{Kusenko, Loveridge,
  and Shaposhnikov}}]{Kusenko:2004yw}
\bibinfo{author}{\bibfnamefont{A.}~\bibnamefont{Kusenko}},
  \bibinfo{author}{\bibfnamefont{L.}~\bibnamefont{Loveridge}},
  \bibnamefont{and}
  \bibinfo{author}{\bibfnamefont{M.}~\bibnamefont{Shaposhnikov}},
  \bibinfo{journal}{Phys.Rev.} \textbf{\bibinfo{volume}{D72}},
  \bibinfo{pages}{025015} (\bibinfo{year}{2005}), \eprint{hep-ph/0405044}.

\bibitem[{\citenamefont{Barnard}(2011)}]{Barnard:2011jw}
\bibinfo{author}{\bibfnamefont{J.}~\bibnamefont{Barnard}},
  \bibinfo{journal}{JHEP} \textbf{\bibinfo{volume}{1108}}, \bibinfo{pages}{058}
  (\bibinfo{year}{2011}), \eprint{1106.1182}.

\bibitem[{\citenamefont{Bernabei et~al.}(2008{\natexlab{a}})}]{Bernabei:2008yi}
\bibinfo{author}{\bibfnamefont{R.}~\bibnamefont{Bernabei}} \bibnamefont{et~al.}
  (\bibinfo{collaboration}{DAMA Collaboration}), \bibinfo{journal}{Eur.Phys.J.}
  \textbf{\bibinfo{volume}{C56}}, \bibinfo{pages}{333}
  (\bibinfo{year}{2008}{\natexlab{a}}), \eprint{0804.2741}.

\bibitem[{\citenamefont{Bernabei et~al.}(2010)}]{Bernabei:2010mq}
\bibinfo{author}{\bibfnamefont{R.}~\bibnamefont{Bernabei}} \bibnamefont{et~al.}
  (\bibinfo{collaboration}{DAMA Collaboration, LIBRA Collaboration}),
  \bibinfo{journal}{Eur.Phys.J.} \textbf{\bibinfo{volume}{C67}},
  \bibinfo{pages}{39} (\bibinfo{year}{2010}), \eprint{1002.1028}.

\bibitem[{\citenamefont{Aalseth et~al.}(2011{\natexlab{a}})}]{Aalseth:2010vx}
\bibinfo{author}{\bibfnamefont{C.}~\bibnamefont{Aalseth}} \bibnamefont{et~al.}
  (\bibinfo{collaboration}{CoGeNT collaboration}),
  \bibinfo{journal}{Phys.Rev.Lett.} \textbf{\bibinfo{volume}{106}},
  \bibinfo{pages}{131301} (\bibinfo{year}{2011}{\natexlab{a}}),
  \eprint{1002.4703}.

\bibitem[{\citenamefont{Aalseth
  et~al.}(2011{\natexlab{b}})\citenamefont{Aalseth, Barbeau, Colaresi, Collar,
  Diaz~Leon et~al.}}]{Aalseth:2011wp}
\bibinfo{author}{\bibfnamefont{C.}~\bibnamefont{Aalseth}},
  \bibinfo{author}{\bibfnamefont{P.}~\bibnamefont{Barbeau}},
  \bibinfo{author}{\bibfnamefont{J.}~\bibnamefont{Colaresi}},
  \bibinfo{author}{\bibfnamefont{J.}~\bibnamefont{Collar}},
  \bibinfo{author}{\bibfnamefont{J.}~\bibnamefont{Diaz~Leon}},
  \bibnamefont{et~al.}, \bibinfo{journal}{Phys.Rev.Lett.}
  \textbf{\bibinfo{volume}{107}}, \bibinfo{pages}{141301}
  (\bibinfo{year}{2011}{\natexlab{b}}), \eprint{1106.0650}.

\bibitem[{\citenamefont{Angloher et~al.}(2012)\citenamefont{Angloher, Bauer,
  Bavykina, Bento, Bucci et~al.}}]{Angloher:2011uu}
\bibinfo{author}{\bibfnamefont{G.}~\bibnamefont{Angloher}},
  \bibinfo{author}{\bibfnamefont{M.}~\bibnamefont{Bauer}},
  \bibinfo{author}{\bibfnamefont{I.}~\bibnamefont{Bavykina}},
  \bibinfo{author}{\bibfnamefont{A.}~\bibnamefont{Bento}},
  \bibinfo{author}{\bibfnamefont{C.}~\bibnamefont{Bucci}},
  \bibnamefont{et~al.}, \bibinfo{journal}{Eur.Phys.J.}
  \textbf{\bibinfo{volume}{C72}}, \bibinfo{pages}{1971} (\bibinfo{year}{2012}),
  \eprint{1109.0702}.

\bibitem[{\citenamefont{Agnese et~al.}(2013)}]{Agnese:2013rvf}
\bibinfo{author}{\bibfnamefont{R.}~\bibnamefont{Agnese}} \bibnamefont{et~al.}
  (\bibinfo{collaboration}{CDMS Collaboration}) (\bibinfo{year}{2013}),
  \eprint{1304.4279}.

\bibitem[{\citenamefont{Foot}(2004)}]{Foot:2003iv}
\bibinfo{author}{\bibfnamefont{R.}~\bibnamefont{Foot}},
  \bibinfo{journal}{Phys.Rev.} \textbf{\bibinfo{volume}{D69}},
  \bibinfo{pages}{036001} (\bibinfo{year}{2004}), \eprint{hep-ph/0308254}.

\bibitem[{\citenamefont{Foot}(2010)}]{Foot:2010hu}
\bibinfo{author}{\bibfnamefont{R.}~\bibnamefont{Foot}},
  \bibinfo{journal}{Phys.Rev.} \textbf{\bibinfo{volume}{D82}},
  \bibinfo{pages}{095001} (\bibinfo{year}{2010}), \eprint{1008.0685}.

\bibitem[{\citenamefont{Foot}(2011)}]{Foot:2011pi}
\bibinfo{author}{\bibfnamefont{R.}~\bibnamefont{Foot}},
  \bibinfo{journal}{Phys.Lett.} \textbf{\bibinfo{volume}{B703}},
  \bibinfo{pages}{7} (\bibinfo{year}{2011}), \eprint{1106.2688}.

\bibitem[{\citenamefont{Foot}(2013{\natexlab{a}})}]{Foot:2013msa}
\bibinfo{author}{\bibfnamefont{R.}~\bibnamefont{Foot}}
  (\bibinfo{year}{2013}{\natexlab{a}}), \eprint{1305.4316}.

\bibitem[{\citenamefont{Foot}(2012{\natexlab{a}})}]{Foot:2012cs}
\bibinfo{author}{\bibfnamefont{R.}~\bibnamefont{Foot}}
  (\bibinfo{year}{2012}{\natexlab{a}}), \eprint{1209.5602}.

\bibitem[{\citenamefont{Fornengo et~al.}(2011)\citenamefont{Fornengo, Panci,
  and Regis}}]{Fornengo:2011sz}
\bibinfo{author}{\bibfnamefont{N.}~\bibnamefont{Fornengo}},
  \bibinfo{author}{\bibfnamefont{P.}~\bibnamefont{Panci}}, \bibnamefont{and}
  \bibinfo{author}{\bibfnamefont{M.}~\bibnamefont{Regis}},
  \bibinfo{journal}{Phys.Rev.} \textbf{\bibinfo{volume}{D84}},
  \bibinfo{pages}{115002} (\bibinfo{year}{2011}), \eprint{1108.4661}.

\bibitem[{\citenamefont{Foot et~al.}(1991)\citenamefont{Foot, Lew, and
  Volkas}}]{Foot:1991bp}
\bibinfo{author}{\bibfnamefont{R.}~\bibnamefont{Foot}},
  \bibinfo{author}{\bibfnamefont{H.}~\bibnamefont{Lew}}, \bibnamefont{and}
  \bibinfo{author}{\bibfnamefont{R.}~\bibnamefont{Volkas}},
  \bibinfo{journal}{Phys.Lett.} \textbf{\bibinfo{volume}{B272}},
  \bibinfo{pages}{67} (\bibinfo{year}{1991}).

\bibitem[{\citenamefont{Foot et~al.}(1992)\citenamefont{Foot, Lew, and
  Volkas}}]{Foot:1991py}
\bibinfo{author}{\bibfnamefont{R.}~\bibnamefont{Foot}},
  \bibinfo{author}{\bibfnamefont{H.}~\bibnamefont{Lew}}, \bibnamefont{and}
  \bibinfo{author}{\bibfnamefont{R.}~\bibnamefont{Volkas}},
  \bibinfo{journal}{Mod.Phys.Lett.} \textbf{\bibinfo{volume}{A7}},
  \bibinfo{pages}{2567} (\bibinfo{year}{1992}).

\bibitem[{\citenamefont{Foot and Volkas}(1995)}]{Foot:1995pa}
\bibinfo{author}{\bibfnamefont{R.}~\bibnamefont{Foot}} \bibnamefont{and}
  \bibinfo{author}{\bibfnamefont{R.}~\bibnamefont{Volkas}},
  \bibinfo{journal}{Phys.Rev.} \textbf{\bibinfo{volume}{D52}},
  \bibinfo{pages}{6595} (\bibinfo{year}{1995}), \eprint{hep-ph/9505359}.

\bibitem[{\citenamefont{Lee and Yang}(1956)}]{Lee:1956qn}
\bibinfo{author}{\bibfnamefont{T.}~\bibnamefont{Lee}} \bibnamefont{and}
  \bibinfo{author}{\bibfnamefont{C.-N.} \bibnamefont{Yang}},
  \bibinfo{journal}{Phys.Rev.} \textbf{\bibinfo{volume}{104}},
  \bibinfo{pages}{254} (\bibinfo{year}{1956}).

\bibitem[{\citenamefont{Kobzarev et~al.}(1966)\citenamefont{Kobzarev, Okun, and
  Pomeranchuk}}]{Kobzarev}
\bibinfo{author}{\bibfnamefont{I.}~\bibnamefont{Kobzarev}},
  \bibinfo{author}{\bibfnamefont{L.}~\bibnamefont{Okun}}, \bibnamefont{and}
  \bibinfo{author}{\bibfnamefont{I.}~\bibnamefont{Pomeranchuk}},
  \bibinfo{journal}{Sov.J.Nucl.Phys.} \textbf{\bibinfo{volume}{3}},
  \bibinfo{pages}{837} (\bibinfo{year}{1966}).

\bibitem[{\citenamefont{Pavsic}(1974)}]{Pavsic:1974rq}
\bibinfo{author}{\bibfnamefont{M.}~\bibnamefont{Pavsic}},
  \bibinfo{journal}{Int.J.Theor.Phys.} \textbf{\bibinfo{volume}{9}},
  \bibinfo{pages}{229} (\bibinfo{year}{1974}), \eprint{hep-ph/0105344}.

\bibitem[{\citenamefont{Blinnikov and Khlopov}(1982)}]{Blinnikov:1982eh}
\bibinfo{author}{\bibfnamefont{S.}~\bibnamefont{Blinnikov}} \bibnamefont{and}
  \bibinfo{author}{\bibfnamefont{M.~Y.} \bibnamefont{Khlopov}},
  \bibinfo{journal}{Sov.J.Nucl.Phys.} \textbf{\bibinfo{volume}{36}},
  \bibinfo{pages}{472} (\bibinfo{year}{1982}).

\bibitem[{\citenamefont{Blinnikov and Khlopov}(1983)}]{Blinnikov:1983gh}
\bibinfo{author}{\bibfnamefont{S.}~\bibnamefont{Blinnikov}} \bibnamefont{and}
  \bibinfo{author}{\bibfnamefont{M.}~\bibnamefont{Khlopov}},
  \bibinfo{journal}{Sov.Astron.} \textbf{\bibinfo{volume}{27}},
  \bibinfo{pages}{371} (\bibinfo{year}{1983}).

\bibitem[{\citenamefont{Berezhiani et~al.}(2001)\citenamefont{Berezhiani,
  Comelli, and Villante}}]{Berezhiani:2000gw}
\bibinfo{author}{\bibfnamefont{Z.}~\bibnamefont{Berezhiani}},
  \bibinfo{author}{\bibfnamefont{D.}~\bibnamefont{Comelli}}, \bibnamefont{and}
  \bibinfo{author}{\bibfnamefont{F.~L.} \bibnamefont{Villante}},
  \bibinfo{journal}{Phys.Lett.} \textbf{\bibinfo{volume}{B503}},
  \bibinfo{pages}{362} (\bibinfo{year}{2001}), \eprint{hep-ph/0008105}.

\bibitem[{\citenamefont{Ignatiev and Volkas}(2003)}]{Ignatiev:2003js}
\bibinfo{author}{\bibfnamefont{A.~Y.} \bibnamefont{Ignatiev}} \bibnamefont{and}
  \bibinfo{author}{\bibfnamefont{R.}~\bibnamefont{Volkas}},
  \bibinfo{journal}{Phys.Rev.} \textbf{\bibinfo{volume}{D68}},
  \bibinfo{pages}{023518} (\bibinfo{year}{2003}), \eprint{hep-ph/0304260}.

\bibitem[{\citenamefont{Foot and Volkas}(2003)}]{Foot:2003jt}
\bibinfo{author}{\bibfnamefont{R.}~\bibnamefont{Foot}} \bibnamefont{and}
  \bibinfo{author}{\bibfnamefont{R.}~\bibnamefont{Volkas}},
  \bibinfo{journal}{Phys.Rev.} \textbf{\bibinfo{volume}{D68}},
  \bibinfo{pages}{021304} (\bibinfo{year}{2003}), \eprint{hep-ph/0304261}.

\bibitem[{\citenamefont{Foot and Volkas}(2004{\natexlab{a}})}]{Foot:2004pq}
\bibinfo{author}{\bibfnamefont{R.}~\bibnamefont{Foot}} \bibnamefont{and}
  \bibinfo{author}{\bibfnamefont{R.}~\bibnamefont{Volkas}},
  \bibinfo{journal}{Phys.Rev.} \textbf{\bibinfo{volume}{D69}},
  \bibinfo{pages}{123510} (\bibinfo{year}{2004}{\natexlab{a}}),
  \eprint{hep-ph/0402267}.

\bibitem[{\citenamefont{Berezhiani et~al.}(2005)\citenamefont{Berezhiani,
  Ciarcelluti, Comelli, and Villante}}]{Berezhiani:2003wj}
\bibinfo{author}{\bibfnamefont{Z.}~\bibnamefont{Berezhiani}},
  \bibinfo{author}{\bibfnamefont{P.}~\bibnamefont{Ciarcelluti}},
  \bibinfo{author}{\bibfnamefont{D.}~\bibnamefont{Comelli}}, \bibnamefont{and}
  \bibinfo{author}{\bibfnamefont{F.~L.} \bibnamefont{Villante}},
  \bibinfo{journal}{Int.J.Mod.Phys.} \textbf{\bibinfo{volume}{D14}},
  \bibinfo{pages}{107} (\bibinfo{year}{2005}), \eprint{astro-ph/0312605}.

\bibitem[{\citenamefont{Ciarcelluti}(2005{\natexlab{a}})}]{Ciarcelluti:2004ik}
\bibinfo{author}{\bibfnamefont{P.}~\bibnamefont{Ciarcelluti}},
  \bibinfo{journal}{Int.J.Mod.Phys.} \textbf{\bibinfo{volume}{D14}},
  \bibinfo{pages}{187} (\bibinfo{year}{2005}{\natexlab{a}}),
  \eprint{astro-ph/0409630}.

\bibitem[{\citenamefont{Ciarcelluti}(2005{\natexlab{b}})}]{Ciarcelluti:2004ip}
\bibinfo{author}{\bibfnamefont{P.}~\bibnamefont{Ciarcelluti}},
  \bibinfo{journal}{Int.J.Mod.Phys.} \textbf{\bibinfo{volume}{D14}},
  \bibinfo{pages}{223} (\bibinfo{year}{2005}{\natexlab{b}}),
  \eprint{astro-ph/0409633}.

\bibitem[{\citenamefont{Berezhiani et~al.}(2006)\citenamefont{Berezhiani,
  Cassisi, Ciarcelluti, and Pietrinferni}}]{Berezhiani:2005vv}
\bibinfo{author}{\bibfnamefont{Z.}~\bibnamefont{Berezhiani}},
  \bibinfo{author}{\bibfnamefont{S.}~\bibnamefont{Cassisi}},
  \bibinfo{author}{\bibfnamefont{P.}~\bibnamefont{Ciarcelluti}},
  \bibnamefont{and}
  \bibinfo{author}{\bibfnamefont{A.}~\bibnamefont{Pietrinferni}},
  \bibinfo{journal}{Astropart.Phys.} \textbf{\bibinfo{volume}{24}},
  \bibinfo{pages}{495} (\bibinfo{year}{2006}), \eprint{astro-ph/0507153}.

\bibitem[{\citenamefont{Nussinov}(1985)}]{Nussinov:1985xr}
\bibinfo{author}{\bibfnamefont{S.}~\bibnamefont{Nussinov}},
  \bibinfo{journal}{Phys.Lett.} \textbf{\bibinfo{volume}{B165}},
  \bibinfo{pages}{55} (\bibinfo{year}{1985}).

\bibitem[{\citenamefont{Barr et~al.}(1990)\citenamefont{Barr, Chivukula, and
  Farhi}}]{Barr:1990ca}
\bibinfo{author}{\bibfnamefont{S.~M.} \bibnamefont{Barr}},
  \bibinfo{author}{\bibfnamefont{R.~S.} \bibnamefont{Chivukula}},
  \bibnamefont{and} \bibinfo{author}{\bibfnamefont{E.}~\bibnamefont{Farhi}},
  \bibinfo{journal}{Phys.Lett.} \textbf{\bibinfo{volume}{B241}},
  \bibinfo{pages}{387} (\bibinfo{year}{1990}).

\bibitem[{\citenamefont{Thomas}(1995)}]{Thomas:1995ze}
\bibinfo{author}{\bibfnamefont{S.~D.} \bibnamefont{Thomas}},
  \bibinfo{journal}{Phys.Lett.} \textbf{\bibinfo{volume}{B356}},
  \bibinfo{pages}{256} (\bibinfo{year}{1995}), \eprint{hep-ph/9506274}.

\bibitem[{\citenamefont{Enqvist and McDonald}(1999)}]{Enqvist:1998en}
\bibinfo{author}{\bibfnamefont{K.}~\bibnamefont{Enqvist}} \bibnamefont{and}
  \bibinfo{author}{\bibfnamefont{J.}~\bibnamefont{McDonald}},
  \bibinfo{journal}{Nucl.Phys.} \textbf{\bibinfo{volume}{B538}},
  \bibinfo{pages}{321} (\bibinfo{year}{1999}), \eprint{hep-ph/9803380}.

\bibitem[{\citenamefont{Fujii and Hamaguchi}(2002)}]{Fujii:2002kr}
\bibinfo{author}{\bibfnamefont{M.}~\bibnamefont{Fujii}} \bibnamefont{and}
  \bibinfo{author}{\bibfnamefont{K.}~\bibnamefont{Hamaguchi}},
  \bibinfo{journal}{Phys.Rev.} \textbf{\bibinfo{volume}{D66}},
  \bibinfo{pages}{083501} (\bibinfo{year}{2002}), \eprint{hep-ph/0205044}.

\bibitem[{\citenamefont{Fujii and Yanagida}(2002)}]{Fujii:2002aj}
\bibinfo{author}{\bibfnamefont{M.}~\bibnamefont{Fujii}} \bibnamefont{and}
  \bibinfo{author}{\bibfnamefont{T.}~\bibnamefont{Yanagida}},
  \bibinfo{journal}{Phys.Lett.} \textbf{\bibinfo{volume}{B542}},
  \bibinfo{pages}{80} (\bibinfo{year}{2002}), \eprint{hep-ph/0206066}.

\bibitem[{\citenamefont{Enqvist and Mazumdar}(2003)}]{Enqvist:2003gh}
\bibinfo{author}{\bibfnamefont{K.}~\bibnamefont{Enqvist}} \bibnamefont{and}
  \bibinfo{author}{\bibfnamefont{A.}~\bibnamefont{Mazumdar}},
  \bibinfo{journal}{Phys.Rept.} \textbf{\bibinfo{volume}{380}},
  \bibinfo{pages}{99} (\bibinfo{year}{2003}), \eprint{hep-ph/0209244}.

\bibitem[{\citenamefont{Dine and Kusenko}(2003)}]{Dine:2003ax}
\bibinfo{author}{\bibfnamefont{M.}~\bibnamefont{Dine}} \bibnamefont{and}
  \bibinfo{author}{\bibfnamefont{A.}~\bibnamefont{Kusenko}},
  \bibinfo{journal}{Rev.Mod.Phys.} \textbf{\bibinfo{volume}{76}},
  \bibinfo{pages}{1} (\bibinfo{year}{2003}), \eprint{hep-ph/0303065}.

\bibitem[{\citenamefont{Roszkowski and Seto}(2007)}]{Roszkowski:2006kw}
\bibinfo{author}{\bibfnamefont{L.}~\bibnamefont{Roszkowski}} \bibnamefont{and}
  \bibinfo{author}{\bibfnamefont{O.}~\bibnamefont{Seto}},
  \bibinfo{journal}{Phys.Rev.Lett.} \textbf{\bibinfo{volume}{98}},
  \bibinfo{pages}{161304} (\bibinfo{year}{2007}), \eprint{hep-ph/0608013}.

\bibitem[{\citenamefont{McDonald}(2007)}]{McDonald:2006if}
\bibinfo{author}{\bibfnamefont{J.}~\bibnamefont{McDonald}},
  \bibinfo{journal}{JCAP} \textbf{\bibinfo{volume}{0701}}, \bibinfo{pages}{001}
  (\bibinfo{year}{2007}), \eprint{hep-ph/0609126}.

\bibitem[{\citenamefont{Kitano et~al.}(2008)\citenamefont{Kitano, Murayama, and
  Ratz}}]{Kitano:2008tk}
\bibinfo{author}{\bibfnamefont{R.}~\bibnamefont{Kitano}},
  \bibinfo{author}{\bibfnamefont{H.}~\bibnamefont{Murayama}}, \bibnamefont{and}
  \bibinfo{author}{\bibfnamefont{M.}~\bibnamefont{Ratz}},
  \bibinfo{journal}{Phys.Lett.} \textbf{\bibinfo{volume}{B669}},
  \bibinfo{pages}{145} (\bibinfo{year}{2008}), \eprint{0807.4313}.

\bibitem[{\citenamefont{Shoemaker and Kusenko}(2009)}]{Shoemaker:2009kg}
\bibinfo{author}{\bibfnamefont{I.~M.} \bibnamefont{Shoemaker}}
  \bibnamefont{and} \bibinfo{author}{\bibfnamefont{A.}~\bibnamefont{Kusenko}},
  \bibinfo{journal}{Phys.Rev.} \textbf{\bibinfo{volume}{D80}},
  \bibinfo{pages}{075021} (\bibinfo{year}{2009}), \eprint{0909.3334}.

\bibitem[{\citenamefont{Higashi et~al.}(2011)\citenamefont{Higashi, Ishima, and
  Suematsu}}]{Higashi:2011qq}
\bibinfo{author}{\bibfnamefont{H.}~\bibnamefont{Higashi}},
  \bibinfo{author}{\bibfnamefont{T.}~\bibnamefont{Ishima}}, \bibnamefont{and}
  \bibinfo{author}{\bibfnamefont{D.}~\bibnamefont{Suematsu}},
  \bibinfo{journal}{Int.J.Mod.Phys.} \textbf{\bibinfo{volume}{A26}},
  \bibinfo{pages}{995} (\bibinfo{year}{2011}), \eprint{1101.2704}.

\bibitem[{\citenamefont{Doddato and McDonald}(2011)}]{Doddato:2011fz}
\bibinfo{author}{\bibfnamefont{F.}~\bibnamefont{Doddato}} \bibnamefont{and}
  \bibinfo{author}{\bibfnamefont{J.}~\bibnamefont{McDonald}},
  \bibinfo{journal}{JCAP} \textbf{\bibinfo{volume}{1106}}, \bibinfo{pages}{008}
  (\bibinfo{year}{2011}), \eprint{1101.5328}.

\bibitem[{\citenamefont{Mazumdar}(2011)}]{Mazumdar:2011zd}
\bibinfo{author}{\bibfnamefont{A.}~\bibnamefont{Mazumdar}}
  (\bibinfo{year}{2011}), \eprint{1106.5408}.

\bibitem[{\citenamefont{Kasuya and Kawasaki}(2011)}]{Kasuya:2011ix}
\bibinfo{author}{\bibfnamefont{S.}~\bibnamefont{Kasuya}} \bibnamefont{and}
  \bibinfo{author}{\bibfnamefont{M.}~\bibnamefont{Kawasaki}},
  \bibinfo{journal}{Phys.Rev.} \textbf{\bibinfo{volume}{D84}},
  \bibinfo{pages}{123528} (\bibinfo{year}{2011}), \eprint{1107.0403}.

\bibitem[{\citenamefont{Doddato and McDonald}(2012)}]{Doddato:2011hx}
\bibinfo{author}{\bibfnamefont{F.}~\bibnamefont{Doddato}} \bibnamefont{and}
  \bibinfo{author}{\bibfnamefont{J.}~\bibnamefont{McDonald}},
  \bibinfo{journal}{JCAP} \textbf{\bibinfo{volume}{1206}}, \bibinfo{pages}{031}
  (\bibinfo{year}{2012}), \eprint{1111.2305}.

\bibitem[{\citenamefont{D'Eramo et~al.}(2012)\citenamefont{D'Eramo, Fei, and
  Thaler}}]{D'Eramo:2011ec}
\bibinfo{author}{\bibfnamefont{F.}~\bibnamefont{D'Eramo}},
  \bibinfo{author}{\bibfnamefont{L.}~\bibnamefont{Fei}}, \bibnamefont{and}
  \bibinfo{author}{\bibfnamefont{J.}~\bibnamefont{Thaler}},
  \bibinfo{journal}{JCAP} \textbf{\bibinfo{volume}{1203}}, \bibinfo{pages}{010}
  (\bibinfo{year}{2012}), \eprint{1111.5615}.

\bibitem[{\citenamefont{Gu}(2010)}]{Gu:2010yf}
\bibinfo{author}{\bibfnamefont{P.-H.} \bibnamefont{Gu}},
  \bibinfo{journal}{Phys.Rev.} \textbf{\bibinfo{volume}{D81}},
  \bibinfo{pages}{095002} (\bibinfo{year}{2010}), \eprint{1001.1341}.

\bibitem[{\citenamefont{Renner et~al.}(2010)\citenamefont{Renner, Feng, Jansen,
  and Petschlies}}]{Allahverdi:2010rh}
\bibinfo{author}{\bibfnamefont{D.~B.} \bibnamefont{Renner}},
  \bibinfo{author}{\bibfnamefont{X.}~\bibnamefont{Feng}},
  \bibinfo{author}{\bibfnamefont{K.}~\bibnamefont{Jansen}}, \bibnamefont{and}
  \bibinfo{author}{\bibfnamefont{M.}~\bibnamefont{Petschlies}},
  \bibinfo{journal}{PoS} \textbf{\bibinfo{volume}{LATTICE2010}},
  \bibinfo{pages}{155} (\bibinfo{year}{2010}), \eprint{1011.4231}.

\bibitem[{\citenamefont{McDonald}(2012)}]{McDonald:2012vw}
\bibinfo{author}{\bibfnamefont{J.}~\bibnamefont{McDonald}}
  (\bibinfo{year}{2012}), \eprint{1201.3124}.

\bibitem[{\citenamefont{Unwin}(2012)}]{Unwin:2012rp}
\bibinfo{author}{\bibfnamefont{J.}~\bibnamefont{Unwin}} (\bibinfo{year}{2012}),
  \eprint{1212.1425}.

\bibitem[{\citenamefont{Sakharov}(1967)}]{Sakharov:1967dj}
\bibinfo{author}{\bibfnamefont{A.~D.} \bibnamefont{Sakharov}},
  \bibinfo{journal}{Pisma Zh. Eksp. Teor. Fiz.} \textbf{\bibinfo{volume}{5}},
  \bibinfo{pages}{32} (\bibinfo{year}{1967}).

\bibitem[{\citenamefont{Hooper et~al.}(2005)\citenamefont{Hooper,
  March-Russell, and West}}]{Hooper:2004dc}
\bibinfo{author}{\bibfnamefont{D.}~\bibnamefont{Hooper}},
  \bibinfo{author}{\bibfnamefont{J.}~\bibnamefont{March-Russell}},
  \bibnamefont{and} \bibinfo{author}{\bibfnamefont{S.~M.} \bibnamefont{West}},
  \bibinfo{journal}{Phys.Lett.} \textbf{\bibinfo{volume}{B605}},
  \bibinfo{pages}{228} (\bibinfo{year}{2005}), \eprint{hep-ph/0410114}.

\bibitem[{\citenamefont{Kaplan et~al.}(2009)\citenamefont{Kaplan, Luty, and
  Zurek}}]{Kaplan:2009ag}
\bibinfo{author}{\bibfnamefont{D.~E.} \bibnamefont{Kaplan}},
  \bibinfo{author}{\bibfnamefont{M.~A.} \bibnamefont{Luty}}, \bibnamefont{and}
  \bibinfo{author}{\bibfnamefont{K.~M.} \bibnamefont{Zurek}},
  \bibinfo{journal}{Phys.Rev.} \textbf{\bibinfo{volume}{D79}},
  \bibinfo{pages}{115016} (\bibinfo{year}{2009}), \eprint{0901.4117}.

\bibitem[{\citenamefont{Cohen and Zurek}(2010)}]{Cohen:2009fz}
\bibinfo{author}{\bibfnamefont{T.}~\bibnamefont{Cohen}} \bibnamefont{and}
  \bibinfo{author}{\bibfnamefont{K.~M.} \bibnamefont{Zurek}},
  \bibinfo{journal}{Phys.Rev.Lett.} \textbf{\bibinfo{volume}{104}},
  \bibinfo{pages}{101301} (\bibinfo{year}{2010}), \eprint{0909.2035}.

\bibitem[{\citenamefont{Cai et~al.}(2009)\citenamefont{Cai, Luty, and
  Kaplan}}]{Cai:2009ia}
\bibinfo{author}{\bibfnamefont{Y.}~\bibnamefont{Cai}},
  \bibinfo{author}{\bibfnamefont{M.~A.} \bibnamefont{Luty}}, \bibnamefont{and}
  \bibinfo{author}{\bibfnamefont{D.~E.} \bibnamefont{Kaplan}}
  (\bibinfo{year}{2009}), \eprint{0909.5499}.

\bibitem[{\citenamefont{Blennow et~al.}(2011)\citenamefont{Blennow, Dasgupta,
  Fernandez-Martinez, and Rius}}]{Blennow:2010qp}
\bibinfo{author}{\bibfnamefont{M.}~\bibnamefont{Blennow}},
  \bibinfo{author}{\bibfnamefont{B.}~\bibnamefont{Dasgupta}},
  \bibinfo{author}{\bibfnamefont{E.}~\bibnamefont{Fernandez-Martinez}},
  \bibnamefont{and} \bibinfo{author}{\bibfnamefont{N.}~\bibnamefont{Rius}},
  \bibinfo{journal}{JHEP} \textbf{\bibinfo{volume}{1103}}, \bibinfo{pages}{014}
  (\bibinfo{year}{2011}), \eprint{1009.3159}.

\bibitem[{\citenamefont{Haba et~al.}(2011)\citenamefont{Haba, Matsumoto, and
  Sato}}]{Haba:2011uz}
\bibinfo{author}{\bibfnamefont{N.}~\bibnamefont{Haba}},
  \bibinfo{author}{\bibfnamefont{S.}~\bibnamefont{Matsumoto}},
  \bibnamefont{and} \bibinfo{author}{\bibfnamefont{R.}~\bibnamefont{Sato}},
  \bibinfo{journal}{Phys.Rev.} \textbf{\bibinfo{volume}{D84}},
  \bibinfo{pages}{055016} (\bibinfo{year}{2011}), \eprint{1101.5679}.

\bibitem[{\citenamefont{Servant and Tulin}(2013)}]{Servant:2013uwa}
\bibinfo{author}{\bibfnamefont{G.}~\bibnamefont{Servant}} \bibnamefont{and}
  \bibinfo{author}{\bibfnamefont{S.}~\bibnamefont{Tulin}}
  (\bibinfo{year}{2013}), \eprint{1304.3464}.

\bibitem[{\citenamefont{Dutta and Kumar}(2006)}]{Dutta:2006pt}
\bibinfo{author}{\bibfnamefont{B.}~\bibnamefont{Dutta}} \bibnamefont{and}
  \bibinfo{author}{\bibfnamefont{J.}~\bibnamefont{Kumar}},
  \bibinfo{journal}{Phys.Lett.} \textbf{\bibinfo{volume}{B643}},
  \bibinfo{pages}{284} (\bibinfo{year}{2006}), \eprint{hep-th/0608188}.

\bibitem[{\citenamefont{Buckley and Randall}(2011)}]{Buckley:2010ui}
\bibinfo{author}{\bibfnamefont{M.~R.} \bibnamefont{Buckley}} \bibnamefont{and}
  \bibinfo{author}{\bibfnamefont{L.}~\bibnamefont{Randall}},
  \bibinfo{journal}{JHEP} \textbf{\bibinfo{volume}{1109}}, \bibinfo{pages}{009}
  (\bibinfo{year}{2011}), \eprint{1009.0270}.

\bibitem[{\citenamefont{Dutta and Kumar}(2011)}]{Dutta:2010va}
\bibinfo{author}{\bibfnamefont{B.}~\bibnamefont{Dutta}} \bibnamefont{and}
  \bibinfo{author}{\bibfnamefont{J.}~\bibnamefont{Kumar}},
  \bibinfo{journal}{Phys.Lett.} \textbf{\bibinfo{volume}{B699}},
  \bibinfo{pages}{364} (\bibinfo{year}{2011}), \eprint{1012.1341}.

\bibitem[{\citenamefont{Shelton and Zurek}(2010)}]{Shelton:2010ta}
\bibinfo{author}{\bibfnamefont{J.}~\bibnamefont{Shelton}} \bibnamefont{and}
  \bibinfo{author}{\bibfnamefont{K.~M.} \bibnamefont{Zurek}},
  \bibinfo{journal}{Phys.Rev.} \textbf{\bibinfo{volume}{D82}},
  \bibinfo{pages}{123512} (\bibinfo{year}{2010}), \eprint{1008.1997}.

\bibitem[{\citenamefont{Haba and Matsumoto}(2011)}]{Haba:2010bm}
\bibinfo{author}{\bibfnamefont{N.}~\bibnamefont{Haba}} \bibnamefont{and}
  \bibinfo{author}{\bibfnamefont{S.}~\bibnamefont{Matsumoto}},
  \bibinfo{journal}{Prog.Theor.Phys.} \textbf{\bibinfo{volume}{125}},
  \bibinfo{pages}{1311} (\bibinfo{year}{2011}), \eprint{1008.2487}.

\bibitem[{\citenamefont{Feng et~al.}(2013)\citenamefont{Feng, Mazumdar, and
  Nath}}]{Feng:2013wn}
\bibinfo{author}{\bibfnamefont{W.-Z.} \bibnamefont{Feng}},
  \bibinfo{author}{\bibfnamefont{A.}~\bibnamefont{Mazumdar}}, \bibnamefont{and}
  \bibinfo{author}{\bibfnamefont{P.}~\bibnamefont{Nath}}
  (\bibinfo{year}{2013}), \eprint{1302.0012}.

\bibitem[{\citenamefont{Berezhiani}(2008)}]{Berezhiani:2008zza}
\bibinfo{author}{\bibfnamefont{Z.}~\bibnamefont{Berezhiani}},
  \bibinfo{journal}{Eur.Phys.J.ST} \textbf{\bibinfo{volume}{163}},
  \bibinfo{pages}{271} (\bibinfo{year}{2008}).

\bibitem[{\citenamefont{Falkowski et~al.}(2011)\citenamefont{Falkowski,
  Ruderman, and Volansky}}]{Falkowski:2011xh}
\bibinfo{author}{\bibfnamefont{A.}~\bibnamefont{Falkowski}},
  \bibinfo{author}{\bibfnamefont{J.~T.} \bibnamefont{Ruderman}},
  \bibnamefont{and} \bibinfo{author}{\bibfnamefont{T.}~\bibnamefont{Volansky}},
  \bibinfo{journal}{JHEP} \textbf{\bibinfo{volume}{1105}}, \bibinfo{pages}{106}
  (\bibinfo{year}{2011}), \eprint{1101.4936}.

\bibitem[{\citenamefont{Kuzmin}(1998)}]{Kuzmin:1996he}
\bibinfo{author}{\bibfnamefont{V.~A.} \bibnamefont{Kuzmin}},
  \bibinfo{journal}{Phys. Part. Nucl.} \textbf{\bibinfo{volume}{29}},
  \bibinfo{pages}{257} (\bibinfo{year}{1998}), \eprint{hep-ph/9701269}.

\bibitem[{\citenamefont{Kitano and Low}(2005{\natexlab{a}})}]{Kitano:2004sv}
\bibinfo{author}{\bibfnamefont{R.}~\bibnamefont{Kitano}} \bibnamefont{and}
  \bibinfo{author}{\bibfnamefont{I.}~\bibnamefont{Low}},
  \bibinfo{journal}{Phys.Rev.} \textbf{\bibinfo{volume}{D71}},
  \bibinfo{pages}{023510} (\bibinfo{year}{2005}{\natexlab{a}}),
  \eprint{hep-ph/0411133}.

\bibitem[{\citenamefont{Kitano and Low}(2005{\natexlab{b}})}]{Kitano:2005ge}
\bibinfo{author}{\bibfnamefont{R.}~\bibnamefont{Kitano}} \bibnamefont{and}
  \bibinfo{author}{\bibfnamefont{I.}~\bibnamefont{Low}}
  (\bibinfo{year}{2005}{\natexlab{b}}), \eprint{hep-ph/0503112}.

\bibitem[{\citenamefont{Gu et~al.}(2009)\citenamefont{Gu, Sarkar, and
  Zhang}}]{Gu:2009yy}
\bibinfo{author}{\bibfnamefont{P.-H.} \bibnamefont{Gu}},
  \bibinfo{author}{\bibfnamefont{U.}~\bibnamefont{Sarkar}}, \bibnamefont{and}
  \bibinfo{author}{\bibfnamefont{X.}~\bibnamefont{Zhang}},
  \bibinfo{journal}{Phys.Rev.} \textbf{\bibinfo{volume}{D80}},
  \bibinfo{pages}{076003} (\bibinfo{year}{2009}), \eprint{0906.3103}.

\bibitem[{\citenamefont{Davoudiasl et~al.}(2010)\citenamefont{Davoudiasl,
  Morrissey, Sigurdson, and Tulin}}]{Davoudiasl:2010am}
\bibinfo{author}{\bibfnamefont{H.}~\bibnamefont{Davoudiasl}},
  \bibinfo{author}{\bibfnamefont{D.~E.} \bibnamefont{Morrissey}},
  \bibinfo{author}{\bibfnamefont{K.}~\bibnamefont{Sigurdson}},
  \bibnamefont{and} \bibinfo{author}{\bibfnamefont{S.}~\bibnamefont{Tulin}},
  \bibinfo{journal}{Phys.Rev.Lett.} \textbf{\bibinfo{volume}{105}},
  \bibinfo{pages}{211304} (\bibinfo{year}{2010}), \eprint{1008.2399}.

\bibitem[{\citenamefont{Gu et~al.}(2011)\citenamefont{Gu, Lindner, Sarkar, and
  Zhang}}]{Gu:2010ft}
\bibinfo{author}{\bibfnamefont{P.-H.} \bibnamefont{Gu}},
  \bibinfo{author}{\bibfnamefont{M.}~\bibnamefont{Lindner}},
  \bibinfo{author}{\bibfnamefont{U.}~\bibnamefont{Sarkar}}, \bibnamefont{and}
  \bibinfo{author}{\bibfnamefont{X.}~\bibnamefont{Zhang}},
  \bibinfo{journal}{Phys.Rev.} \textbf{\bibinfo{volume}{D83}},
  \bibinfo{pages}{055008} (\bibinfo{year}{2011}), \eprint{1009.2690}.

\bibitem[{\citenamefont{Heckman and Rey}(2011)}]{Heckman:2011sw}
\bibinfo{author}{\bibfnamefont{J.~J.} \bibnamefont{Heckman}} \bibnamefont{and}
  \bibinfo{author}{\bibfnamefont{S.-J.} \bibnamefont{Rey}},
  \bibinfo{journal}{JHEP} \textbf{\bibinfo{volume}{1106}}, \bibinfo{pages}{120}
  (\bibinfo{year}{2011}), \eprint{1102.5346}.

\bibitem[{\citenamefont{Gelmini et~al.}(1987)\citenamefont{Gelmini, Hall, and
  Lin}}]{Gelmini:1986zz}
\bibinfo{author}{\bibfnamefont{G.}~\bibnamefont{Gelmini}},
  \bibinfo{author}{\bibfnamefont{L.~J.} \bibnamefont{Hall}}, \bibnamefont{and}
  \bibinfo{author}{\bibfnamefont{M.}~\bibnamefont{Lin}},
  \bibinfo{journal}{Nucl.Phys.} \textbf{\bibinfo{volume}{B281}},
  \bibinfo{pages}{726} (\bibinfo{year}{1987}).

\bibitem[{\citenamefont{Oaknin and Zhitnitsky}(2005)}]{Oaknin:2003uv}
\bibinfo{author}{\bibfnamefont{D.~H.} \bibnamefont{Oaknin}} \bibnamefont{and}
  \bibinfo{author}{\bibfnamefont{A.}~\bibnamefont{Zhitnitsky}},
  \bibinfo{journal}{Phys.Rev.} \textbf{\bibinfo{volume}{D71}},
  \bibinfo{pages}{023519} (\bibinfo{year}{2005}), \eprint{hep-ph/0309086}.

\bibitem[{\citenamefont{Farrar and Zaharijas}(2006)}]{Farrar:2005zd}
\bibinfo{author}{\bibfnamefont{G.~R.} \bibnamefont{Farrar}} \bibnamefont{and}
  \bibinfo{author}{\bibfnamefont{G.}~\bibnamefont{Zaharijas}},
  \bibinfo{journal}{Phys.Rev.Lett.} \textbf{\bibinfo{volume}{96}},
  \bibinfo{pages}{041302} (\bibinfo{year}{2006}), \eprint{hep-ph/0510079}.

\bibitem[{\citenamefont{Hall et~al.}(2010{\natexlab{a}})\citenamefont{Hall,
  March-Russell, and West}}]{Hall:2010jx}
\bibinfo{author}{\bibfnamefont{L.~J.} \bibnamefont{Hall}},
  \bibinfo{author}{\bibfnamefont{J.}~\bibnamefont{March-Russell}},
  \bibnamefont{and} \bibinfo{author}{\bibfnamefont{S.~M.} \bibnamefont{West}}
  (\bibinfo{year}{2010}{\natexlab{a}}), \eprint{1010.0245}.

\bibitem[{\citenamefont{Bell et~al.}(2011{\natexlab{a}})\citenamefont{Bell,
  Petraki, Shoemaker, and Volkas}}]{Bell:2011tn}
\bibinfo{author}{\bibfnamefont{N.~F.} \bibnamefont{Bell}},
  \bibinfo{author}{\bibfnamefont{K.}~\bibnamefont{Petraki}},
  \bibinfo{author}{\bibfnamefont{I.~M.} \bibnamefont{Shoemaker}},
  \bibnamefont{and} \bibinfo{author}{\bibfnamefont{R.~R.}
  \bibnamefont{Volkas}}, \bibinfo{journal}{Phys.Rev.}
  \textbf{\bibinfo{volume}{D84}}, \bibinfo{pages}{123505}
  (\bibinfo{year}{2011}{\natexlab{a}}), \eprint{1105.3730}.

\bibitem[{\citenamefont{Cheung and Zurek}(2011)}]{Cheung:2011if}
\bibinfo{author}{\bibfnamefont{C.}~\bibnamefont{Cheung}} \bibnamefont{and}
  \bibinfo{author}{\bibfnamefont{K.~M.} \bibnamefont{Zurek}},
  \bibinfo{journal}{Phys.Rev.} \textbf{\bibinfo{volume}{D84}},
  \bibinfo{pages}{035007} (\bibinfo{year}{2011}), \eprint{1105.4612}.

\bibitem[{\citenamefont{Graesser
  et~al.}(2011{\natexlab{a}})\citenamefont{Graesser, Shoemaker, and
  Vecchi}}]{Graesser:2011vj}
\bibinfo{author}{\bibfnamefont{M.~L.} \bibnamefont{Graesser}},
  \bibinfo{author}{\bibfnamefont{I.~M.} \bibnamefont{Shoemaker}},
  \bibnamefont{and} \bibinfo{author}{\bibfnamefont{L.}~\bibnamefont{Vecchi}}
  (\bibinfo{year}{2011}{\natexlab{a}}), \eprint{1107.2666}.

\bibitem[{\citenamefont{von Harling et~al.}(2012)\citenamefont{von Harling,
  Petraki, and Volkas}}]{vonHarling:2012yn}
\bibinfo{author}{\bibfnamefont{B.}~\bibnamefont{von Harling}},
  \bibinfo{author}{\bibfnamefont{K.}~\bibnamefont{Petraki}}, \bibnamefont{and}
  \bibinfo{author}{\bibfnamefont{R.~R.} \bibnamefont{Volkas}},
  \bibinfo{journal}{JCAP} \textbf{\bibinfo{volume}{1205}}, \bibinfo{pages}{021}
  (\bibinfo{year}{2012}), \eprint{1201.2200}.

\bibitem[{\citenamefont{March-Russell and
  McCullough}(2012)}]{MarchRussell:2011fi}
\bibinfo{author}{\bibfnamefont{J.}~\bibnamefont{March-Russell}}
  \bibnamefont{and}
  \bibinfo{author}{\bibfnamefont{M.}~\bibnamefont{McCullough}},
  \bibinfo{journal}{JCAP} \textbf{\bibinfo{volume}{1203}}, \bibinfo{pages}{019}
  (\bibinfo{year}{2012}), \eprint{1106.4319}.

\bibitem[{\citenamefont{Petraki et~al.}(2012)\citenamefont{Petraki, Trodden,
  and Volkas}}]{Petraki:2011mv}
\bibinfo{author}{\bibfnamefont{K.}~\bibnamefont{Petraki}},
  \bibinfo{author}{\bibfnamefont{M.}~\bibnamefont{Trodden}}, \bibnamefont{and}
  \bibinfo{author}{\bibfnamefont{R.~R.} \bibnamefont{Volkas}},
  \bibinfo{journal}{JCAP} \textbf{\bibinfo{volume}{1202}}, \bibinfo{pages}{044}
  (\bibinfo{year}{2012}), \eprint{1111.4786}.

\bibitem[{\citenamefont{Dodelson and
  Widrow}(1990{\natexlab{a}})}]{Dodelson:1989ii}
\bibinfo{author}{\bibfnamefont{S.}~\bibnamefont{Dodelson}} \bibnamefont{and}
  \bibinfo{author}{\bibfnamefont{L.~M.} \bibnamefont{Widrow}},
  \bibinfo{journal}{Phys. Rev. Lett.} \textbf{\bibinfo{volume}{64}},
  \bibinfo{pages}{340} (\bibinfo{year}{1990}{\natexlab{a}}).

\bibitem[{\citenamefont{Dodelson and
  Widrow}(1990{\natexlab{b}})}]{Dodelson:1989cq}
\bibinfo{author}{\bibfnamefont{S.}~\bibnamefont{Dodelson}} \bibnamefont{and}
  \bibinfo{author}{\bibfnamefont{L.~M.} \bibnamefont{Widrow}},
  \bibinfo{journal}{Phys.Rev.} \textbf{\bibinfo{volume}{D42}},
  \bibinfo{pages}{326} (\bibinfo{year}{1990}{\natexlab{b}}).

\bibitem[{\citenamefont{Cui et~al.}(2011)\citenamefont{Cui, Randall, and
  Shuve}}]{Cui:2011qe}
\bibinfo{author}{\bibfnamefont{Y.}~\bibnamefont{Cui}},
  \bibinfo{author}{\bibfnamefont{L.}~\bibnamefont{Randall}}, \bibnamefont{and}
  \bibinfo{author}{\bibfnamefont{B.}~\bibnamefont{Shuve}},
  \bibinfo{journal}{JHEP} \textbf{\bibinfo{volume}{1108}}, \bibinfo{pages}{073}
  (\bibinfo{year}{2011}), \eprint{1106.4834}.

\bibitem[{\citenamefont{Barr}(1991)}]{Barr:1991qn}
\bibinfo{author}{\bibfnamefont{S.~M.} \bibnamefont{Barr}},
  \bibinfo{journal}{Phys.Rev.} \textbf{\bibinfo{volume}{D44}},
  \bibinfo{pages}{3062} (\bibinfo{year}{1991}).

\bibitem[{\citenamefont{Kribs et~al.}(2010)\citenamefont{Kribs, Roy, Terning,
  and Zurek}}]{Kribs:2009fy}
\bibinfo{author}{\bibfnamefont{G.~D.} \bibnamefont{Kribs}},
  \bibinfo{author}{\bibfnamefont{T.~S.} \bibnamefont{Roy}},
  \bibinfo{author}{\bibfnamefont{J.}~\bibnamefont{Terning}}, \bibnamefont{and}
  \bibinfo{author}{\bibfnamefont{K.~M.} \bibnamefont{Zurek}},
  \bibinfo{journal}{Phys.Rev.} \textbf{\bibinfo{volume}{D81}},
  \bibinfo{pages}{095001} (\bibinfo{year}{2010}), \eprint{0909.2034}.

\bibitem[{\citenamefont{Kang et~al.}(2011)\citenamefont{Kang, Li, Li, Liu, and
  Yang}}]{Kang:2011wb}
\bibinfo{author}{\bibfnamefont{Z.}~\bibnamefont{Kang}},
  \bibinfo{author}{\bibfnamefont{J.}~\bibnamefont{Li}},
  \bibinfo{author}{\bibfnamefont{T.}~\bibnamefont{Li}},
  \bibinfo{author}{\bibfnamefont{T.}~\bibnamefont{Liu}}, \bibnamefont{and}
  \bibinfo{author}{\bibfnamefont{J.}~\bibnamefont{Yang}}
  (\bibinfo{year}{2011}), \eprint{1102.5644}.

\bibitem[{\citenamefont{Weinberg}(1979)}]{Weinberg:1979bt}
\bibinfo{author}{\bibfnamefont{S.}~\bibnamefont{Weinberg}},
  \bibinfo{journal}{Phys.Rev.Lett.} \textbf{\bibinfo{volume}{42}},
  \bibinfo{pages}{850} (\bibinfo{year}{1979}).

\bibitem[{\citenamefont{Fukugita and Yanagida}(1986)}]{Fukugita:1986hr}
\bibinfo{author}{\bibfnamefont{M.}~\bibnamefont{Fukugita}} \bibnamefont{and}
  \bibinfo{author}{\bibfnamefont{T.}~\bibnamefont{Yanagida}},
  \bibinfo{journal}{Phys.Lett.} \textbf{\bibinfo{volume}{B174}},
  \bibinfo{pages}{45} (\bibinfo{year}{1986}).

\bibitem[{\citenamefont{Affleck and Dine}(1985)}]{Affleck:1984fy}
\bibinfo{author}{\bibfnamefont{I.}~\bibnamefont{Affleck}} \bibnamefont{and}
  \bibinfo{author}{\bibfnamefont{M.}~\bibnamefont{Dine}},
  \bibinfo{journal}{Nucl.Phys.} \textbf{\bibinfo{volume}{B249}},
  \bibinfo{pages}{361} (\bibinfo{year}{1985}).

\bibitem[{\citenamefont{Dine et~al.}(1996)\citenamefont{Dine, Randall, and
  Thomas}}]{Dine:1995kz}
\bibinfo{author}{\bibfnamefont{M.}~\bibnamefont{Dine}},
  \bibinfo{author}{\bibfnamefont{L.}~\bibnamefont{Randall}}, \bibnamefont{and}
  \bibinfo{author}{\bibfnamefont{S.~D.} \bibnamefont{Thomas}},
  \bibinfo{journal}{Nucl.Phys.} \textbf{\bibinfo{volume}{B458}},
  \bibinfo{pages}{291} (\bibinfo{year}{1996}), \eprint{hep-ph/9507453}.

\bibitem[{\citenamefont{Allahverdi and Mazumdar}(2012)}]{Allahverdi:2012ju}
\bibinfo{author}{\bibfnamefont{R.}~\bibnamefont{Allahverdi}} \bibnamefont{and}
  \bibinfo{author}{\bibfnamefont{A.}~\bibnamefont{Mazumdar}},
  \bibinfo{journal}{New J.Phys.} \textbf{\bibinfo{volume}{14}},
  \bibinfo{pages}{125013} (\bibinfo{year}{2012}).

\bibitem[{\citenamefont{Riotto and Trodden}(1999)}]{Riotto:1999yt}
\bibinfo{author}{\bibfnamefont{A.}~\bibnamefont{Riotto}} \bibnamefont{and}
  \bibinfo{author}{\bibfnamefont{M.}~\bibnamefont{Trodden}},
  \bibinfo{journal}{Ann.Rev.Nucl.Part.Sci.} \textbf{\bibinfo{volume}{49}},
  \bibinfo{pages}{35} (\bibinfo{year}{1999}), \eprint{hep-ph/9901362}.

\bibitem[{\citenamefont{Cohen and Kaplan}(1987)}]{Cohen:1987vi}
\bibinfo{author}{\bibfnamefont{A.~G.} \bibnamefont{Cohen}} \bibnamefont{and}
  \bibinfo{author}{\bibfnamefont{D.~B.} \bibnamefont{Kaplan}},
  \bibinfo{journal}{Phys.Lett.} \textbf{\bibinfo{volume}{B199}},
  \bibinfo{pages}{251} (\bibinfo{year}{1987}).

\bibitem[{\citenamefont{Cohen and Kaplan}(1988)}]{Cohen:1988kt}
\bibinfo{author}{\bibfnamefont{A.~G.} \bibnamefont{Cohen}} \bibnamefont{and}
  \bibinfo{author}{\bibfnamefont{D.~B.} \bibnamefont{Kaplan}},
  \bibinfo{journal}{Nucl.Phys.} \textbf{\bibinfo{volume}{B308}},
  \bibinfo{pages}{913} (\bibinfo{year}{1988}).

\bibitem[{\citenamefont{Blanchet et~al.}(2009)\citenamefont{Blanchet, Chacko,
  and Mohapatra}}]{Blanchet:2008zg}
\bibinfo{author}{\bibfnamefont{S.}~\bibnamefont{Blanchet}},
  \bibinfo{author}{\bibfnamefont{Z.}~\bibnamefont{Chacko}}, \bibnamefont{and}
  \bibinfo{author}{\bibfnamefont{R.~N.} \bibnamefont{Mohapatra}},
  \bibinfo{journal}{Phys.Rev.} \textbf{\bibinfo{volume}{D80}},
  \bibinfo{pages}{085002} (\bibinfo{year}{2009}), \eprint{0812.3837}.

\bibitem[{\citenamefont{Baldes et~al.}(2013)\citenamefont{Baldes, Bell,
  Petraki, and Volkas}}]{Baldes:2013eva}
\bibinfo{author}{\bibfnamefont{I.}~\bibnamefont{Baldes}},
  \bibinfo{author}{\bibfnamefont{N.~F.} \bibnamefont{Bell}},
  \bibinfo{author}{\bibfnamefont{K.}~\bibnamefont{Petraki}}, \bibnamefont{and}
  \bibinfo{author}{\bibfnamefont{R.~R.} \bibnamefont{Volkas}}
  (\bibinfo{year}{2013}), \eprint{1304.6162}.

\bibitem[{\citenamefont{Dodelson et~al.}(1992)\citenamefont{Dodelson, Greene,
  and Widrow}}]{Dodelson:1991iv}
\bibinfo{author}{\bibfnamefont{S.}~\bibnamefont{Dodelson}},
  \bibinfo{author}{\bibfnamefont{B.~R.} \bibnamefont{Greene}},
  \bibnamefont{and} \bibinfo{author}{\bibfnamefont{L.~M.}
  \bibnamefont{Widrow}}, \bibinfo{journal}{Nucl.Phys.}
  \textbf{\bibinfo{volume}{B372}}, \bibinfo{pages}{467} (\bibinfo{year}{1992}).

\bibitem[{\citenamefont{Agashe and Servant}(2005)}]{Agashe:2004bm}
\bibinfo{author}{\bibfnamefont{K.}~\bibnamefont{Agashe}} \bibnamefont{and}
  \bibinfo{author}{\bibfnamefont{G.}~\bibnamefont{Servant}},
  \bibinfo{journal}{JCAP} \textbf{\bibinfo{volume}{0502}}, \bibinfo{pages}{002}
  (\bibinfo{year}{2005}), \eprint{hep-ph/0411254}.

\bibitem[{\citenamefont{Cosme et~al.}(2005)\citenamefont{Cosme, Lopez~Honorez,
  and Tytgat}}]{Cosme:2005sb}
\bibinfo{author}{\bibfnamefont{N.}~\bibnamefont{Cosme}},
  \bibinfo{author}{\bibfnamefont{L.}~\bibnamefont{Lopez~Honorez}},
  \bibnamefont{and} \bibinfo{author}{\bibfnamefont{M.~H.}
  \bibnamefont{Tytgat}}, \bibinfo{journal}{Phys.Rev.}
  \textbf{\bibinfo{volume}{D72}}, \bibinfo{pages}{043505}
  (\bibinfo{year}{2005}), \eprint{hep-ph/0506320}.

\bibitem[{\citenamefont{Kaplan et~al.}(2010)\citenamefont{Kaplan, Krnjaic,
  Rehermann, and Wells}}]{Kaplan:2009de}
\bibinfo{author}{\bibfnamefont{D.~E.} \bibnamefont{Kaplan}},
  \bibinfo{author}{\bibfnamefont{G.~Z.} \bibnamefont{Krnjaic}},
  \bibinfo{author}{\bibfnamefont{K.~R.} \bibnamefont{Rehermann}},
  \bibnamefont{and} \bibinfo{author}{\bibfnamefont{C.~M.} \bibnamefont{Wells}},
  \bibinfo{journal}{JCAP} \textbf{\bibinfo{volume}{1005}}, \bibinfo{pages}{021}
  (\bibinfo{year}{2010}), \eprint{0909.0753}.

\bibitem[{\citenamefont{An et~al.}(2010)\citenamefont{An, Chen, Mohapatra, and
  Zhang}}]{An:2009vq}
\bibinfo{author}{\bibfnamefont{H.}~\bibnamefont{An}},
  \bibinfo{author}{\bibfnamefont{S.-L.} \bibnamefont{Chen}},
  \bibinfo{author}{\bibfnamefont{R.~N.} \bibnamefont{Mohapatra}},
  \bibnamefont{and} \bibinfo{author}{\bibfnamefont{Y.}~\bibnamefont{Zhang}},
  \bibinfo{journal}{JHEP} \textbf{\bibinfo{volume}{1003}}, \bibinfo{pages}{124}
  (\bibinfo{year}{2010}), \eprint{0911.4463}.

\bibitem[{\citenamefont{Dulaney et~al.}(2011)\citenamefont{Dulaney,
  Fileviez~Perez, and Wise}}]{Dulaney:2010dj}
\bibinfo{author}{\bibfnamefont{T.~R.} \bibnamefont{Dulaney}},
  \bibinfo{author}{\bibfnamefont{P.}~\bibnamefont{Fileviez~Perez}},
  \bibnamefont{and} \bibinfo{author}{\bibfnamefont{M.~B.} \bibnamefont{Wise}},
  \bibinfo{journal}{Phys.Rev.} \textbf{\bibinfo{volume}{D83}},
  \bibinfo{pages}{023520} (\bibinfo{year}{2011}), \eprint{1005.0617}.

\bibitem[{\citenamefont{Chun}(2011{\natexlab{a}})}]{Chun:2010hz}
\bibinfo{author}{\bibfnamefont{E.~J.} \bibnamefont{Chun}},
  \bibinfo{journal}{Phys.Rev.} \textbf{\bibinfo{volume}{D83}},
  \bibinfo{pages}{053004} (\bibinfo{year}{2011}{\natexlab{a}}),
  \eprint{1009.0983}.

\bibitem[{\citenamefont{Arina and Sahu}(2012)}]{Arina:2011cu}
\bibinfo{author}{\bibfnamefont{C.}~\bibnamefont{Arina}} \bibnamefont{and}
  \bibinfo{author}{\bibfnamefont{N.}~\bibnamefont{Sahu}},
  \bibinfo{journal}{Nucl.Phys.} \textbf{\bibinfo{volume}{B854}},
  \bibinfo{pages}{666} (\bibinfo{year}{2012}), \eprint{1108.3967}.

\bibitem[{\citenamefont{Barr}(2012)}]{Barr:2011cz}
\bibinfo{author}{\bibfnamefont{S.}~\bibnamefont{Barr}},
  \bibinfo{journal}{Phys.Rev.} \textbf{\bibinfo{volume}{D85}},
  \bibinfo{pages}{013001} (\bibinfo{year}{2012}), \eprint{1109.2562}.

\bibitem[{\citenamefont{Chun}(2011{\natexlab{b}})}]{Chun:2011cc}
\bibinfo{author}{\bibfnamefont{E.~J.} \bibnamefont{Chun}},
  \bibinfo{journal}{JHEP} \textbf{\bibinfo{volume}{1103}}, \bibinfo{pages}{098}
  (\bibinfo{year}{2011}{\natexlab{b}}), \eprint{1102.3455}.

\bibitem[{\citenamefont{Blum et~al.}(2012)\citenamefont{Blum, Efrati, Grossman,
  Nir, and Riotto}}]{Blum:2012nf}
\bibinfo{author}{\bibfnamefont{K.}~\bibnamefont{Blum}},
  \bibinfo{author}{\bibfnamefont{A.}~\bibnamefont{Efrati}},
  \bibinfo{author}{\bibfnamefont{Y.}~\bibnamefont{Grossman}},
  \bibinfo{author}{\bibfnamefont{Y.}~\bibnamefont{Nir}}, \bibnamefont{and}
  \bibinfo{author}{\bibfnamefont{A.}~\bibnamefont{Riotto}},
  \bibinfo{journal}{Phys.Rev.Lett.} \textbf{\bibinfo{volume}{109}},
  \bibinfo{pages}{051302} (\bibinfo{year}{2012}), \eprint{1201.2699}.

\bibitem[{\citenamefont{Arina et~al.}(2012)\citenamefont{Arina, Gong, and
  Sahu}}]{Arina:2012fb}
\bibinfo{author}{\bibfnamefont{C.}~\bibnamefont{Arina}},
  \bibinfo{author}{\bibfnamefont{J.-O.} \bibnamefont{Gong}}, \bibnamefont{and}
  \bibinfo{author}{\bibfnamefont{N.}~\bibnamefont{Sahu}},
  \bibinfo{journal}{Nucl.Phys.} \textbf{\bibinfo{volume}{B865}},
  \bibinfo{pages}{430} (\bibinfo{year}{2012}), \eprint{1206.0009}.

\bibitem[{\citenamefont{Arina et~al.}(2013)\citenamefont{Arina, Mohapatra, and
  Sahu}}]{Arina:2012aj}
\bibinfo{author}{\bibfnamefont{C.}~\bibnamefont{Arina}},
  \bibinfo{author}{\bibfnamefont{R.~N.} \bibnamefont{Mohapatra}},
  \bibnamefont{and} \bibinfo{author}{\bibfnamefont{N.}~\bibnamefont{Sahu}},
  \bibinfo{journal}{Phys.Lett.} \textbf{\bibinfo{volume}{B720}},
  \bibinfo{pages}{130} (\bibinfo{year}{2013}), \eprint{1211.0435}.

\bibitem[{\citenamefont{Kuismanen and Vilja}(2013)}]{Kuismanen:2012iz}
\bibinfo{author}{\bibfnamefont{H.}~\bibnamefont{Kuismanen}} \bibnamefont{and}
  \bibinfo{author}{\bibfnamefont{I.}~\bibnamefont{Vilja}},
  \bibinfo{journal}{Phys.Rev.} \textbf{\bibinfo{volume}{D87}},
  \bibinfo{pages}{015005} (\bibinfo{year}{2013}), \eprint{1210.4335}.

\bibitem[{\citenamefont{Fileviez~Perez and Wise}(2013)}]{Perez:2013nra}
\bibinfo{author}{\bibfnamefont{P.}~\bibnamefont{Fileviez~Perez}}
  \bibnamefont{and} \bibinfo{author}{\bibfnamefont{M.~B.} \bibnamefont{Wise}}
  (\bibinfo{year}{2013}), \eprint{1303.1452}.

\bibitem[{\citenamefont{Ignatiev et~al.}(1978)\citenamefont{Ignatiev,
  Krasnikov, Kuzmin, and Tavkhelidze}}]{Ignatiev:1978uf}
\bibinfo{author}{\bibfnamefont{A.~Y.} \bibnamefont{Ignatiev}},
  \bibinfo{author}{\bibfnamefont{N.}~\bibnamefont{Krasnikov}},
  \bibinfo{author}{\bibfnamefont{V.}~\bibnamefont{Kuzmin}}, \bibnamefont{and}
  \bibinfo{author}{\bibfnamefont{A.}~\bibnamefont{Tavkhelidze}},
  \bibinfo{journal}{Phys.Lett.} \textbf{\bibinfo{volume}{B76}},
  \bibinfo{pages}{436} (\bibinfo{year}{1978}).

\bibitem[{\citenamefont{Suematsu}(2006{\natexlab{a}})}]{Suematsu:2005kp}
\bibinfo{author}{\bibfnamefont{D.}~\bibnamefont{Suematsu}},
  \bibinfo{journal}{Astropart. Phys.} \textbf{\bibinfo{volume}{24}},
  \bibinfo{pages}{511} (\bibinfo{year}{2006}{\natexlab{a}}),
  \eprint{hep-ph/0510251}.

\bibitem[{\citenamefont{Suematsu}(2006{\natexlab{b}})}]{Suematsu:2005zc}
\bibinfo{author}{\bibfnamefont{D.}~\bibnamefont{Suematsu}},
  \bibinfo{journal}{JCAP} \textbf{\bibinfo{volume}{0601}}, \bibinfo{pages}{026}
  (\bibinfo{year}{2006}{\natexlab{b}}), \eprint{astro-ph/0511742}.

\bibitem[{\citenamefont{Kane et~al.}(2011)\citenamefont{Kane, Shao, Watson, and
  Yu}}]{Kane:2011ih}
\bibinfo{author}{\bibfnamefont{G.}~\bibnamefont{Kane}},
  \bibinfo{author}{\bibfnamefont{J.}~\bibnamefont{Shao}},
  \bibinfo{author}{\bibfnamefont{S.}~\bibnamefont{Watson}}, \bibnamefont{and}
  \bibinfo{author}{\bibfnamefont{H.-B.} \bibnamefont{Yu}},
  \bibinfo{journal}{JCAP} \textbf{\bibinfo{volume}{1111}}, \bibinfo{pages}{012}
  (\bibinfo{year}{2011}), \eprint{1108.5178}.

\bibitem[{\citenamefont{Choi and Seto}(2013)}]{Choi:2013fva}
\bibinfo{author}{\bibfnamefont{K.-Y.} \bibnamefont{Choi}} \bibnamefont{and}
  \bibinfo{author}{\bibfnamefont{O.}~\bibnamefont{Seto}}
  (\bibinfo{year}{2013}), \eprint{1305.4322}.

\bibitem[{\citenamefont{Kaplan}(1992)}]{Kaplan:1991ah}
\bibinfo{author}{\bibfnamefont{D.~B.} \bibnamefont{Kaplan}},
  \bibinfo{journal}{Phys.Rev.Lett.} \textbf{\bibinfo{volume}{68}},
  \bibinfo{pages}{741} (\bibinfo{year}{1992}).

\bibitem[{\citenamefont{Walker}(2012{\natexlab{a}})}]{Walker:2012ka}
\bibinfo{author}{\bibfnamefont{D.~G.} \bibnamefont{Walker}}
  (\bibinfo{year}{2012}{\natexlab{a}}), \eprint{1202.2348}.

\bibitem[{\citenamefont{McDonald}(2002)}]{McDonald:2001vt}
\bibinfo{author}{\bibfnamefont{J.}~\bibnamefont{McDonald}},
  \bibinfo{journal}{Phys.Rev.Lett.} \textbf{\bibinfo{volume}{88}},
  \bibinfo{pages}{091304} (\bibinfo{year}{2002}), \eprint{hep-ph/0106249}.

\bibitem[{\citenamefont{Hall et~al.}(2010{\natexlab{b}})\citenamefont{Hall,
  Jedamzik, March-Russell, and West}}]{Hall:2009bx}
\bibinfo{author}{\bibfnamefont{L.~J.} \bibnamefont{Hall}},
  \bibinfo{author}{\bibfnamefont{K.}~\bibnamefont{Jedamzik}},
  \bibinfo{author}{\bibfnamefont{J.}~\bibnamefont{March-Russell}},
  \bibnamefont{and} \bibinfo{author}{\bibfnamefont{S.~M.} \bibnamefont{West}},
  \bibinfo{journal}{JHEP} \textbf{\bibinfo{volume}{1003}}, \bibinfo{pages}{080}
  (\bibinfo{year}{2010}{\natexlab{b}}), \eprint{0911.1120}.

\bibitem[{\citenamefont{Banks et~al.}(2006)\citenamefont{Banks, Echols, and
  Jones}}]{Banks:2006xr}
\bibinfo{author}{\bibfnamefont{T.}~\bibnamefont{Banks}},
  \bibinfo{author}{\bibfnamefont{S.}~\bibnamefont{Echols}}, \bibnamefont{and}
  \bibinfo{author}{\bibfnamefont{J.}~\bibnamefont{Jones}},
  \bibinfo{journal}{JHEP} \textbf{\bibinfo{volume}{0611}}, \bibinfo{pages}{046}
  (\bibinfo{year}{2006}), \eprint{hep-ph/0608104}.

\bibitem[{\citenamefont{Kamada and Yamaguchi}(2012)}]{Kamada:2012ht}
\bibinfo{author}{\bibfnamefont{K.}~\bibnamefont{Kamada}} \bibnamefont{and}
  \bibinfo{author}{\bibfnamefont{M.}~\bibnamefont{Yamaguchi}},
  \bibinfo{journal}{Phys.Rev.} \textbf{\bibinfo{volume}{D85}},
  \bibinfo{pages}{103530} (\bibinfo{year}{2012}), \eprint{1201.2636}.

\bibitem[{\citenamefont{Steigman et~al.}(2012)\citenamefont{Steigman, Dasgupta,
  and Beacom}}]{Steigman:2012nb}
\bibinfo{author}{\bibfnamefont{G.}~\bibnamefont{Steigman}},
  \bibinfo{author}{\bibfnamefont{B.}~\bibnamefont{Dasgupta}}, \bibnamefont{and}
  \bibinfo{author}{\bibfnamefont{J.~F.} \bibnamefont{Beacom}},
  \bibinfo{journal}{Phys.Rev.} \textbf{\bibinfo{volume}{D86}},
  \bibinfo{pages}{023506} (\bibinfo{year}{2012}), \eprint{1204.3622}.

\bibitem[{\citenamefont{Kolb and Turner}(1990)}]{Kolb:1990vq}
\bibinfo{author}{\bibfnamefont{E.~W.} \bibnamefont{Kolb}} \bibnamefont{and}
  \bibinfo{author}{\bibfnamefont{M.~S.} \bibnamefont{Turner}},
  \bibinfo{journal}{Front.Phys.} \textbf{\bibinfo{volume}{69}},
  \bibinfo{pages}{1} (\bibinfo{year}{1990}).

\bibitem[{\citenamefont{Graesser
  et~al.}(2011{\natexlab{b}})\citenamefont{Graesser, Shoemaker, and
  Vecchi}}]{Graesser:2011wi}
\bibinfo{author}{\bibfnamefont{M.~L.} \bibnamefont{Graesser}},
  \bibinfo{author}{\bibfnamefont{I.~M.} \bibnamefont{Shoemaker}},
  \bibnamefont{and} \bibinfo{author}{\bibfnamefont{L.}~\bibnamefont{Vecchi}},
  \bibinfo{journal}{JHEP} \textbf{\bibinfo{volume}{1110}}, \bibinfo{pages}{110}
  (\bibinfo{year}{2011}{\natexlab{b}}), \eprint{1103.2771}.

\bibitem[{\citenamefont{Iminniyaz et~al.}(2011)\citenamefont{Iminniyaz, Drees,
  and Chen}}]{Iminniyaz:2011yp}
\bibinfo{author}{\bibfnamefont{H.}~\bibnamefont{Iminniyaz}},
  \bibinfo{author}{\bibfnamefont{M.}~\bibnamefont{Drees}}, \bibnamefont{and}
  \bibinfo{author}{\bibfnamefont{X.}~\bibnamefont{Chen}},
  \bibinfo{journal}{JCAP} \textbf{\bibinfo{volume}{1107}}, \bibinfo{pages}{003}
  (\bibinfo{year}{2011}), \eprint{1104.5548}.

\bibitem[{\citenamefont{Okada and Seto}(2012)}]{Okada:2012rm}
\bibinfo{author}{\bibfnamefont{N.}~\bibnamefont{Okada}} \bibnamefont{and}
  \bibinfo{author}{\bibfnamefont{O.}~\bibnamefont{Seto}},
  \bibinfo{journal}{Phys.Rev.} \textbf{\bibinfo{volume}{D86}},
  \bibinfo{pages}{063525} (\bibinfo{year}{2012}), \eprint{1205.2844}.

\bibitem[{\citenamefont{Buckley and Profumo}(2012)}]{Buckley:2011ye}
\bibinfo{author}{\bibfnamefont{M.~R.} \bibnamefont{Buckley}} \bibnamefont{and}
  \bibinfo{author}{\bibfnamefont{S.}~\bibnamefont{Profumo}},
  \bibinfo{journal}{Phys.Rev.Lett.} \textbf{\bibinfo{volume}{108}},
  \bibinfo{pages}{011301} (\bibinfo{year}{2012}), \eprint{1109.2164}.

\bibitem[{\citenamefont{Cirelli et~al.}(2012)\citenamefont{Cirelli, Panci,
  Servant, and Zaharijas}}]{Cirelli:2011ac}
\bibinfo{author}{\bibfnamefont{M.}~\bibnamefont{Cirelli}},
  \bibinfo{author}{\bibfnamefont{P.}~\bibnamefont{Panci}},
  \bibinfo{author}{\bibfnamefont{G.}~\bibnamefont{Servant}}, \bibnamefont{and}
  \bibinfo{author}{\bibfnamefont{G.}~\bibnamefont{Zaharijas}},
  \bibinfo{journal}{JCAP} \textbf{\bibinfo{volume}{1203}}, \bibinfo{pages}{015}
  (\bibinfo{year}{2012}), \eprint{1110.3809}.

\bibitem[{\citenamefont{Tulin et~al.}(2012)\citenamefont{Tulin, Yu, and
  Zurek}}]{Tulin:2012re}
\bibinfo{author}{\bibfnamefont{S.}~\bibnamefont{Tulin}},
  \bibinfo{author}{\bibfnamefont{H.-B.} \bibnamefont{Yu}}, \bibnamefont{and}
  \bibinfo{author}{\bibfnamefont{K.~M.} \bibnamefont{Zurek}},
  \bibinfo{journal}{JCAP} \textbf{\bibinfo{volume}{1205}}, \bibinfo{pages}{013}
  (\bibinfo{year}{2012}), \eprint{1202.0283}.

\bibitem[{\citenamefont{Ivanov and Keus}(2012)}]{Ivanov:2012hc}
\bibinfo{author}{\bibfnamefont{I.}~\bibnamefont{Ivanov}} \bibnamefont{and}
  \bibinfo{author}{\bibfnamefont{V.}~\bibnamefont{Keus}},
  \bibinfo{journal}{Phys.Rev.} \textbf{\bibinfo{volume}{D86}},
  \bibinfo{pages}{016004} (\bibinfo{year}{2012}), \eprint{1203.3426}.

\bibitem[{\citenamefont{Gelmini et~al.}(2013)\citenamefont{Gelmini, Huh, and
  Rehagen}}]{Gelmini:2013awa}
\bibinfo{author}{\bibfnamefont{G.~B.} \bibnamefont{Gelmini}},
  \bibinfo{author}{\bibfnamefont{J.-H.} \bibnamefont{Huh}}, \bibnamefont{and}
  \bibinfo{author}{\bibfnamefont{T.}~\bibnamefont{Rehagen}}
  (\bibinfo{year}{2013}), \eprint{1304.3679}.

\bibitem[{\citenamefont{March-Russell et~al.}(2012)\citenamefont{March-Russell,
  Unwin, and West}}]{MarchRussell:2012hi}
\bibinfo{author}{\bibfnamefont{J.}~\bibnamefont{March-Russell}},
  \bibinfo{author}{\bibfnamefont{J.}~\bibnamefont{Unwin}}, \bibnamefont{and}
  \bibinfo{author}{\bibfnamefont{S.~M.} \bibnamefont{West}},
  \bibinfo{journal}{JHEP} \textbf{\bibinfo{volume}{1208}}, \bibinfo{pages}{029}
  (\bibinfo{year}{2012}), \eprint{1203.4854}.

\bibitem[{\citenamefont{Holdom}(1986)}]{Holdom:1985ag}
\bibinfo{author}{\bibfnamefont{B.}~\bibnamefont{Holdom}},
  \bibinfo{journal}{Phys.Lett.} \textbf{\bibinfo{volume}{B166}},
  \bibinfo{pages}{196} (\bibinfo{year}{1986}).

\bibitem[{\citenamefont{Foot and He}(1991)}]{Foot:1991kb}
\bibinfo{author}{\bibfnamefont{R.}~\bibnamefont{Foot}} \bibnamefont{and}
  \bibinfo{author}{\bibfnamefont{X.-G.} \bibnamefont{He}},
  \bibinfo{journal}{Phys.Lett.} \textbf{\bibinfo{volume}{B267}},
  \bibinfo{pages}{509} (\bibinfo{year}{1991}).

\bibitem[{\citenamefont{Cyburt et~al.}(2005)\citenamefont{Cyburt, Fields,
  Olive, and Skillman}}]{Cyburt:2004yc}
\bibinfo{author}{\bibfnamefont{R.~H.} \bibnamefont{Cyburt}},
  \bibinfo{author}{\bibfnamefont{B.~D.} \bibnamefont{Fields}},
  \bibinfo{author}{\bibfnamefont{K.~A.} \bibnamefont{Olive}}, \bibnamefont{and}
  \bibinfo{author}{\bibfnamefont{E.}~\bibnamefont{Skillman}},
  \bibinfo{journal}{Astropart.Phys.} \textbf{\bibinfo{volume}{23}},
  \bibinfo{pages}{313} (\bibinfo{year}{2005}), \eprint{astro-ph/0408033}.

\bibitem[{\citenamefont{Mangano and Serpico}(2011)}]{Mangano:2011ar}
\bibinfo{author}{\bibfnamefont{G.}~\bibnamefont{Mangano}} \bibnamefont{and}
  \bibinfo{author}{\bibfnamefont{P.~D.} \bibnamefont{Serpico}},
  \bibinfo{journal}{Phys.Lett.} \textbf{\bibinfo{volume}{B701}},
  \bibinfo{pages}{296} (\bibinfo{year}{2011}), \eprint{1103.1261}.

\bibitem[{\citenamefont{Harvey and Turner}(1990)}]{Harvey:1990qw}
\bibinfo{author}{\bibfnamefont{J.~A.} \bibnamefont{Harvey}} \bibnamefont{and}
  \bibinfo{author}{\bibfnamefont{M.~S.} \bibnamefont{Turner}},
  \bibinfo{journal}{Phys.Rev.} \textbf{\bibinfo{volume}{D42}},
  \bibinfo{pages}{3344} (\bibinfo{year}{1990}).

\bibitem[{\citenamefont{Ibe et~al.}(2012)\citenamefont{Ibe, Matsumoto, and
  Yanagida}}]{Ibe:2011hq}
\bibinfo{author}{\bibfnamefont{M.}~\bibnamefont{Ibe}},
  \bibinfo{author}{\bibfnamefont{S.}~\bibnamefont{Matsumoto}},
  \bibnamefont{and} \bibinfo{author}{\bibfnamefont{T.~T.}
  \bibnamefont{Yanagida}}, \bibinfo{journal}{Phys.Lett.}
  \textbf{\bibinfo{volume}{B708}}, \bibinfo{pages}{112} (\bibinfo{year}{2012}),
  \eprint{1110.5452}.

\bibitem[{\citenamefont{Kusenko}(1997{\natexlab{a}})}]{Kusenko:1997ad}
\bibinfo{author}{\bibfnamefont{A.}~\bibnamefont{Kusenko}},
  \bibinfo{journal}{Phys.Lett.} \textbf{\bibinfo{volume}{B404}},
  \bibinfo{pages}{285} (\bibinfo{year}{1997}{\natexlab{a}}),
  \eprint{hep-th/9704073}.

\bibitem[{\citenamefont{Kusenko}(1997{\natexlab{b}})}]{Kusenko:1997zq}
\bibinfo{author}{\bibfnamefont{A.}~\bibnamefont{Kusenko}},
  \bibinfo{journal}{Phys.Lett.} \textbf{\bibinfo{volume}{B405}},
  \bibinfo{pages}{108} (\bibinfo{year}{1997}{\natexlab{b}}),
  \eprint{hep-ph/9704273}.

\bibitem[{\citenamefont{Cohen et~al.}(2010)\citenamefont{Cohen, Phalen, Pierce,
  and Zurek}}]{Cohen:2010kn}
\bibinfo{author}{\bibfnamefont{T.}~\bibnamefont{Cohen}},
  \bibinfo{author}{\bibfnamefont{D.~J.} \bibnamefont{Phalen}},
  \bibinfo{author}{\bibfnamefont{A.}~\bibnamefont{Pierce}}, \bibnamefont{and}
  \bibinfo{author}{\bibfnamefont{K.~M.} \bibnamefont{Zurek}},
  \bibinfo{journal}{Phys.Rev.} \textbf{\bibinfo{volume}{D82}},
  \bibinfo{pages}{056001} (\bibinfo{year}{2010}), \eprint{1005.1655}.

\bibitem[{\citenamefont{Khlopov}(2006)}]{Khlopov:2005ew}
\bibinfo{author}{\bibfnamefont{M.~Y.} \bibnamefont{Khlopov}},
  \bibinfo{journal}{Pisma Zh.Eksp.Teor.Fiz.} \textbf{\bibinfo{volume}{83}},
  \bibinfo{pages}{3} (\bibinfo{year}{2006}), \eprint{astro-ph/0511796}.

\bibitem[{\citenamefont{Gudnason
  et~al.}(2006{\natexlab{a}})\citenamefont{Gudnason, Kouvaris, and
  Sannino}}]{Gudnason:2006ug}
\bibinfo{author}{\bibfnamefont{S.~B.} \bibnamefont{Gudnason}},
  \bibinfo{author}{\bibfnamefont{C.}~\bibnamefont{Kouvaris}}, \bibnamefont{and}
  \bibinfo{author}{\bibfnamefont{F.}~\bibnamefont{Sannino}},
  \bibinfo{journal}{Phys.Rev.} \textbf{\bibinfo{volume}{D73}},
  \bibinfo{pages}{115003} (\bibinfo{year}{2006}{\natexlab{a}}),
  \eprint{hep-ph/0603014}.

\bibitem[{\citenamefont{Gudnason
  et~al.}(2006{\natexlab{b}})\citenamefont{Gudnason, Kouvaris, and
  Sannino}}]{Gudnason:2006yj}
\bibinfo{author}{\bibfnamefont{S.~B.} \bibnamefont{Gudnason}},
  \bibinfo{author}{\bibfnamefont{C.}~\bibnamefont{Kouvaris}}, \bibnamefont{and}
  \bibinfo{author}{\bibfnamefont{F.}~\bibnamefont{Sannino}},
  \bibinfo{journal}{Phys.Rev.} \textbf{\bibinfo{volume}{D74}},
  \bibinfo{pages}{095008} (\bibinfo{year}{2006}{\natexlab{b}}),
  \eprint{hep-ph/0608055}.

\bibitem[{\citenamefont{Khlopov and Kouvaris}(2008)}]{Khlopov:2008ty}
\bibinfo{author}{\bibfnamefont{M.~Y.} \bibnamefont{Khlopov}} \bibnamefont{and}
  \bibinfo{author}{\bibfnamefont{C.}~\bibnamefont{Kouvaris}},
  \bibinfo{journal}{Phys.Rev.} \textbf{\bibinfo{volume}{D78}},
  \bibinfo{pages}{065040} (\bibinfo{year}{2008}), \eprint{0806.1191}.

\bibitem[{\citenamefont{Ryttov and Sannino}(2008)}]{Ryttov:2008xe}
\bibinfo{author}{\bibfnamefont{T.~A.} \bibnamefont{Ryttov}} \bibnamefont{and}
  \bibinfo{author}{\bibfnamefont{F.}~\bibnamefont{Sannino}},
  \bibinfo{journal}{Phys.Rev.} \textbf{\bibinfo{volume}{D78}},
  \bibinfo{pages}{115010} (\bibinfo{year}{2008}), \eprint{0809.0713}.

\bibitem[{\citenamefont{Foadi et~al.}(2009)\citenamefont{Foadi, Frandsen, and
  Sannino}}]{Foadi:2008qv}
\bibinfo{author}{\bibfnamefont{R.}~\bibnamefont{Foadi}},
  \bibinfo{author}{\bibfnamefont{M.~T.} \bibnamefont{Frandsen}},
  \bibnamefont{and} \bibinfo{author}{\bibfnamefont{F.}~\bibnamefont{Sannino}},
  \bibinfo{journal}{Phys.Rev.} \textbf{\bibinfo{volume}{D80}},
  \bibinfo{pages}{037702} (\bibinfo{year}{2009}), \eprint{0812.3406}.

\bibitem[{\citenamefont{Frandsen and Sannino}(2010)}]{Frandsen:2009mi}
\bibinfo{author}{\bibfnamefont{M.~T.} \bibnamefont{Frandsen}} \bibnamefont{and}
  \bibinfo{author}{\bibfnamefont{F.}~\bibnamefont{Sannino}},
  \bibinfo{journal}{Phys.Rev.} \textbf{\bibinfo{volume}{D81}},
  \bibinfo{pages}{097704} (\bibinfo{year}{2010}), \eprint{0911.1570}.

\bibitem[{\citenamefont{Frandsen et~al.}(2011)\citenamefont{Frandsen, Sarkar,
  and Schmidt-Hoberg}}]{Frandsen:2011kt}
\bibinfo{author}{\bibfnamefont{M.~T.} \bibnamefont{Frandsen}},
  \bibinfo{author}{\bibfnamefont{S.}~\bibnamefont{Sarkar}}, \bibnamefont{and}
  \bibinfo{author}{\bibfnamefont{K.}~\bibnamefont{Schmidt-Hoberg}},
  \bibinfo{journal}{Phys.Rev.} \textbf{\bibinfo{volume}{D84}},
  \bibinfo{pages}{051703} (\bibinfo{year}{2011}), \eprint{1103.4350}.

\bibitem[{\citenamefont{Khlopov}(2011)}]{Khlopov:2011tn}
\bibinfo{author}{\bibfnamefont{M.~Y.} \bibnamefont{Khlopov}},
  \bibinfo{journal}{Mod.Phys.Lett.} \textbf{\bibinfo{volume}{A26}},
  \bibinfo{pages}{2823} (\bibinfo{year}{2011}), \eprint{1111.2838}.

\bibitem[{\citenamefont{Lewis et~al.}(2012)\citenamefont{Lewis, Pica, and
  Sannino}}]{Lewis:2011zb}
\bibinfo{author}{\bibfnamefont{R.}~\bibnamefont{Lewis}},
  \bibinfo{author}{\bibfnamefont{C.}~\bibnamefont{Pica}}, \bibnamefont{and}
  \bibinfo{author}{\bibfnamefont{F.}~\bibnamefont{Sannino}},
  \bibinfo{journal}{Phys.Rev.} \textbf{\bibinfo{volume}{D85}},
  \bibinfo{pages}{014504} (\bibinfo{year}{2012}), \eprint{1109.3513}.

\bibitem[{\citenamefont{Foot et~al.}(2000)\citenamefont{Foot, Lew, and
  Volkas}}]{Foot:2000tp}
\bibinfo{author}{\bibfnamefont{R.}~\bibnamefont{Foot}},
  \bibinfo{author}{\bibfnamefont{H.}~\bibnamefont{Lew}}, \bibnamefont{and}
  \bibinfo{author}{\bibfnamefont{R.}~\bibnamefont{Volkas}},
  \bibinfo{journal}{JHEP} \textbf{\bibinfo{volume}{0007}}, \bibinfo{pages}{032}
  (\bibinfo{year}{2000}), \eprint{hep-ph/0006027}.

\bibitem[{\citenamefont{Berezhiani and Mohapatra}(1995)}]{Berezhiani:1995yi}
\bibinfo{author}{\bibfnamefont{Z.~G.} \bibnamefont{Berezhiani}}
  \bibnamefont{and} \bibinfo{author}{\bibfnamefont{R.~N.}
  \bibnamefont{Mohapatra}}, \bibinfo{journal}{Phys.Rev.}
  \textbf{\bibinfo{volume}{D52}}, \bibinfo{pages}{6607} (\bibinfo{year}{1995}),
  \eprint{hep-ph/9505385}.

\bibitem[{\citenamefont{Cui et~al.}(2012{\natexlab{b}})\citenamefont{Cui, He,
  Lu, and Yin}}]{Cui:2011wk}
\bibinfo{author}{\bibfnamefont{J.-W.} \bibnamefont{Cui}},
  \bibinfo{author}{\bibfnamefont{H.-J.} \bibnamefont{He}},
  \bibinfo{author}{\bibfnamefont{L.-C.} \bibnamefont{Lu}}, \bibnamefont{and}
  \bibinfo{author}{\bibfnamefont{F.-R.} \bibnamefont{Yin}},
  \bibinfo{journal}{Phys.Rev.} \textbf{\bibinfo{volume}{D85}},
  \bibinfo{pages}{096003} (\bibinfo{year}{2012}{\natexlab{b}}),
  \eprint{1110.6893}.

\bibitem[{\citenamefont{Gu}(2013{\natexlab{a}})}]{Gu:2012fg}
\bibinfo{author}{\bibfnamefont{P.-H.} \bibnamefont{Gu}},
  \bibinfo{journal}{Nucl.Phys.} \textbf{\bibinfo{volume}{B872}},
  \bibinfo{pages}{38} (\bibinfo{year}{2013}{\natexlab{a}}), \eprint{1209.4579}.

\bibitem[{\citenamefont{Gu}(2013{\natexlab{b}})}]{Gu:2013nya}
\bibinfo{author}{\bibfnamefont{P.-H.} \bibnamefont{Gu}}
  (\bibinfo{year}{2013}{\natexlab{b}}), \eprint{1303.6545}.

\bibitem[{\citenamefont{Foot and Volkas}(2004{\natexlab{b}})}]{Foot:2004wz}
\bibinfo{author}{\bibfnamefont{R.}~\bibnamefont{Foot}} \bibnamefont{and}
  \bibinfo{author}{\bibfnamefont{R.}~\bibnamefont{Volkas}},
  \bibinfo{journal}{Phys.Rev.} \textbf{\bibinfo{volume}{D70}},
  \bibinfo{pages}{123508} (\bibinfo{year}{2004}{\natexlab{b}}),
  \eprint{astro-ph/0407522}.

\bibitem[{\citenamefont{Gherghetta et~al.}(1996)\citenamefont{Gherghetta,
  Kolda, and Martin}}]{Gherghetta:1995dv}
\bibinfo{author}{\bibfnamefont{T.}~\bibnamefont{Gherghetta}},
  \bibinfo{author}{\bibfnamefont{C.~F.} \bibnamefont{Kolda}}, \bibnamefont{and}
  \bibinfo{author}{\bibfnamefont{S.~P.} \bibnamefont{Martin}},
  \bibinfo{journal}{Nucl.Phys.} \textbf{\bibinfo{volume}{B468}},
  \bibinfo{pages}{37} (\bibinfo{year}{1996}), \eprint{hep-ph/9510370}.

\bibitem[{\citenamefont{Sievers et~al.}(2013)\citenamefont{Sievers, Hlozek,
  Nolta, Acquaviva, Addison et~al.}}]{Sievers:2013ica}
\bibinfo{author}{\bibfnamefont{J.~L.} \bibnamefont{Sievers}},
  \bibinfo{author}{\bibfnamefont{R.~A.} \bibnamefont{Hlozek}},
  \bibinfo{author}{\bibfnamefont{M.~R.} \bibnamefont{Nolta}},
  \bibinfo{author}{\bibfnamefont{V.}~\bibnamefont{Acquaviva}},
  \bibinfo{author}{\bibfnamefont{G.~E.} \bibnamefont{Addison}},
  \bibnamefont{et~al.} (\bibinfo{year}{2013}), \eprint{1301.0824}.

\bibitem[{\citenamefont{Hou et~al.}(2012)\citenamefont{Hou, Reichardt, Story,
  Follin, Keisler et~al.}}]{Hou:2012xq}
\bibinfo{author}{\bibfnamefont{Z.}~\bibnamefont{Hou}},
  \bibinfo{author}{\bibfnamefont{C.}~\bibnamefont{Reichardt}},
  \bibinfo{author}{\bibfnamefont{K.}~\bibnamefont{Story}},
  \bibinfo{author}{\bibfnamefont{B.}~\bibnamefont{Follin}},
  \bibinfo{author}{\bibfnamefont{R.}~\bibnamefont{Keisler}},
  \bibnamefont{et~al.} (\bibinfo{year}{2012}), \eprint{1212.6267}.

\bibitem[{\citenamefont{Padmanabhan et~al.}(2012)\citenamefont{Padmanabhan, Xu,
  Eisenstein, Scalzo, Cuesta et~al.}}]{Padmanabhan:2012hf}
\bibinfo{author}{\bibfnamefont{N.}~\bibnamefont{Padmanabhan}},
  \bibinfo{author}{\bibfnamefont{X.}~\bibnamefont{Xu}},
  \bibinfo{author}{\bibfnamefont{D.~J.} \bibnamefont{Eisenstein}},
  \bibinfo{author}{\bibfnamefont{R.}~\bibnamefont{Scalzo}},
  \bibinfo{author}{\bibfnamefont{A.~J.} \bibnamefont{Cuesta}},
  \bibnamefont{et~al.} (\bibinfo{year}{2012}), \eprint{1202.0090}.

\bibitem[{\citenamefont{Anderson et~al.}(2013)\citenamefont{Anderson, Aubourg,
  Bailey, Bizyaev, Blanton et~al.}}]{Anderson:2012sa}
\bibinfo{author}{\bibfnamefont{L.}~\bibnamefont{Anderson}},
  \bibinfo{author}{\bibfnamefont{E.}~\bibnamefont{Aubourg}},
  \bibinfo{author}{\bibfnamefont{S.}~\bibnamefont{Bailey}},
  \bibinfo{author}{\bibfnamefont{D.}~\bibnamefont{Bizyaev}},
  \bibinfo{author}{\bibfnamefont{M.}~\bibnamefont{Blanton}},
  \bibnamefont{et~al.}, \bibinfo{journal}{Mon.Not.Roy.Astron.Soc.}
  \textbf{\bibinfo{volume}{427}}, \bibinfo{pages}{3435} (\bibinfo{year}{2013}),
  \eprint{1203.6594}.

\bibitem[{\citenamefont{Blake et~al.}(2011)\citenamefont{Blake, Davis, Poole,
  Parkinson, Brough et~al.}}]{Blake:2011wn}
\bibinfo{author}{\bibfnamefont{C.}~\bibnamefont{Blake}},
  \bibinfo{author}{\bibfnamefont{T.}~\bibnamefont{Davis}},
  \bibinfo{author}{\bibfnamefont{G.}~\bibnamefont{Poole}},
  \bibinfo{author}{\bibfnamefont{D.}~\bibnamefont{Parkinson}},
  \bibinfo{author}{\bibfnamefont{S.}~\bibnamefont{Brough}},
  \bibnamefont{et~al.}, \bibinfo{journal}{Mon.Not.Roy.Astron.Soc.}
  \textbf{\bibinfo{volume}{415}}, \bibinfo{pages}{2892} (\bibinfo{year}{2011}),
  \eprint{1105.2862}.

\bibitem[{\citenamefont{Blennow et~al.}(2012)\citenamefont{Blennow,
  Fernandez-Martinez, Mena, Redondo, and Serra}}]{Blennow:2012de}
\bibinfo{author}{\bibfnamefont{M.}~\bibnamefont{Blennow}},
  \bibinfo{author}{\bibfnamefont{E.}~\bibnamefont{Fernandez-Martinez}},
  \bibinfo{author}{\bibfnamefont{O.}~\bibnamefont{Mena}},
  \bibinfo{author}{\bibfnamefont{J.}~\bibnamefont{Redondo}}, \bibnamefont{and}
  \bibinfo{author}{\bibfnamefont{P.}~\bibnamefont{Serra}},
  \bibinfo{journal}{JCAP} \textbf{\bibinfo{volume}{1207}}, \bibinfo{pages}{022}
  (\bibinfo{year}{2012}), \eprint{1203.5803}.

\bibitem[{\citenamefont{Foot}(2013{\natexlab{b}})}]{Foot:2012ai}
\bibinfo{author}{\bibfnamefont{R.}~\bibnamefont{Foot}},
  \bibinfo{journal}{Phys.Lett.} \textbf{\bibinfo{volume}{B718}},
  \bibinfo{pages}{745} (\bibinfo{year}{2013}{\natexlab{b}}),
  \eprint{1208.6022}.

\bibitem[{\citenamefont{Cyr-Racine and Sigurdson}(2012)}]{CyrRacine:2012fz}
\bibinfo{author}{\bibfnamefont{F.-Y.} \bibnamefont{Cyr-Racine}}
  \bibnamefont{and} \bibinfo{author}{\bibfnamefont{K.}~\bibnamefont{Sigurdson}}
  (\bibinfo{year}{2012}), \eprint{1209.5752}.

\bibitem[{\citenamefont{Gilmore
  et~al.}(2007{\natexlab{a}})\citenamefont{Gilmore, Wilkinson, Kleyna, Koch,
  Evans et~al.}}]{Gilmore:2006iy}
\bibinfo{author}{\bibfnamefont{G.}~\bibnamefont{Gilmore}},
  \bibinfo{author}{\bibfnamefont{M.}~\bibnamefont{Wilkinson}},
  \bibinfo{author}{\bibfnamefont{J.}~\bibnamefont{Kleyna}},
  \bibinfo{author}{\bibfnamefont{A.}~\bibnamefont{Koch}},
  \bibinfo{author}{\bibfnamefont{N.~W.} \bibnamefont{Evans}},
  \bibnamefont{et~al.}, \bibinfo{journal}{Nucl.Phys.Proc.Suppl.}
  \textbf{\bibinfo{volume}{173}}, \bibinfo{pages}{15}
  (\bibinfo{year}{2007}{\natexlab{a}}), \eprint{astro-ph/0608528}.

\bibitem[{\citenamefont{Gilmore et~al.}(2007{\natexlab{b}})}]{Gilmore:2007fy}
\bibinfo{author}{\bibfnamefont{G.}~\bibnamefont{Gilmore}} \bibnamefont{et~al.},
  \bibinfo{journal}{Astrophys. J.} \textbf{\bibinfo{volume}{663}},
  \bibinfo{pages}{948} (\bibinfo{year}{2007}{\natexlab{b}}),
  \eprint{astro-ph/0703308}.

\bibitem[{\citenamefont{Gilmore et~al.}(2008)\citenamefont{Gilmore, Zucker,
  Wilkinson, Wyse, Belokurov et~al.}}]{Gilmore:2008yp}
\bibinfo{author}{\bibfnamefont{G.}~\bibnamefont{Gilmore}},
  \bibinfo{author}{\bibfnamefont{D.}~\bibnamefont{Zucker}},
  \bibinfo{author}{\bibfnamefont{M.}~\bibnamefont{Wilkinson}},
  \bibinfo{author}{\bibfnamefont{R.~F.} \bibnamefont{Wyse}},
  \bibinfo{author}{\bibfnamefont{V.}~\bibnamefont{Belokurov}},
  \bibnamefont{et~al.} (\bibinfo{year}{2008}), \eprint{0804.1919}.

\bibitem[{\citenamefont{de~Blok}(2010)}]{deBlok:2009sp}
\bibinfo{author}{\bibfnamefont{W.}~\bibnamefont{de~Blok}},
  \bibinfo{journal}{Adv.Astron.} \textbf{\bibinfo{volume}{2010}},
  \bibinfo{pages}{789293} (\bibinfo{year}{2010}), \eprint{0910.3538}.

\bibitem[{\citenamefont{Walker and Penarrubia}(2011)}]{Walker:2011zu}
\bibinfo{author}{\bibfnamefont{M.~G.} \bibnamefont{Walker}} \bibnamefont{and}
  \bibinfo{author}{\bibfnamefont{J.}~\bibnamefont{Penarrubia}},
  \bibinfo{journal}{Astrophys.J.} \textbf{\bibinfo{volume}{742}},
  \bibinfo{pages}{20} (\bibinfo{year}{2011}), \eprint{1108.2404}.

\bibitem[{\citenamefont{de~Naray and Spekkens}(2011)}]{deNaray:2011hy}
\bibinfo{author}{\bibfnamefont{R.~K.} \bibnamefont{de~Naray}} \bibnamefont{and}
  \bibinfo{author}{\bibfnamefont{K.}~\bibnamefont{Spekkens}},
  \bibinfo{journal}{Astrophys.J.} \textbf{\bibinfo{volume}{741}},
  \bibinfo{pages}{L29} (\bibinfo{year}{2011}), \eprint{1109.1288}.

\bibitem[{\citenamefont{Klypin et~al.}(1999)\citenamefont{Klypin, Kravtsov,
  Valenzuela, and Prada}}]{Klypin:1999uc}
\bibinfo{author}{\bibfnamefont{A.~A.} \bibnamefont{Klypin}},
  \bibinfo{author}{\bibfnamefont{A.~V.} \bibnamefont{Kravtsov}},
  \bibinfo{author}{\bibfnamefont{O.}~\bibnamefont{Valenzuela}},
  \bibnamefont{and} \bibinfo{author}{\bibfnamefont{F.}~\bibnamefont{Prada}},
  \bibinfo{journal}{Astrophys. J.} \textbf{\bibinfo{volume}{522}},
  \bibinfo{pages}{82} (\bibinfo{year}{1999}), \eprint{astro-ph/9901240}.

\bibitem[{\citenamefont{Moore et~al.}(1999)\citenamefont{Moore, Ghigna,
  Governato, Lake, Quinn et~al.}}]{Moore:1999nt}
\bibinfo{author}{\bibfnamefont{B.}~\bibnamefont{Moore}},
  \bibinfo{author}{\bibfnamefont{S.}~\bibnamefont{Ghigna}},
  \bibinfo{author}{\bibfnamefont{F.}~\bibnamefont{Governato}},
  \bibinfo{author}{\bibfnamefont{G.}~\bibnamefont{Lake}},
  \bibinfo{author}{\bibfnamefont{T.~R.} \bibnamefont{Quinn}},
  \bibnamefont{et~al.}, \bibinfo{journal}{Astrophys.J.}
  \textbf{\bibinfo{volume}{524}}, \bibinfo{pages}{L19} (\bibinfo{year}{1999}),
  \eprint{astro-ph/9907411}.

\bibitem[{\citenamefont{Kravtsov}(2010)}]{Kravtsov:2009gi}
\bibinfo{author}{\bibfnamefont{A.~V.} \bibnamefont{Kravtsov}},
  \bibinfo{journal}{Adv.Astron.} \textbf{\bibinfo{volume}{2010}},
  \bibinfo{pages}{281913} (\bibinfo{year}{2010}), \eprint{0906.3295}.

\bibitem[{\citenamefont{Boylan-Kolchin
  et~al.}(2011)\citenamefont{Boylan-Kolchin, Bullock, and
  Kaplinghat}}]{BoylanKolchin:2011de}
\bibinfo{author}{\bibfnamefont{M.}~\bibnamefont{Boylan-Kolchin}},
  \bibinfo{author}{\bibfnamefont{J.~S.} \bibnamefont{Bullock}},
  \bibnamefont{and}
  \bibinfo{author}{\bibfnamefont{M.}~\bibnamefont{Kaplinghat}},
  \bibinfo{journal}{Mon.Not.Roy.Astron.Soc.} \textbf{\bibinfo{volume}{415}},
  \bibinfo{pages}{L40} (\bibinfo{year}{2011}), \eprint{1103.0007}.

\bibitem[{\citenamefont{Mac~Low and Ferrara}(1999)}]{MacLow:1998wv}
\bibinfo{author}{\bibfnamefont{M.-M.} \bibnamefont{Mac~Low}} \bibnamefont{and}
  \bibinfo{author}{\bibfnamefont{A.}~\bibnamefont{Ferrara}},
  \bibinfo{journal}{Astrophys.J.} \textbf{\bibinfo{volume}{513}},
  \bibinfo{pages}{142} (\bibinfo{year}{1999}), \eprint{astro-ph/9801237}.

\bibitem[{\citenamefont{Barkana and Loeb}(1999)}]{Barkana:1999er}
\bibinfo{author}{\bibfnamefont{R.}~\bibnamefont{Barkana}} \bibnamefont{and}
  \bibinfo{author}{\bibfnamefont{A.}~\bibnamefont{Loeb}},
  \bibinfo{journal}{Astrophys.J.} \textbf{\bibinfo{volume}{523}},
  \bibinfo{pages}{54} (\bibinfo{year}{1999}), \eprint{astro-ph/9906398}.

\bibitem[{\citenamefont{Shapiro et~al.}(2004)\citenamefont{Shapiro, Iliev, and
  Raga}}]{Shapiro:2003gxa}
\bibinfo{author}{\bibfnamefont{P.~R.} \bibnamefont{Shapiro}},
  \bibinfo{author}{\bibfnamefont{I.~T.} \bibnamefont{Iliev}}, \bibnamefont{and}
  \bibinfo{author}{\bibfnamefont{A.~C.} \bibnamefont{Raga}},
  \bibinfo{journal}{Mon.Not.Roy.Astron.Soc.} \textbf{\bibinfo{volume}{348}},
  \bibinfo{pages}{753} (\bibinfo{year}{2004}), \eprint{astro-ph/0307266}.

\bibitem[{\citenamefont{Hoeft et~al.}(2006)\citenamefont{Hoeft, Yepes,
  Gottlober, and Springel}}]{Hoeft:2005jn}
\bibinfo{author}{\bibfnamefont{M.}~\bibnamefont{Hoeft}},
  \bibinfo{author}{\bibfnamefont{G.}~\bibnamefont{Yepes}},
  \bibinfo{author}{\bibfnamefont{S.}~\bibnamefont{Gottlober}},
  \bibnamefont{and} \bibinfo{author}{\bibfnamefont{V.}~\bibnamefont{Springel}},
  \bibinfo{journal}{Mon.Not.Roy.Astron.Soc.} \textbf{\bibinfo{volume}{371}},
  \bibinfo{pages}{401} (\bibinfo{year}{2006}), \eprint{astro-ph/0501304}.

\bibitem[{\citenamefont{Guo et~al.}(2010)\citenamefont{Guo, White, Li, and
  Boylan-Kolchin}}]{Guo:2009fn}
\bibinfo{author}{\bibfnamefont{Q.}~\bibnamefont{Guo}},
  \bibinfo{author}{\bibfnamefont{S.}~\bibnamefont{White}},
  \bibinfo{author}{\bibfnamefont{C.}~\bibnamefont{Li}}, \bibnamefont{and}
  \bibinfo{author}{\bibfnamefont{M.}~\bibnamefont{Boylan-Kolchin}},
  \bibinfo{journal}{Mon.Not.Roy.Astron.Soc.} \textbf{\bibinfo{volume}{404}},
  \bibinfo{pages}{1111} (\bibinfo{year}{2010}), \eprint{0909.4305}.

\bibitem[{\citenamefont{Zwaan et~al.}(2009)\citenamefont{Zwaan, Meyer, and
  Staveley-Smith}}]{Zwaan:2009dz}
\bibinfo{author}{\bibfnamefont{M.~A.} \bibnamefont{Zwaan}},
  \bibinfo{author}{\bibfnamefont{M.~J.} \bibnamefont{Meyer}}, \bibnamefont{and}
  \bibinfo{author}{\bibfnamefont{L.}~\bibnamefont{Staveley-Smith}}
  (\bibinfo{year}{2009}), \eprint{0912.1754}.

\bibitem[{\citenamefont{Pontzen and Governato}(2011)}]{Pontzen:2011ty}
\bibinfo{author}{\bibfnamefont{A.}~\bibnamefont{Pontzen}} \bibnamefont{and}
  \bibinfo{author}{\bibfnamefont{F.}~\bibnamefont{Governato}}
  (\bibinfo{year}{2011}), \eprint{1106.0499}.

\bibitem[{\citenamefont{Hopkins et~al.}(2012)\citenamefont{Hopkins, Quataert,
  and Murray}}]{Hopkins:2011rj}
\bibinfo{author}{\bibfnamefont{P.~F.} \bibnamefont{Hopkins}},
  \bibinfo{author}{\bibfnamefont{E.}~\bibnamefont{Quataert}}, \bibnamefont{and}
  \bibinfo{author}{\bibfnamefont{N.}~\bibnamefont{Murray}},
  \bibinfo{journal}{Mon.Not.Roy.Astron.Soc.} \textbf{\bibinfo{volume}{421}},
  \bibinfo{pages}{3522} (\bibinfo{year}{2012}), \eprint{1110.4638}.

\bibitem[{\citenamefont{Breddels et~al.}(2012)\citenamefont{Breddels, Helmi,
  van~den Bosch, van~de Ven, and Battaglia}}]{Breddels:2012cq}
\bibinfo{author}{\bibfnamefont{M.~A.} \bibnamefont{Breddels}},
  \bibinfo{author}{\bibfnamefont{A.}~\bibnamefont{Helmi}},
  \bibinfo{author}{\bibfnamefont{R.}~\bibnamefont{van~den Bosch}},
  \bibinfo{author}{\bibfnamefont{G.}~\bibnamefont{van~de Ven}},
  \bibnamefont{and} \bibinfo{author}{\bibfnamefont{G.}~\bibnamefont{Battaglia}}
  (\bibinfo{year}{2012}), \eprint{1205.4712}.

\bibitem[{\citenamefont{Vera-Ciro et~al.}(2012)\citenamefont{Vera-Ciro, Helmi,
  Starkenburg, and Breddels}}]{VeraCiro:2012na}
\bibinfo{author}{\bibfnamefont{C.~A.} \bibnamefont{Vera-Ciro}},
  \bibinfo{author}{\bibfnamefont{A.}~\bibnamefont{Helmi}},
  \bibinfo{author}{\bibfnamefont{E.}~\bibnamefont{Starkenburg}},
  \bibnamefont{and} \bibinfo{author}{\bibfnamefont{M.~A.}
  \bibnamefont{Breddels}} (\bibinfo{year}{2012}), \eprint{1202.6061}.

\bibitem[{\citenamefont{Loeb and Zaldarriaga}(2005)}]{Loeb:2005pm}
\bibinfo{author}{\bibfnamefont{A.}~\bibnamefont{Loeb}} \bibnamefont{and}
  \bibinfo{author}{\bibfnamefont{M.}~\bibnamefont{Zaldarriaga}},
  \bibinfo{journal}{Phys.Rev.} \textbf{\bibinfo{volume}{D71}},
  \bibinfo{pages}{103520} (\bibinfo{year}{2005}), \eprint{astro-ph/0504112}.

\bibitem[{\citenamefont{Bertschinger}(2006)}]{Bertschinger:2006nq}
\bibinfo{author}{\bibfnamefont{E.}~\bibnamefont{Bertschinger}},
  \bibinfo{journal}{Phys.Rev.} \textbf{\bibinfo{volume}{D74}},
  \bibinfo{pages}{063509} (\bibinfo{year}{2006}), \eprint{astro-ph/0607319}.

\bibitem[{\citenamefont{Silk}(1968)}]{Silk:1967kq}
\bibinfo{author}{\bibfnamefont{J.}~\bibnamefont{Silk}},
  \bibinfo{journal}{Astrophys.J.} \textbf{\bibinfo{volume}{151}},
  \bibinfo{pages}{459} (\bibinfo{year}{1968}).

\bibitem[{\citenamefont{Bode et~al.}(2001)\citenamefont{Bode, Ostriker, and
  Turok}}]{Bode:2000gq}
\bibinfo{author}{\bibfnamefont{P.}~\bibnamefont{Bode}},
  \bibinfo{author}{\bibfnamefont{J.~P.} \bibnamefont{Ostriker}},
  \bibnamefont{and} \bibinfo{author}{\bibfnamefont{N.}~\bibnamefont{Turok}},
  \bibinfo{journal}{Astrophys.J.} \textbf{\bibinfo{volume}{556}},
  \bibinfo{pages}{93} (\bibinfo{year}{2001}), \eprint{astro-ph/0010389}.

\bibitem[{\citenamefont{Dalcanton and Hogan}(2001)}]{Dalcanton:2000hn}
\bibinfo{author}{\bibfnamefont{J.~J.} \bibnamefont{Dalcanton}}
  \bibnamefont{and} \bibinfo{author}{\bibfnamefont{C.~J.} \bibnamefont{Hogan}},
  \bibinfo{journal}{Astrophys.J.} \textbf{\bibinfo{volume}{561}},
  \bibinfo{pages}{35} (\bibinfo{year}{2001}), \eprint{astro-ph/0004381}.

\bibitem[{\citenamefont{Boyanovsky and Wu}(2011)}]{Boyanovsky:2010pw}
\bibinfo{author}{\bibfnamefont{D.}~\bibnamefont{Boyanovsky}} \bibnamefont{and}
  \bibinfo{author}{\bibfnamefont{J.}~\bibnamefont{Wu}},
  \bibinfo{journal}{Phys.Rev.} \textbf{\bibinfo{volume}{D83}},
  \bibinfo{pages}{043524} (\bibinfo{year}{2011}), \eprint{1008.0992}.

\bibitem[{\citenamefont{Feng et~al.}(2009)\citenamefont{Feng, Kaplinghat, Tu,
  and Yu}}]{Feng:2009mn}
\bibinfo{author}{\bibfnamefont{J.~L.} \bibnamefont{Feng}},
  \bibinfo{author}{\bibfnamefont{M.}~\bibnamefont{Kaplinghat}},
  \bibinfo{author}{\bibfnamefont{H.}~\bibnamefont{Tu}}, \bibnamefont{and}
  \bibinfo{author}{\bibfnamefont{H.-B.} \bibnamefont{Yu}},
  \bibinfo{journal}{JCAP} \textbf{\bibinfo{volume}{0907}}, \bibinfo{pages}{004}
  (\bibinfo{year}{2009}), \eprint{0905.3039}.

\bibitem[{\citenamefont{Shoemaker}(2013)}]{Shoemaker:2013tda}
\bibinfo{author}{\bibfnamefont{I.~M.} \bibnamefont{Shoemaker}}
  (\bibinfo{year}{2013}), \eprint{1305.1936}.

\bibitem[{\citenamefont{van~den Aarssen et~al.}(2012)\citenamefont{van~den
  Aarssen, Bringmann, and Pfrommer}}]{Aarssen:2012fx}
\bibinfo{author}{\bibfnamefont{L.~G.} \bibnamefont{van~den Aarssen}},
  \bibinfo{author}{\bibfnamefont{T.}~\bibnamefont{Bringmann}},
  \bibnamefont{and} \bibinfo{author}{\bibfnamefont{C.}~\bibnamefont{Pfrommer}},
  \bibinfo{journal}{Phys.Rev.Lett.} \textbf{\bibinfo{volume}{109}},
  \bibinfo{pages}{231301} (\bibinfo{year}{2012}), \eprint{1205.5809}.

\bibitem[{\citenamefont{Laha et~al.}(2013)\citenamefont{Laha, Dasgupta, and
  Beacom}}]{Laha:2013xua}
\bibinfo{author}{\bibfnamefont{R.}~\bibnamefont{Laha}},
  \bibinfo{author}{\bibfnamefont{B.}~\bibnamefont{Dasgupta}}, \bibnamefont{and}
  \bibinfo{author}{\bibfnamefont{J.~F.} \bibnamefont{Beacom}}
  (\bibinfo{year}{2013}), \eprint{1304.3460}.

\bibitem[{\citenamefont{Spergel and Steinhardt}(2000)}]{Spergel:1999mh}
\bibinfo{author}{\bibfnamefont{D.~N.} \bibnamefont{Spergel}} \bibnamefont{and}
  \bibinfo{author}{\bibfnamefont{P.~J.} \bibnamefont{Steinhardt}},
  \bibinfo{journal}{Phys.Rev.Lett.} \textbf{\bibinfo{volume}{84}},
  \bibinfo{pages}{3760} (\bibinfo{year}{2000}), \eprint{astro-ph/9909386}.

\bibitem[{\citenamefont{Feng et~al.}(2010)\citenamefont{Feng, Kaplinghat, and
  Yu}}]{Feng:2009hw}
\bibinfo{author}{\bibfnamefont{J.~L.} \bibnamefont{Feng}},
  \bibinfo{author}{\bibfnamefont{M.}~\bibnamefont{Kaplinghat}},
  \bibnamefont{and} \bibinfo{author}{\bibfnamefont{H.-B.} \bibnamefont{Yu}},
  \bibinfo{journal}{Phys.Rev.Lett.} \textbf{\bibinfo{volume}{104}},
  \bibinfo{pages}{151301} (\bibinfo{year}{2010}), \eprint{0911.0422}.

\bibitem[{\citenamefont{Markevitch et~al.}(2004)}]{Markevitch:2003at}
\bibinfo{author}{\bibfnamefont{M.}~\bibnamefont{Markevitch}}
  \bibnamefont{et~al.}, \bibinfo{journal}{Astrophys. J.}
  \textbf{\bibinfo{volume}{606}}, \bibinfo{pages}{819} (\bibinfo{year}{2004}),
  \eprint{astro-ph/0309303}.

\bibitem[{\citenamefont{Miralda-Escude}(2000)}]{MiraldaEscude:2000qt}
\bibinfo{author}{\bibfnamefont{J.}~\bibnamefont{Miralda-Escude}}
  (\bibinfo{year}{2000}), \eprint{astro-ph/0002050}.

\bibitem[{\citenamefont{Rocha et~al.}(2013)\citenamefont{Rocha, Peter, Bullock,
  Kaplinghat, Garrison-Kimmel et~al.}}]{Rocha:2012jg}
\bibinfo{author}{\bibfnamefont{M.}~\bibnamefont{Rocha}},
  \bibinfo{author}{\bibfnamefont{A.~H.} \bibnamefont{Peter}},
  \bibinfo{author}{\bibfnamefont{J.~S.} \bibnamefont{Bullock}},
  \bibinfo{author}{\bibfnamefont{M.}~\bibnamefont{Kaplinghat}},
  \bibinfo{author}{\bibfnamefont{S.}~\bibnamefont{Garrison-Kimmel}},
  \bibnamefont{et~al.}, \bibinfo{journal}{Mon.Not.Roy.Astron.Soc.}
  \textbf{\bibinfo{volume}{430}}, \bibinfo{pages}{81} (\bibinfo{year}{2013}),
  \eprint{1208.3025}.

\bibitem[{\citenamefont{Peter et~al.}(2012)\citenamefont{Peter, Rocha, Bullock,
  and Kaplinghat}}]{Peter:2012jh}
\bibinfo{author}{\bibfnamefont{A.~H.} \bibnamefont{Peter}},
  \bibinfo{author}{\bibfnamefont{M.}~\bibnamefont{Rocha}},
  \bibinfo{author}{\bibfnamefont{J.~S.} \bibnamefont{Bullock}},
  \bibnamefont{and}
  \bibinfo{author}{\bibfnamefont{M.}~\bibnamefont{Kaplinghat}}
  (\bibinfo{year}{2012}), \eprint{1208.3026}.

\bibitem[{\citenamefont{Vogelsberger and Zavala}(2013)}]{Vogelsberger:2012sa}
\bibinfo{author}{\bibfnamefont{M.}~\bibnamefont{Vogelsberger}}
  \bibnamefont{and} \bibinfo{author}{\bibfnamefont{J.}~\bibnamefont{Zavala}},
  \bibinfo{journal}{Mon.Not.Roy.Astron.Soc.} \textbf{\bibinfo{volume}{430}},
  \bibinfo{pages}{1722} (\bibinfo{year}{2013}), \eprint{1211.1377}.

\bibitem[{\citenamefont{Vogelsberger et~al.}(2012)\citenamefont{Vogelsberger,
  Zavala, and Loeb}}]{Vogelsberger:2012ku}
\bibinfo{author}{\bibfnamefont{M.}~\bibnamefont{Vogelsberger}},
  \bibinfo{author}{\bibfnamefont{J.}~\bibnamefont{Zavala}}, \bibnamefont{and}
  \bibinfo{author}{\bibfnamefont{A.}~\bibnamefont{Loeb}},
  \bibinfo{journal}{Mon.Not.Roy.Astron.Soc.} \textbf{\bibinfo{volume}{423}},
  \bibinfo{pages}{3740} (\bibinfo{year}{2012}), \eprint{1201.5892}.

\bibitem[{\citenamefont{Zavala et~al.}(2012)\citenamefont{Zavala, Vogelsberger,
  and Walker}}]{Zavala:2012us}
\bibinfo{author}{\bibfnamefont{J.}~\bibnamefont{Zavala}},
  \bibinfo{author}{\bibfnamefont{M.}~\bibnamefont{Vogelsberger}},
  \bibnamefont{and} \bibinfo{author}{\bibfnamefont{M.~G.} \bibnamefont{Walker}}
  (\bibinfo{year}{2012}), \eprint{1211.6426}.

\bibitem[{\citenamefont{Loeb and Weiner}(2011)}]{Loeb:2010gj}
\bibinfo{author}{\bibfnamefont{A.}~\bibnamefont{Loeb}} \bibnamefont{and}
  \bibinfo{author}{\bibfnamefont{N.}~\bibnamefont{Weiner}},
  \bibinfo{journal}{Phys.Rev.Lett.} \textbf{\bibinfo{volume}{106}},
  \bibinfo{pages}{171302} (\bibinfo{year}{2011}), \eprint{1011.6374}.

\bibitem[{\citenamefont{Tulin et~al.}(2013)\citenamefont{Tulin, Yu, and
  Zurek}}]{Tulin:2013teo}
\bibinfo{author}{\bibfnamefont{S.}~\bibnamefont{Tulin}},
  \bibinfo{author}{\bibfnamefont{H.-B.} \bibnamefont{Yu}}, \bibnamefont{and}
  \bibinfo{author}{\bibfnamefont{K.~M.} \bibnamefont{Zurek}}
  (\bibinfo{year}{2013}), \eprint{1302.3898}.

\bibitem[{\citenamefont{Foot}(2013{\natexlab{c}})}]{Foot:2013vna}
\bibinfo{author}{\bibfnamefont{R.}~\bibnamefont{Foot}}
  (\bibinfo{year}{2013}{\natexlab{c}}), \eprint{1304.4717}.

\bibitem[{\citenamefont{Foot}(2013{\natexlab{d}})}]{Foot:2013uxa}
\bibinfo{author}{\bibfnamefont{R.}~\bibnamefont{Foot}}
  (\bibinfo{year}{2013}{\natexlab{d}}), \eprint{1303.1727}.

\bibitem[{\citenamefont{Burkert}(1996)}]{Burkert:1995yz}
\bibinfo{author}{\bibfnamefont{A.}~\bibnamefont{Burkert}},
  \bibinfo{journal}{IAU Symp.} \textbf{\bibinfo{volume}{171}},
  \bibinfo{pages}{175} (\bibinfo{year}{1996}), \eprint{astro-ph/9504041}.

\bibitem[{\citenamefont{Salucci and Burkert}(2000)}]{Salucci:2000ps}
\bibinfo{author}{\bibfnamefont{P.}~\bibnamefont{Salucci}} \bibnamefont{and}
  \bibinfo{author}{\bibfnamefont{A.}~\bibnamefont{Burkert}},
  \bibinfo{journal}{Astrophys.J.} \textbf{\bibinfo{volume}{537}},
  \bibinfo{pages}{L9} (\bibinfo{year}{2000}), \eprint{astro-ph/0004397}.

\bibitem[{\citenamefont{Fan et~al.}(2013{\natexlab{a}})\citenamefont{Fan, Katz,
  Randall, and Reece}}]{Fan:2013yva}
\bibinfo{author}{\bibfnamefont{J.}~\bibnamefont{Fan}},
  \bibinfo{author}{\bibfnamefont{A.}~\bibnamefont{Katz}},
  \bibinfo{author}{\bibfnamefont{L.}~\bibnamefont{Randall}}, \bibnamefont{and}
  \bibinfo{author}{\bibfnamefont{M.}~\bibnamefont{Reece}}
  (\bibinfo{year}{2013}{\natexlab{a}}), \eprint{1303.1521}.

\bibitem[{\citenamefont{Fan et~al.}(2013{\natexlab{b}})\citenamefont{Fan, Katz,
  Randall, and Reece}}]{Fan:2013tia}
\bibinfo{author}{\bibfnamefont{J.}~\bibnamefont{Fan}},
  \bibinfo{author}{\bibfnamefont{A.}~\bibnamefont{Katz}},
  \bibinfo{author}{\bibfnamefont{L.}~\bibnamefont{Randall}}, \bibnamefont{and}
  \bibinfo{author}{\bibfnamefont{M.}~\bibnamefont{Reece}}
  (\bibinfo{year}{2013}{\natexlab{b}}), \eprint{1303.3271}.

\bibitem[{\citenamefont{Foot and Silagadze}(2013)}]{Foot:2013nea}
\bibinfo{author}{\bibfnamefont{R.}~\bibnamefont{Foot}} \bibnamefont{and}
  \bibinfo{author}{\bibfnamefont{Z.}~\bibnamefont{Silagadze}}
  (\bibinfo{year}{2013}), \eprint{1306.1305}.

\bibitem[{\citenamefont{Ibata et~al.}(2013)\citenamefont{Ibata, Lewis, Conn,
  Irwin, McConnachie et~al.}}]{Ibata:2013rh}
\bibinfo{author}{\bibfnamefont{R.~A.} \bibnamefont{Ibata}},
  \bibinfo{author}{\bibfnamefont{G.~F.} \bibnamefont{Lewis}},
  \bibinfo{author}{\bibfnamefont{A.~R.} \bibnamefont{Conn}},
  \bibinfo{author}{\bibfnamefont{M.~J.} \bibnamefont{Irwin}},
  \bibinfo{author}{\bibfnamefont{A.~W.} \bibnamefont{McConnachie}},
  \bibnamefont{et~al.}, \bibinfo{journal}{Nature}
  \textbf{\bibinfo{volume}{493}}, \bibinfo{pages}{62} (\bibinfo{year}{2013}),
  \eprint{1301.0446}.

\bibitem[{\citenamefont{Pawlowski et~al.}(2012)\citenamefont{Pawlowski,
  Pflamm-Altenburg, and Kroupa}}]{Pawlowski:2012vz}
\bibinfo{author}{\bibfnamefont{M.}~\bibnamefont{Pawlowski}},
  \bibinfo{author}{\bibfnamefont{J.}~\bibnamefont{Pflamm-Altenburg}},
  \bibnamefont{and} \bibinfo{author}{\bibfnamefont{P.}~\bibnamefont{Kroupa}}
  (\bibinfo{year}{2012}), \eprint{1204.5176}.

\bibitem[{\citenamefont{Walker}(2012{\natexlab{b}})}]{Walker:2012td}
\bibinfo{author}{\bibfnamefont{M.~G.} \bibnamefont{Walker}}
  (\bibinfo{year}{2012}{\natexlab{b}}), \eprint{1205.0311}.

\bibitem[{\citenamefont{Kaplinghat}(2005)}]{Kaplinghat:2005sy}
\bibinfo{author}{\bibfnamefont{M.}~\bibnamefont{Kaplinghat}},
  \bibinfo{journal}{Phys.Rev.} \textbf{\bibinfo{volume}{D72}},
  \bibinfo{pages}{063510} (\bibinfo{year}{2005}), \eprint{astro-ph/0507300}.

\bibitem[{\citenamefont{Abdelqader and Melia}(2008)}]{Abdelqader2008wa}
\bibinfo{author}{\bibfnamefont{M.}~\bibnamefont{Abdelqader}} \bibnamefont{and}
  \bibinfo{author}{\bibfnamefont{F.}~\bibnamefont{Melia}},
  \bibinfo{journal}{Mon.Not.Roy.Astron.Soc.} \textbf{\bibinfo{volume}{388}},
  \bibinfo{pages}{1869} (\bibinfo{year}{2008}), \eprint{0806.0602}.

\bibitem[{\citenamefont{Bell et~al.}(2010)\citenamefont{Bell, Galea, and
  Petraki}}]{Bell:2010fk}
\bibinfo{author}{\bibfnamefont{N.~F.} \bibnamefont{Bell}},
  \bibinfo{author}{\bibfnamefont{A.~J.} \bibnamefont{Galea}}, \bibnamefont{and}
  \bibinfo{author}{\bibfnamefont{K.}~\bibnamefont{Petraki}},
  \bibinfo{journal}{Phys.Rev.} \textbf{\bibinfo{volume}{D82}},
  \bibinfo{pages}{023514} (\bibinfo{year}{2010}), \eprint{1004.1008}.

\bibitem[{\citenamefont{Bell et~al.}(2011{\natexlab{b}})\citenamefont{Bell,
  Galea, and Volkas}}]{Bell:2010qt}
\bibinfo{author}{\bibfnamefont{N.~F.} \bibnamefont{Bell}},
  \bibinfo{author}{\bibfnamefont{A.~J.} \bibnamefont{Galea}}, \bibnamefont{and}
  \bibinfo{author}{\bibfnamefont{R.~R.} \bibnamefont{Volkas}},
  \bibinfo{journal}{Phys.Rev.} \textbf{\bibinfo{volume}{D83}},
  \bibinfo{pages}{063504} (\bibinfo{year}{2011}{\natexlab{b}}),
  \eprint{1012.0067}.

\bibitem[{\citenamefont{Peter}(2010)}]{Peter:2010au}
\bibinfo{author}{\bibfnamefont{A.~H.} \bibnamefont{Peter}},
  \bibinfo{journal}{Phys.Rev.} \textbf{\bibinfo{volume}{D81}},
  \bibinfo{pages}{083511} (\bibinfo{year}{2010}), \eprint{1001.3870}.

\bibitem[{\citenamefont{Peter et~al.}(2010)\citenamefont{Peter, Moody, and
  Kamionkowski}}]{Peter:2010jy}
\bibinfo{author}{\bibfnamefont{A.~H.} \bibnamefont{Peter}},
  \bibinfo{author}{\bibfnamefont{C.~E.} \bibnamefont{Moody}}, \bibnamefont{and}
  \bibinfo{author}{\bibfnamefont{M.}~\bibnamefont{Kamionkowski}},
  \bibinfo{journal}{Phys.Rev.} \textbf{\bibinfo{volume}{D81}},
  \bibinfo{pages}{103501} (\bibinfo{year}{2010}), \eprint{1003.0419}.

\bibitem[{\citenamefont{Aprile et~al.}(2011)}]{Aprile:2011hi}
\bibinfo{author}{\bibfnamefont{E.}~\bibnamefont{Aprile}} \bibnamefont{et~al.}
  (\bibinfo{collaboration}{XENON100 Collaboration}),
  \bibinfo{journal}{Phys.Rev.Lett.} \textbf{\bibinfo{volume}{107}},
  \bibinfo{pages}{131302} (\bibinfo{year}{2011}), \eprint{1104.2549}.

\bibitem[{\citenamefont{Angle et~al.}(2011)}]{Angle:2011th}
\bibinfo{author}{\bibfnamefont{J.}~\bibnamefont{Angle}} \bibnamefont{et~al.}
  (\bibinfo{collaboration}{XENON10 Collaboration}),
  \bibinfo{journal}{Phys.Rev.Lett.} \textbf{\bibinfo{volume}{107}},
  \bibinfo{pages}{051301} (\bibinfo{year}{2011}), \eprint{1104.3088}.

\bibitem[{\citenamefont{Baudis}(2012)}]{Baudis:2012zs}
\bibinfo{author}{\bibfnamefont{L.}~\bibnamefont{Baudis}}
  (\bibinfo{collaboration}{XENON Collaboration}) (\bibinfo{year}{2012}),
  \eprint{1203.1589}.

\bibitem[{\citenamefont{Essig et~al.}(2012)\citenamefont{Essig, Manalaysay,
  Mardon, Sorensen, and Volansky}}]{Essig:2012yx}
\bibinfo{author}{\bibfnamefont{R.}~\bibnamefont{Essig}},
  \bibinfo{author}{\bibfnamefont{A.}~\bibnamefont{Manalaysay}},
  \bibinfo{author}{\bibfnamefont{J.}~\bibnamefont{Mardon}},
  \bibinfo{author}{\bibfnamefont{P.}~\bibnamefont{Sorensen}}, \bibnamefont{and}
  \bibinfo{author}{\bibfnamefont{T.}~\bibnamefont{Volansky}},
  \bibinfo{journal}{Phys.Rev.Lett.} \textbf{\bibinfo{volume}{109}},
  \bibinfo{pages}{021301} (\bibinfo{year}{2012}), \eprint{1206.2644}.

\bibitem[{\citenamefont{Belli et~al.}(2002)\citenamefont{Belli, Cerulli,
  Fornengo, and Scopel}}]{Belli:2002yt}
\bibinfo{author}{\bibfnamefont{P.}~\bibnamefont{Belli}},
  \bibinfo{author}{\bibfnamefont{R.}~\bibnamefont{Cerulli}},
  \bibinfo{author}{\bibfnamefont{N.}~\bibnamefont{Fornengo}}, \bibnamefont{and}
  \bibinfo{author}{\bibfnamefont{S.}~\bibnamefont{Scopel}},
  \bibinfo{journal}{Phys.Rev.} \textbf{\bibinfo{volume}{D66}},
  \bibinfo{pages}{043503} (\bibinfo{year}{2002}), \eprint{hep-ph/0203242}.

\bibitem[{\citenamefont{Friedland and Shoemaker}(2012)}]{Friedland:2012fa}
\bibinfo{author}{\bibfnamefont{A.}~\bibnamefont{Friedland}} \bibnamefont{and}
  \bibinfo{author}{\bibfnamefont{I.~M.} \bibnamefont{Shoemaker}}
  (\bibinfo{year}{2012}), \eprint{1212.4139}.

\bibitem[{\citenamefont{Bernabei
  et~al.}(2008{\natexlab{b}})\citenamefont{Bernabei, Belli, Incicchitti, and
  Prosperi}}]{Bernabei:2008jm}
\bibinfo{author}{\bibfnamefont{R.}~\bibnamefont{Bernabei}},
  \bibinfo{author}{\bibfnamefont{P.}~\bibnamefont{Belli}},
  \bibinfo{author}{\bibfnamefont{A.}~\bibnamefont{Incicchitti}},
  \bibnamefont{and} \bibinfo{author}{\bibfnamefont{D.}~\bibnamefont{Prosperi}}
  (\bibinfo{year}{2008}{\natexlab{b}}), \eprint{0806.0011}.

\bibitem[{\citenamefont{Manzur et~al.}(2010)\citenamefont{Manzur, Curioni,
  Kastens, McKinsey, Ni et~al.}}]{Manzur:2009hp}
\bibinfo{author}{\bibfnamefont{A.}~\bibnamefont{Manzur}},
  \bibinfo{author}{\bibfnamefont{A.}~\bibnamefont{Curioni}},
  \bibinfo{author}{\bibfnamefont{L.}~\bibnamefont{Kastens}},
  \bibinfo{author}{\bibfnamefont{D.}~\bibnamefont{McKinsey}},
  \bibinfo{author}{\bibfnamefont{K.}~\bibnamefont{Ni}}, \bibnamefont{et~al.},
  \bibinfo{journal}{Phys.Rev.} \textbf{\bibinfo{volume}{C81}},
  \bibinfo{pages}{025808} (\bibinfo{year}{2010}), \eprint{0909.1063}.

\bibitem[{\citenamefont{Collar}(2011{\natexlab{a}})}]{Collar:2011kf}
\bibinfo{author}{\bibfnamefont{J.}~\bibnamefont{Collar}}
  (\bibinfo{year}{2011}{\natexlab{a}}), \eprint{1103.3481}.

\bibitem[{\citenamefont{Collar}(2011{\natexlab{b}})}]{Collar:2011wq}
\bibinfo{author}{\bibfnamefont{J.}~\bibnamefont{Collar}}
  (\bibinfo{year}{2011}{\natexlab{b}}), \eprint{1106.0653}.

\bibitem[{\citenamefont{Del~Nobile
  et~al.}(2012{\natexlab{a}})\citenamefont{Del~Nobile, Kouvaris, Panci,
  Sannino, and Virkajarvi}}]{DelNobile:2012tx}
\bibinfo{author}{\bibfnamefont{E.}~\bibnamefont{Del~Nobile}},
  \bibinfo{author}{\bibfnamefont{C.}~\bibnamefont{Kouvaris}},
  \bibinfo{author}{\bibfnamefont{P.}~\bibnamefont{Panci}},
  \bibinfo{author}{\bibfnamefont{F.}~\bibnamefont{Sannino}}, \bibnamefont{and}
  \bibinfo{author}{\bibfnamefont{J.}~\bibnamefont{Virkajarvi}},
  \bibinfo{journal}{JCAP} \textbf{\bibinfo{volume}{1208}}, \bibinfo{pages}{010}
  (\bibinfo{year}{2012}{\natexlab{a}}), \eprint{1203.6652}.

\bibitem[{\citenamefont{Feng et~al.}(2011)\citenamefont{Feng, Kumar, Marfatia,
  and Sanford}}]{Feng:2011vu}
\bibinfo{author}{\bibfnamefont{J.~L.} \bibnamefont{Feng}},
  \bibinfo{author}{\bibfnamefont{J.}~\bibnamefont{Kumar}},
  \bibinfo{author}{\bibfnamefont{D.}~\bibnamefont{Marfatia}}, \bibnamefont{and}
  \bibinfo{author}{\bibfnamefont{D.}~\bibnamefont{Sanford}},
  \bibinfo{journal}{Phys.Lett.} \textbf{\bibinfo{volume}{B703}},
  \bibinfo{pages}{124} (\bibinfo{year}{2011}), \eprint{1102.4331}.

\bibitem[{\citenamefont{Chang et~al.}(2010)\citenamefont{Chang, Liu, Pierce,
  Weiner, and Yavin}}]{Chang:2010yk}
\bibinfo{author}{\bibfnamefont{S.}~\bibnamefont{Chang}},
  \bibinfo{author}{\bibfnamefont{J.}~\bibnamefont{Liu}},
  \bibinfo{author}{\bibfnamefont{A.}~\bibnamefont{Pierce}},
  \bibinfo{author}{\bibfnamefont{N.}~\bibnamefont{Weiner}}, \bibnamefont{and}
  \bibinfo{author}{\bibfnamefont{I.}~\bibnamefont{Yavin}},
  \bibinfo{journal}{JCAP} \textbf{\bibinfo{volume}{1008}}, \bibinfo{pages}{018}
  (\bibinfo{year}{2010}), \eprint{1004.0697}.

\bibitem[{\citenamefont{Del~Nobile
  et~al.}(2012{\natexlab{b}})\citenamefont{Del~Nobile, Kouvaris, Sannino, and
  Virkajarvi}}]{DelNobile:2011yb}
\bibinfo{author}{\bibfnamefont{E.}~\bibnamefont{Del~Nobile}},
  \bibinfo{author}{\bibfnamefont{C.}~\bibnamefont{Kouvaris}},
  \bibinfo{author}{\bibfnamefont{F.}~\bibnamefont{Sannino}}, \bibnamefont{and}
  \bibinfo{author}{\bibfnamefont{J.}~\bibnamefont{Virkajarvi}},
  \bibinfo{journal}{Mod.Phys.Lett.} \textbf{\bibinfo{volume}{A27}},
  \bibinfo{pages}{1250108} (\bibinfo{year}{2012}{\natexlab{b}}),
  \eprint{1111.1902}.

\bibitem[{\citenamefont{Del~Nobile et~al.}(2011)\citenamefont{Del~Nobile,
  Kouvaris, and Sannino}}]{DelNobile:2011je}
\bibinfo{author}{\bibfnamefont{E.}~\bibnamefont{Del~Nobile}},
  \bibinfo{author}{\bibfnamefont{C.}~\bibnamefont{Kouvaris}}, \bibnamefont{and}
  \bibinfo{author}{\bibfnamefont{F.}~\bibnamefont{Sannino}},
  \bibinfo{journal}{Phys.Rev.} \textbf{\bibinfo{volume}{D84}},
  \bibinfo{pages}{027301} (\bibinfo{year}{2011}), \eprint{1105.5431}.

\bibitem[{\citenamefont{Okada and Seto}(2013)}]{Okada:2013cba}
\bibinfo{author}{\bibfnamefont{N.}~\bibnamefont{Okada}} \bibnamefont{and}
  \bibinfo{author}{\bibfnamefont{O.}~\bibnamefont{Seto}}
  (\bibinfo{year}{2013}), \eprint{1304.6791}.

\bibitem[{\citenamefont{Foot}(2012{\natexlab{b}})}]{Foot:2011fh}
\bibinfo{author}{\bibfnamefont{R.}~\bibnamefont{Foot}}, \bibinfo{journal}{JCAP}
  \textbf{\bibinfo{volume}{1204}}, \bibinfo{pages}{014}
  (\bibinfo{year}{2012}{\natexlab{b}}), \eprint{1110.2908}.

\bibitem[{\citenamefont{Lin et~al.}(2012)\citenamefont{Lin, Yu, and
  Zurek}}]{Lin:2011gj}
\bibinfo{author}{\bibfnamefont{T.}~\bibnamefont{Lin}},
  \bibinfo{author}{\bibfnamefont{H.-B.} \bibnamefont{Yu}}, \bibnamefont{and}
  \bibinfo{author}{\bibfnamefont{K.~M.} \bibnamefont{Zurek}},
  \bibinfo{journal}{Phys.Rev.} \textbf{\bibinfo{volume}{D85}},
  \bibinfo{pages}{063503} (\bibinfo{year}{2012}), \eprint{1111.0293}.

\bibitem[{\citenamefont{Davoudiasl et~al.}(2011)\citenamefont{Davoudiasl,
  Morrissey, Sigurdson, and Tulin}}]{Davoudiasl:2011fj}
\bibinfo{author}{\bibfnamefont{H.}~\bibnamefont{Davoudiasl}},
  \bibinfo{author}{\bibfnamefont{D.~E.} \bibnamefont{Morrissey}},
  \bibinfo{author}{\bibfnamefont{K.}~\bibnamefont{Sigurdson}},
  \bibnamefont{and} \bibinfo{author}{\bibfnamefont{S.}~\bibnamefont{Tulin}},
  \bibinfo{journal}{Phys.Rev.} \textbf{\bibinfo{volume}{D84}},
  \bibinfo{pages}{096008} (\bibinfo{year}{2011}), \eprint{1106.4320}.

\bibitem[{\citenamefont{Chang and Goodenough}(2011)}]{Chang:2011xn}
\bibinfo{author}{\bibfnamefont{S.}~\bibnamefont{Chang}} \bibnamefont{and}
  \bibinfo{author}{\bibfnamefont{L.}~\bibnamefont{Goodenough}},
  \bibinfo{journal}{Phys.Rev.} \textbf{\bibinfo{volume}{D84}},
  \bibinfo{pages}{023524} (\bibinfo{year}{2011}), \eprint{1105.3976}.

\bibitem[{\citenamefont{Masina and Sannino}(2011)}]{Masina:2011hu}
\bibinfo{author}{\bibfnamefont{I.}~\bibnamefont{Masina}} \bibnamefont{and}
  \bibinfo{author}{\bibfnamefont{F.}~\bibnamefont{Sannino}},
  \bibinfo{journal}{JCAP} \textbf{\bibinfo{volume}{1109}}, \bibinfo{pages}{021}
  (\bibinfo{year}{2011}), \eprint{1106.3353}.

\bibitem[{\citenamefont{Masina and Sannino}(2013)}]{Masina:2013yea}
\bibinfo{author}{\bibfnamefont{I.}~\bibnamefont{Masina}} \bibnamefont{and}
  \bibinfo{author}{\bibfnamefont{F.}~\bibnamefont{Sannino}}
  (\bibinfo{year}{2013}), \eprint{1304.2800}.

\bibitem[{\citenamefont{Feng and Kang}(2013)}]{Feng:2013vva}
\bibinfo{author}{\bibfnamefont{L.}~\bibnamefont{Feng}} \bibnamefont{and}
  \bibinfo{author}{\bibfnamefont{Z.}~\bibnamefont{Kang}}
  (\bibinfo{year}{2013}), \eprint{1304.7492}.

\bibitem[{\citenamefont{Pearce and Kusenko}(2013)}]{Pearce:2013ola}
\bibinfo{author}{\bibfnamefont{L.}~\bibnamefont{Pearce}} \bibnamefont{and}
  \bibinfo{author}{\bibfnamefont{A.}~\bibnamefont{Kusenko}}
  (\bibinfo{year}{2013}), \eprint{1303.7294}.

\bibitem[{\citenamefont{Shepherd et~al.}(2009)\citenamefont{Shepherd, Tait, and
  Zaharijas}}]{Shepherd:2009sa}
\bibinfo{author}{\bibfnamefont{W.}~\bibnamefont{Shepherd}},
  \bibinfo{author}{\bibfnamefont{T.~M.} \bibnamefont{Tait}}, \bibnamefont{and}
  \bibinfo{author}{\bibfnamefont{G.}~\bibnamefont{Zaharijas}},
  \bibinfo{journal}{Phys.Rev.} \textbf{\bibinfo{volume}{D79}},
  \bibinfo{pages}{055022} (\bibinfo{year}{2009}), \eprint{0901.2125}.

\bibitem[{\citenamefont{Cirelli et~al.}(2005)\citenamefont{Cirelli, Fornengo,
  Montaruli, Sokalski, Strumia et~al.}}]{Cirelli:2005gh}
\bibinfo{author}{\bibfnamefont{M.}~\bibnamefont{Cirelli}},
  \bibinfo{author}{\bibfnamefont{N.}~\bibnamefont{Fornengo}},
  \bibinfo{author}{\bibfnamefont{T.}~\bibnamefont{Montaruli}},
  \bibinfo{author}{\bibfnamefont{I.~A.} \bibnamefont{Sokalski}},
  \bibinfo{author}{\bibfnamefont{A.}~\bibnamefont{Strumia}},
  \bibnamefont{et~al.}, \bibinfo{journal}{Nucl.Phys.}
  \textbf{\bibinfo{volume}{B727}}, \bibinfo{pages}{99} (\bibinfo{year}{2005}),
  \eprint{hep-ph/0506298}.

\bibitem[{\citenamefont{Blennow et~al.}(2008)\citenamefont{Blennow, Edsjo, and
  Ohlsson}}]{Blennow:2007tw}
\bibinfo{author}{\bibfnamefont{M.}~\bibnamefont{Blennow}},
  \bibinfo{author}{\bibfnamefont{J.}~\bibnamefont{Edsjo}}, \bibnamefont{and}
  \bibinfo{author}{\bibfnamefont{T.}~\bibnamefont{Ohlsson}},
  \bibinfo{journal}{JCAP} \textbf{\bibinfo{volume}{0801}}, \bibinfo{pages}{021}
  (\bibinfo{year}{2008}), \eprint{0709.3898}.

\bibitem[{\citenamefont{Bell and Petraki}(2011)}]{Bell:2011sn}
\bibinfo{author}{\bibfnamefont{N.~F.} \bibnamefont{Bell}} \bibnamefont{and}
  \bibinfo{author}{\bibfnamefont{K.}~\bibnamefont{Petraki}},
  \bibinfo{journal}{JCAP} \textbf{\bibinfo{volume}{1104}}, \bibinfo{pages}{003}
  (\bibinfo{year}{2011}), \eprint{1102.2958}.

\bibitem[{\citenamefont{Bell et~al.}(2012)\citenamefont{Bell, Brennan, and
  Jacques}}]{Bell:2012dk}
\bibinfo{author}{\bibfnamefont{N.~F.} \bibnamefont{Bell}},
  \bibinfo{author}{\bibfnamefont{A.~J.} \bibnamefont{Brennan}},
  \bibnamefont{and} \bibinfo{author}{\bibfnamefont{T.~D.}
  \bibnamefont{Jacques}}, \bibinfo{journal}{JCAP}
  \textbf{\bibinfo{volume}{1210}}, \bibinfo{pages}{045} (\bibinfo{year}{2012}),
  \eprint{1206.2977}.

\bibitem[{\citenamefont{Bertone and Fairbairn}(2008)}]{Bertone:2007ae}
\bibinfo{author}{\bibfnamefont{G.}~\bibnamefont{Bertone}} \bibnamefont{and}
  \bibinfo{author}{\bibfnamefont{M.}~\bibnamefont{Fairbairn}},
  \bibinfo{journal}{Phys.Rev.} \textbf{\bibinfo{volume}{D77}},
  \bibinfo{pages}{043515} (\bibinfo{year}{2008}), \eprint{0709.1485}.

\bibitem[{\citenamefont{Kouvaris}(2008)}]{Kouvaris:2007ay}
\bibinfo{author}{\bibfnamefont{C.}~\bibnamefont{Kouvaris}},
  \bibinfo{journal}{Phys.Rev.} \textbf{\bibinfo{volume}{D77}},
  \bibinfo{pages}{023006} (\bibinfo{year}{2008}), \eprint{0708.2362}.

\bibitem[{\citenamefont{Frandsen and Sarkar}(2010)}]{Frandsen:2010yj}
\bibinfo{author}{\bibfnamefont{M.~T.} \bibnamefont{Frandsen}} \bibnamefont{and}
  \bibinfo{author}{\bibfnamefont{S.}~\bibnamefont{Sarkar}},
  \bibinfo{journal}{Phys.Rev.Lett.} \textbf{\bibinfo{volume}{105}},
  \bibinfo{pages}{011301} (\bibinfo{year}{2010}), \eprint{1003.4505}.

\bibitem[{\citenamefont{Cumberbatch et~al.}(2010)\citenamefont{Cumberbatch,
  Guzik, Silk, Watson, and West}}]{Cumberbatch:2010hh}
\bibinfo{author}{\bibfnamefont{D.~T.} \bibnamefont{Cumberbatch}},
  \bibinfo{author}{\bibfnamefont{J.}~\bibnamefont{Guzik}},
  \bibinfo{author}{\bibfnamefont{J.}~\bibnamefont{Silk}},
  \bibinfo{author}{\bibfnamefont{L.~S.} \bibnamefont{Watson}},
  \bibnamefont{and} \bibinfo{author}{\bibfnamefont{S.~M.} \bibnamefont{West}},
  \bibinfo{journal}{Phys.Rev.} \textbf{\bibinfo{volume}{D82}},
  \bibinfo{pages}{103503} (\bibinfo{year}{2010}), \eprint{1005.5102}.

\bibitem[{\citenamefont{Taoso et~al.}(2010)\citenamefont{Taoso, Iocco, Meynet,
  Bertone, and Eggenberger}}]{Taoso:2010tg}
\bibinfo{author}{\bibfnamefont{M.}~\bibnamefont{Taoso}},
  \bibinfo{author}{\bibfnamefont{F.}~\bibnamefont{Iocco}},
  \bibinfo{author}{\bibfnamefont{G.}~\bibnamefont{Meynet}},
  \bibinfo{author}{\bibfnamefont{G.}~\bibnamefont{Bertone}}, \bibnamefont{and}
  \bibinfo{author}{\bibfnamefont{P.}~\bibnamefont{Eggenberger}},
  \bibinfo{journal}{Phys.Rev.} \textbf{\bibinfo{volume}{D82}},
  \bibinfo{pages}{083509} (\bibinfo{year}{2010}), \eprint{1005.5711}.

\bibitem[{\citenamefont{Casanellas and Lopes}(2011)}]{Casanellas:2010he}
\bibinfo{author}{\bibfnamefont{J.}~\bibnamefont{Casanellas}} \bibnamefont{and}
  \bibinfo{author}{\bibfnamefont{I.}~\bibnamefont{Lopes}},
  \bibinfo{journal}{Mon.Not.Roy.Astron.Soc.} \textbf{\bibinfo{volume}{410}},
  \bibinfo{pages}{535} (\bibinfo{year}{2011}), \eprint{1008.0646}.

\bibitem[{\citenamefont{Iocco et~al.}(2012)\citenamefont{Iocco, Taoso,
  Leclercq, and Meynet}}]{Iocco:2012wk}
\bibinfo{author}{\bibfnamefont{F.}~\bibnamefont{Iocco}},
  \bibinfo{author}{\bibfnamefont{M.}~\bibnamefont{Taoso}},
  \bibinfo{author}{\bibfnamefont{F.}~\bibnamefont{Leclercq}}, \bibnamefont{and}
  \bibinfo{author}{\bibfnamefont{G.}~\bibnamefont{Meynet}},
  \bibinfo{journal}{Phys.Rev.Lett.} \textbf{\bibinfo{volume}{108}},
  \bibinfo{pages}{061301} (\bibinfo{year}{2012}), \eprint{1201.5387}.

\bibitem[{\citenamefont{Zentner and Hearin}(2011)}]{Zentner:2011wx}
\bibinfo{author}{\bibfnamefont{A.~R.} \bibnamefont{Zentner}} \bibnamefont{and}
  \bibinfo{author}{\bibfnamefont{A.~P.} \bibnamefont{Hearin}},
  \bibinfo{journal}{Phys.Rev.} \textbf{\bibinfo{volume}{D84}},
  \bibinfo{pages}{101302} (\bibinfo{year}{2011}), \eprint{1110.5919}.

\bibitem[{\citenamefont{Lopes and Silk}(2012)}]{Lopes:2012af}
\bibinfo{author}{\bibfnamefont{I.}~\bibnamefont{Lopes}} \bibnamefont{and}
  \bibinfo{author}{\bibfnamefont{J.}~\bibnamefont{Silk}},
  \bibinfo{journal}{Astrophys.J.} \textbf{\bibinfo{volume}{757}},
  \bibinfo{pages}{130} (\bibinfo{year}{2012}), \eprint{1209.3631}.

\bibitem[{\citenamefont{Casanellas and Lopes}(2013)}]{Casanellas:2012jp}
\bibinfo{author}{\bibfnamefont{J.}~\bibnamefont{Casanellas}} \bibnamefont{and}
  \bibinfo{author}{\bibfnamefont{I.}~\bibnamefont{Lopes}},
  \bibinfo{journal}{Astrophys.J.} \textbf{\bibinfo{volume}{L21}}
  (\bibinfo{year}{2013}), \eprint{1212.2985}.

\bibitem[{\citenamefont{Kouvaris and
  Tinyakov}(2011{\natexlab{a}})}]{Kouvaris:2010jy}
\bibinfo{author}{\bibfnamefont{C.}~\bibnamefont{Kouvaris}} \bibnamefont{and}
  \bibinfo{author}{\bibfnamefont{P.}~\bibnamefont{Tinyakov}},
  \bibinfo{journal}{Phys.Rev.} \textbf{\bibinfo{volume}{D83}},
  \bibinfo{pages}{083512} (\bibinfo{year}{2011}{\natexlab{a}}),
  \eprint{1012.2039}.

\bibitem[{\citenamefont{Kouvaris}(2012)}]{Kouvaris:2011gb}
\bibinfo{author}{\bibfnamefont{C.}~\bibnamefont{Kouvaris}},
  \bibinfo{journal}{Phys.Rev.Lett.} \textbf{\bibinfo{volume}{108}},
  \bibinfo{pages}{191301} (\bibinfo{year}{2012}), \eprint{1111.4364}.

\bibitem[{\citenamefont{Kouvaris and
  Tinyakov}(2011{\natexlab{b}})}]{Kouvaris:2011fi}
\bibinfo{author}{\bibfnamefont{C.}~\bibnamefont{Kouvaris}} \bibnamefont{and}
  \bibinfo{author}{\bibfnamefont{P.}~\bibnamefont{Tinyakov}},
  \bibinfo{journal}{Phys.Rev.Lett.} \textbf{\bibinfo{volume}{107}},
  \bibinfo{pages}{091301} (\bibinfo{year}{2011}{\natexlab{b}}),
  \eprint{1104.0382}.

\bibitem[{\citenamefont{McDermott et~al.}(2012)\citenamefont{McDermott, Yu, and
  Zurek}}]{McDermott:2011jp}
\bibinfo{author}{\bibfnamefont{S.~D.} \bibnamefont{McDermott}},
  \bibinfo{author}{\bibfnamefont{H.-B.} \bibnamefont{Yu}}, \bibnamefont{and}
  \bibinfo{author}{\bibfnamefont{K.~M.} \bibnamefont{Zurek}},
  \bibinfo{journal}{Phys.Rev.} \textbf{\bibinfo{volume}{D85}},
  \bibinfo{pages}{023519} (\bibinfo{year}{2012}), \eprint{1103.5472}.

\bibitem[{\citenamefont{Guver et~al.}(2012)\citenamefont{Guver, Erkoca, Reno,
  and Sarcevic}}]{Guver:2012ba}
\bibinfo{author}{\bibfnamefont{T.}~\bibnamefont{Guver}},
  \bibinfo{author}{\bibfnamefont{A.~E.} \bibnamefont{Erkoca}},
  \bibinfo{author}{\bibfnamefont{M.~H.} \bibnamefont{Reno}}, \bibnamefont{and}
  \bibinfo{author}{\bibfnamefont{I.}~\bibnamefont{Sarcevic}}
  (\bibinfo{year}{2012}), \eprint{1201.2400}.

\bibitem[{\citenamefont{Kouvaris and Tinyakov}(2012)}]{Kouvaris:2012dz}
\bibinfo{author}{\bibfnamefont{C.}~\bibnamefont{Kouvaris}} \bibnamefont{and}
  \bibinfo{author}{\bibfnamefont{P.}~\bibnamefont{Tinyakov}}
  (\bibinfo{year}{2012}), \eprint{1212.4075}.

\bibitem[{\citenamefont{Bramante et~al.}(2013)\citenamefont{Bramante,
  Fukushima, and Kumar}}]{Bramante:2013hn}
\bibinfo{author}{\bibfnamefont{J.}~\bibnamefont{Bramante}},
  \bibinfo{author}{\bibfnamefont{K.}~\bibnamefont{Fukushima}},
  \bibnamefont{and} \bibinfo{author}{\bibfnamefont{J.}~\bibnamefont{Kumar}},
  \bibinfo{journal}{Phys.Rev.} \textbf{\bibinfo{volume}{D87}},
  \bibinfo{pages}{055012} (\bibinfo{year}{2013}), \eprint{1301.0036}.

\bibitem[{\citenamefont{Bell et~al.}(2013)\citenamefont{Bell, Melatos, and
  Petraki}}]{Bell:2013xk}
\bibinfo{author}{\bibfnamefont{N.~F.} \bibnamefont{Bell}},
  \bibinfo{author}{\bibfnamefont{A.}~\bibnamefont{Melatos}}, \bibnamefont{and}
  \bibinfo{author}{\bibfnamefont{K.}~\bibnamefont{Petraki}},
  \bibinfo{journal}{Phys. Rev. D 87,} \textbf{\bibinfo{volume}{123507}}
  (\bibinfo{year}{2013}), \eprint{1301.6811}.

\bibitem[{\citenamefont{Sandin and Ciarcelluti}(2009)}]{Sandin:2008db}
\bibinfo{author}{\bibfnamefont{F.}~\bibnamefont{Sandin}} \bibnamefont{and}
  \bibinfo{author}{\bibfnamefont{P.}~\bibnamefont{Ciarcelluti}},
  \bibinfo{journal}{Astropart.Phys.} \textbf{\bibinfo{volume}{32}},
  \bibinfo{pages}{278} (\bibinfo{year}{2009}), \eprint{0809.2942}.

\bibitem[{\citenamefont{Press and Spergel}(1985)}]{Press:1985ug}
\bibinfo{author}{\bibfnamefont{W.~H.} \bibnamefont{Press}} \bibnamefont{and}
  \bibinfo{author}{\bibfnamefont{D.~N.} \bibnamefont{Spergel}},
  \bibinfo{journal}{Astrophys. J.} \textbf{\bibinfo{volume}{296}},
  \bibinfo{pages}{679} (\bibinfo{year}{1985}).

\bibitem[{\citenamefont{Gould}(1987{\natexlab{a}})}]{Gould:1987ir}
\bibinfo{author}{\bibfnamefont{A.}~\bibnamefont{Gould}},
  \bibinfo{journal}{Astrophys. J.} \textbf{\bibinfo{volume}{321}},
  \bibinfo{pages}{571} (\bibinfo{year}{1987}{\natexlab{a}}).

\bibitem[{\citenamefont{Zentner}(2009)}]{Zentner:2009is}
\bibinfo{author}{\bibfnamefont{A.~R.} \bibnamefont{Zentner}},
  \bibinfo{journal}{Phys.Rev.} \textbf{\bibinfo{volume}{D80}},
  \bibinfo{pages}{063501} (\bibinfo{year}{2009}), \eprint{0907.3448}.

\bibitem[{\citenamefont{Griest and Seckel}(1987)}]{Griest:1986yu}
\bibinfo{author}{\bibfnamefont{K.}~\bibnamefont{Griest}} \bibnamefont{and}
  \bibinfo{author}{\bibfnamefont{D.}~\bibnamefont{Seckel}},
  \bibinfo{journal}{Nucl.Phys.} \textbf{\bibinfo{volume}{B283}},
  \bibinfo{pages}{681} (\bibinfo{year}{1987}).

\bibitem[{\citenamefont{Gould}(1987{\natexlab{b}})}]{Gould:1987ju}
\bibinfo{author}{\bibfnamefont{A.}~\bibnamefont{Gould}},
  \bibinfo{journal}{Astrophys.J.} \textbf{\bibinfo{volume}{321}},
  \bibinfo{pages}{560} (\bibinfo{year}{1987}{\natexlab{b}}).

\bibitem[{\citenamefont{{Gould}}(1990)}]{Gould:1990}
\bibinfo{author}{\bibfnamefont{A.}~\bibnamefont{{Gould}}},
  \bibinfo{journal}{Astrophys.J.} \textbf{\bibinfo{volume}{356}},
  \bibinfo{pages}{302} (\bibinfo{year}{1990}).

\bibitem[{\citenamefont{Asplund et~al.}(2006)\citenamefont{Asplund, Grevesse,
  and Sauval}}]{Asplund:2004eu}
\bibinfo{author}{\bibfnamefont{M.}~\bibnamefont{Asplund}},
  \bibinfo{author}{\bibfnamefont{N.}~\bibnamefont{Grevesse}}, \bibnamefont{and}
  \bibinfo{author}{\bibfnamefont{J.}~\bibnamefont{Sauval}},
  \bibinfo{journal}{Nucl.Phys.} \textbf{\bibinfo{volume}{A777}},
  \bibinfo{pages}{1} (\bibinfo{year}{2006}), \eprint{astro-ph/0410214}.

\bibitem[{\citenamefont{Asplund et~al.}(2009)\citenamefont{Asplund, Grevesse,
  Sauval, and Scott}}]{Asplund:2009fu}
\bibinfo{author}{\bibfnamefont{M.}~\bibnamefont{Asplund}},
  \bibinfo{author}{\bibfnamefont{N.}~\bibnamefont{Grevesse}},
  \bibinfo{author}{\bibfnamefont{A.~J.} \bibnamefont{Sauval}},
  \bibnamefont{and} \bibinfo{author}{\bibfnamefont{P.}~\bibnamefont{Scott}},
  \bibinfo{journal}{Ann.Rev.Astron.Astrophys.} \textbf{\bibinfo{volume}{47}},
  \bibinfo{pages}{481} (\bibinfo{year}{2009}), \eprint{0909.0948}.

\bibitem[{\citenamefont{Pena-Garay and Serenelli}(2008)}]{PenaGaray:2008qe}
\bibinfo{author}{\bibfnamefont{C.}~\bibnamefont{Pena-Garay}} \bibnamefont{and}
  \bibinfo{author}{\bibfnamefont{A.}~\bibnamefont{Serenelli}}
  (\bibinfo{year}{2008}), \eprint{0811.2424}.

\bibitem[{\citenamefont{Goldman and Nussinov}(1989)}]{Goldman:1989nd}
\bibinfo{author}{\bibfnamefont{I.}~\bibnamefont{Goldman}} \bibnamefont{and}
  \bibinfo{author}{\bibfnamefont{S.}~\bibnamefont{Nussinov}},
  \bibinfo{journal}{Phys.Rev.} \textbf{\bibinfo{volume}{D40}},
  \bibinfo{pages}{3221} (\bibinfo{year}{1989}).

\bibitem[{\citenamefont{Ruffini and Bonazzola}(1969)}]{Ruffini:1969qy}
\bibinfo{author}{\bibfnamefont{R.}~\bibnamefont{Ruffini}} \bibnamefont{and}
  \bibinfo{author}{\bibfnamefont{S.}~\bibnamefont{Bonazzola}},
  \bibinfo{journal}{Phys.Rev.} \textbf{\bibinfo{volume}{187}},
  \bibinfo{pages}{1767} (\bibinfo{year}{1969}).

\bibitem[{\citenamefont{Colpi et~al.}(1986)\citenamefont{Colpi, Shapiro, and
  Wasserman}}]{Colpi:1986ye}
\bibinfo{author}{\bibfnamefont{M.}~\bibnamefont{Colpi}},
  \bibinfo{author}{\bibfnamefont{S.}~\bibnamefont{Shapiro}}, \bibnamefont{and}
  \bibinfo{author}{\bibfnamefont{I.}~\bibnamefont{Wasserman}},
  \bibinfo{journal}{Phys.Rev.Lett.} \textbf{\bibinfo{volume}{57}},
  \bibinfo{pages}{2485} (\bibinfo{year}{1986}).

\bibitem[{\citenamefont{Ho et~al.}(1999)\citenamefont{Ho, Kim, and
  Lee}}]{Ho:1999hs}
\bibinfo{author}{\bibfnamefont{J.}~\bibnamefont{Ho}},
  \bibinfo{author}{\bibfnamefont{S.-j.} \bibnamefont{Kim}}, \bibnamefont{and}
  \bibinfo{author}{\bibfnamefont{B.-H.} \bibnamefont{Lee}}
  (\bibinfo{year}{1999}), \eprint{gr-qc/9902040}.

\bibitem[{\citenamefont{{Shapiro} and {Teukolsky}}(1983)}]{ShapiroTeukolsky}
\bibinfo{author}{\bibfnamefont{S.~L.} \bibnamefont{{Shapiro}}}
  \bibnamefont{and} \bibinfo{author}{\bibfnamefont{S.~A.}
  \bibnamefont{{Teukolsky}}}, \emph{\bibinfo{title}{{Black holes, white dwarfs,
  and neutron stars: The physics of compact objects}}}
  (\bibinfo{publisher}{Wiley-VCH}, \bibinfo{year}{1983}).

\bibitem[{\citenamefont{de~Gouvea et~al.}(1997)\citenamefont{de~Gouvea, Moroi,
  and Murayama}}]{deGouvea:1997tn}
\bibinfo{author}{\bibfnamefont{A.}~\bibnamefont{de~Gouvea}},
  \bibinfo{author}{\bibfnamefont{T.}~\bibnamefont{Moroi}}, \bibnamefont{and}
  \bibinfo{author}{\bibfnamefont{H.}~\bibnamefont{Murayama}},
  \bibinfo{journal}{Phys.Rev.} \textbf{\bibinfo{volume}{D56}},
  \bibinfo{pages}{1281} (\bibinfo{year}{1997}), \eprint{hep-ph/9701244}.

\bibitem[{\citenamefont{Enqvist and McDonald}(1998)}]{Enqvist:1997si}
\bibinfo{author}{\bibfnamefont{K.}~\bibnamefont{Enqvist}} \bibnamefont{and}
  \bibinfo{author}{\bibfnamefont{J.}~\bibnamefont{McDonald}},
  \bibinfo{journal}{Phys.Lett.} \textbf{\bibinfo{volume}{B425}},
  \bibinfo{pages}{309} (\bibinfo{year}{1998}), \eprint{hep-ph/9711514}.

\bibitem[{\citenamefont{Goldman}(2011)}]{Goldman:2011aa}
\bibinfo{author}{\bibfnamefont{I.}~\bibnamefont{Goldman}},
  \bibinfo{journal}{Acta Phys.Polon.} \textbf{\bibinfo{volume}{B42}},
  \bibinfo{pages}{2203} (\bibinfo{year}{2011}), \eprint{1112.1505}.

\bibitem[{\citenamefont{Goldman et~al.}(2013)\citenamefont{Goldman, Mohapatra,
  Nussinov, Rosenbaum, and Teplitz}}]{Goldman:2013}
\bibinfo{author}{\bibfnamefont{I.}~\bibnamefont{Goldman}},
  \bibinfo{author}{\bibfnamefont{R.}~\bibnamefont{Mohapatra}},
  \bibinfo{author}{\bibfnamefont{S.}~\bibnamefont{Nussinov}},
  \bibinfo{author}{\bibfnamefont{D.}~\bibnamefont{Rosenbaum}},
  \bibnamefont{and} \bibinfo{author}{\bibfnamefont{V.}~\bibnamefont{Teplitz}}
  (\bibinfo{year}{2013}), \eprint{1305.6908}.

\bibitem[{\citenamefont{Li et~al.}(2012)\citenamefont{Li, Huang, and
  Xu}}]{Li:2012ii}
\bibinfo{author}{\bibfnamefont{A.}~\bibnamefont{Li}},
  \bibinfo{author}{\bibfnamefont{F.}~\bibnamefont{Huang}}, \bibnamefont{and}
  \bibinfo{author}{\bibfnamefont{R.-X.} \bibnamefont{Xu}}
  (\bibinfo{year}{2012}), \eprint{1208.3722}.

\bibitem[{\citenamefont{Leung et~al.}(2013)\citenamefont{Leung, Chu, Lin, and
  Wong}}]{Leung:2013pra}
\bibinfo{author}{\bibfnamefont{S.~C.} \bibnamefont{Leung}},
  \bibinfo{author}{\bibfnamefont{M.~C.} \bibnamefont{Chu}},
  \bibinfo{author}{\bibfnamefont{L.~M.} \bibnamefont{Lin}}, \bibnamefont{and}
  \bibinfo{author}{\bibfnamefont{K.~W.} \bibnamefont{Wong}}
  (\bibinfo{year}{2013}), \eprint{1305.6142}.

\bibitem[{\citenamefont{Demorest et~al.}(2010)\citenamefont{Demorest, Pennucci,
  Ransom, Roberts, and Hessels}}]{Demorest:2010bx}
\bibinfo{author}{\bibfnamefont{P.}~\bibnamefont{Demorest}},
  \bibinfo{author}{\bibfnamefont{T.}~\bibnamefont{Pennucci}},
  \bibinfo{author}{\bibfnamefont{S.}~\bibnamefont{Ransom}},
  \bibinfo{author}{\bibfnamefont{M.}~\bibnamefont{Roberts}}, \bibnamefont{and}
  \bibinfo{author}{\bibfnamefont{J.}~\bibnamefont{Hessels}},
  \bibinfo{journal}{Nature} \textbf{\bibinfo{volume}{467}},
  \bibinfo{pages}{1081} (\bibinfo{year}{2010}), \eprint{1010.5788}.

\bibitem[{\citenamefont{Petriello et~al.}(2008)\citenamefont{Petriello,
  Quackenbush, and Zurek}}]{Petriello:2008pu}
\bibinfo{author}{\bibfnamefont{F.~J.} \bibnamefont{Petriello}},
  \bibinfo{author}{\bibfnamefont{S.}~\bibnamefont{Quackenbush}},
  \bibnamefont{and} \bibinfo{author}{\bibfnamefont{K.~M.} \bibnamefont{Zurek}},
  \bibinfo{journal}{Phys.Rev.} \textbf{\bibinfo{volume}{D77}},
  \bibinfo{pages}{115020} (\bibinfo{year}{2008}), \eprint{0803.4005}.

\bibitem[{\citenamefont{Gershtein et~al.}(2008)\citenamefont{Gershtein,
  Petriello, Quackenbush, and Zurek}}]{Gershtein:2008bf}
\bibinfo{author}{\bibfnamefont{Y.}~\bibnamefont{Gershtein}},
  \bibinfo{author}{\bibfnamefont{F.}~\bibnamefont{Petriello}},
  \bibinfo{author}{\bibfnamefont{S.}~\bibnamefont{Quackenbush}},
  \bibnamefont{and} \bibinfo{author}{\bibfnamefont{K.~M.} \bibnamefont{Zurek}},
  \bibinfo{journal}{Phys.Rev.} \textbf{\bibinfo{volume}{D78}},
  \bibinfo{pages}{095002} (\bibinfo{year}{2008}), \eprint{0809.2849}.

\bibitem[{\citenamefont{Petriello and Quackenbush}(2008)}]{Petriello:2008zr}
\bibinfo{author}{\bibfnamefont{F.}~\bibnamefont{Petriello}} \bibnamefont{and}
  \bibinfo{author}{\bibfnamefont{S.}~\bibnamefont{Quackenbush}},
  \bibinfo{journal}{Phys.Rev.} \textbf{\bibinfo{volume}{D77}},
  \bibinfo{pages}{115004} (\bibinfo{year}{2008}), \eprint{0801.4389}.

\bibitem[{\citenamefont{Carlson and Glashow}(1987)}]{Carlson:1987si}
\bibinfo{author}{\bibfnamefont{E.~D.} \bibnamefont{Carlson}} \bibnamefont{and}
  \bibinfo{author}{\bibfnamefont{S.}~\bibnamefont{Glashow}},
  \bibinfo{journal}{Phys.Lett.} \textbf{\bibinfo{volume}{B193}},
  \bibinfo{pages}{168} (\bibinfo{year}{1987}).

\bibitem[{\citenamefont{Foot et~al.}(2001)\citenamefont{Foot, Ignatiev, and
  Volkas}}]{Foot:2000vy}
\bibinfo{author}{\bibfnamefont{R.}~\bibnamefont{Foot}},
  \bibinfo{author}{\bibfnamefont{A.~Y.} \bibnamefont{Ignatiev}},
  \bibnamefont{and} \bibinfo{author}{\bibfnamefont{R.}~\bibnamefont{Volkas}},
  \bibinfo{journal}{Phys.Lett.} \textbf{\bibinfo{volume}{B503}},
  \bibinfo{pages}{355} (\bibinfo{year}{2001}), \eprint{astro-ph/0011156}.

\bibitem[{\citenamefont{Ciarcelluti and Foot}(2009)}]{Ciarcelluti:2008qk}
\bibinfo{author}{\bibfnamefont{P.}~\bibnamefont{Ciarcelluti}} \bibnamefont{and}
  \bibinfo{author}{\bibfnamefont{R.}~\bibnamefont{Foot}},
  \bibinfo{journal}{Phys.Lett.} \textbf{\bibinfo{volume}{B679}},
  \bibinfo{pages}{278} (\bibinfo{year}{2009}), \eprint{0809.4438}.

\bibitem[{\citenamefont{Williams et~al.}(1971)\citenamefont{Williams, Faller,
  and Hill}}]{Williams:1971ms}
\bibinfo{author}{\bibfnamefont{E.}~\bibnamefont{Williams}},
  \bibinfo{author}{\bibfnamefont{J.}~\bibnamefont{Faller}}, \bibnamefont{and}
  \bibinfo{author}{\bibfnamefont{H.}~\bibnamefont{Hill}},
  \bibinfo{journal}{Phys.Rev.Lett.} \textbf{\bibinfo{volume}{26}},
  \bibinfo{pages}{721} (\bibinfo{year}{1971}).

\bibitem[{\citenamefont{Bartlett and Loegl}(1988)}]{Bartlett:1988yy}
\bibinfo{author}{\bibfnamefont{D.}~\bibnamefont{Bartlett}} \bibnamefont{and}
  \bibinfo{author}{\bibfnamefont{S.}~\bibnamefont{Loegl}},
  \bibinfo{journal}{Phys.Rev.Lett.} \textbf{\bibinfo{volume}{61}},
  \bibinfo{pages}{2285} (\bibinfo{year}{1988}).

\bibitem[{\citenamefont{Jaeckel et~al.}(2008)\citenamefont{Jaeckel, Redondo,
  and Ringwald}}]{Jaeckel:2008fi}
\bibinfo{author}{\bibfnamefont{J.}~\bibnamefont{Jaeckel}},
  \bibinfo{author}{\bibfnamefont{J.}~\bibnamefont{Redondo}}, \bibnamefont{and}
  \bibinfo{author}{\bibfnamefont{A.}~\bibnamefont{Ringwald}},
  \bibinfo{journal}{Phys.Rev.Lett.} \textbf{\bibinfo{volume}{101}},
  \bibinfo{pages}{131801} (\bibinfo{year}{2008}), \eprint{0804.4157}.

\bibitem[{\citenamefont{Mirizzi et~al.}(2009)\citenamefont{Mirizzi, Redondo,
  and Sigl}}]{Mirizzi:2009iz}
\bibinfo{author}{\bibfnamefont{A.}~\bibnamefont{Mirizzi}},
  \bibinfo{author}{\bibfnamefont{J.}~\bibnamefont{Redondo}}, \bibnamefont{and}
  \bibinfo{author}{\bibfnamefont{G.}~\bibnamefont{Sigl}},
  \bibinfo{journal}{JCAP} \textbf{\bibinfo{volume}{0903}}, \bibinfo{pages}{026}
  (\bibinfo{year}{2009}), \eprint{0901.0014}.

\bibitem[{\citenamefont{Ruoso et~al.}(1992)\citenamefont{Ruoso, Cameron,
  Cantatore, Melissinos, Semertzidis et~al.}}]{Ruoso:1992nx}
\bibinfo{author}{\bibfnamefont{G.}~\bibnamefont{Ruoso}},
  \bibinfo{author}{\bibfnamefont{R.}~\bibnamefont{Cameron}},
  \bibinfo{author}{\bibfnamefont{G.}~\bibnamefont{Cantatore}},
  \bibinfo{author}{\bibfnamefont{A.}~\bibnamefont{Melissinos}},
  \bibinfo{author}{\bibfnamefont{Y.}~\bibnamefont{Semertzidis}},
  \bibnamefont{et~al.}, \bibinfo{journal}{Zeit.Phys.}
  \textbf{\bibinfo{volume}{C56}}, \bibinfo{pages}{505} (\bibinfo{year}{1992}).

\bibitem[{\citenamefont{Cameron et~al.}(1993)\citenamefont{Cameron, Cantatore,
  Melissinos, Ruoso, Semertzidis et~al.}}]{Cameron:1993mr}
\bibinfo{author}{\bibfnamefont{R.}~\bibnamefont{Cameron}},
  \bibinfo{author}{\bibfnamefont{G.}~\bibnamefont{Cantatore}},
  \bibinfo{author}{\bibfnamefont{A.}~\bibnamefont{Melissinos}},
  \bibinfo{author}{\bibfnamefont{G.}~\bibnamefont{Ruoso}},
  \bibinfo{author}{\bibfnamefont{Y.}~\bibnamefont{Semertzidis}},
  \bibnamefont{et~al.}, \bibinfo{journal}{Phys.Rev.}
  \textbf{\bibinfo{volume}{D47}}, \bibinfo{pages}{3707} (\bibinfo{year}{1993}).

\bibitem[{\citenamefont{Robilliard et~al.}(2007)\citenamefont{Robilliard,
  Battesti, Fouche, Mauchain, Sautivet et~al.}}]{Robilliard:2007bq}
\bibinfo{author}{\bibfnamefont{C.}~\bibnamefont{Robilliard}},
  \bibinfo{author}{\bibfnamefont{R.}~\bibnamefont{Battesti}},
  \bibinfo{author}{\bibfnamefont{M.}~\bibnamefont{Fouche}},
  \bibinfo{author}{\bibfnamefont{J.}~\bibnamefont{Mauchain}},
  \bibinfo{author}{\bibfnamefont{A.-M.} \bibnamefont{Sautivet}},
  \bibnamefont{et~al.}, \bibinfo{journal}{Phys.Rev.Lett.}
  \textbf{\bibinfo{volume}{99}}, \bibinfo{pages}{190403}
  (\bibinfo{year}{2007}), \eprint{0707.1296}.

\bibitem[{\citenamefont{Ahlers et~al.}(2007)\citenamefont{Ahlers, Gies,
  Jaeckel, Redondo, and Ringwald}}]{Ahlers:2007rd}
\bibinfo{author}{\bibfnamefont{M.}~\bibnamefont{Ahlers}},
  \bibinfo{author}{\bibfnamefont{H.}~\bibnamefont{Gies}},
  \bibinfo{author}{\bibfnamefont{J.}~\bibnamefont{Jaeckel}},
  \bibinfo{author}{\bibfnamefont{J.}~\bibnamefont{Redondo}}, \bibnamefont{and}
  \bibinfo{author}{\bibfnamefont{A.}~\bibnamefont{Ringwald}},
  \bibinfo{journal}{Phys.Rev.} \textbf{\bibinfo{volume}{D76}},
  \bibinfo{pages}{115005} (\bibinfo{year}{2007}), \eprint{0706.2836}.

\bibitem[{\citenamefont{Chou et~al.}(2008)}]{Chou:2007zzc}
\bibinfo{author}{\bibfnamefont{A.~S.} \bibnamefont{Chou}} \bibnamefont{et~al.}
  (\bibinfo{collaboration}{GammeV (T-969) Collaboration}),
  \bibinfo{journal}{Phys.Rev.Lett.} \textbf{\bibinfo{volume}{100}},
  \bibinfo{pages}{080402} (\bibinfo{year}{2008}), \eprint{0710.3783}.

\bibitem[{\citenamefont{Ahlers et~al.}(2008)\citenamefont{Ahlers, Gies,
  Jaeckel, Redondo, and Ringwald}}]{Ahlers:2007qf}
\bibinfo{author}{\bibfnamefont{M.}~\bibnamefont{Ahlers}},
  \bibinfo{author}{\bibfnamefont{H.}~\bibnamefont{Gies}},
  \bibinfo{author}{\bibfnamefont{J.}~\bibnamefont{Jaeckel}},
  \bibinfo{author}{\bibfnamefont{J.}~\bibnamefont{Redondo}}, \bibnamefont{and}
  \bibinfo{author}{\bibfnamefont{A.}~\bibnamefont{Ringwald}},
  \bibinfo{journal}{Phys.Rev.} \textbf{\bibinfo{volume}{D77}},
  \bibinfo{pages}{095001} (\bibinfo{year}{2008}), \eprint{0711.4991}.

\bibitem[{\citenamefont{Afanasev et~al.}(2008)\citenamefont{Afanasev, Baker,
  Beard, Biallas, Boyce et~al.}}]{Afanasev:2008jt}
\bibinfo{author}{\bibfnamefont{A.}~\bibnamefont{Afanasev}},
  \bibinfo{author}{\bibfnamefont{O.}~\bibnamefont{Baker}},
  \bibinfo{author}{\bibfnamefont{K.}~\bibnamefont{Beard}},
  \bibinfo{author}{\bibfnamefont{G.}~\bibnamefont{Biallas}},
  \bibinfo{author}{\bibfnamefont{J.}~\bibnamefont{Boyce}},
  \bibnamefont{et~al.}, \bibinfo{journal}{Phys.Rev.Lett.}
  \textbf{\bibinfo{volume}{101}}, \bibinfo{pages}{120401}
  (\bibinfo{year}{2008}), \eprint{0806.2631}.

\bibitem[{\citenamefont{Fouche et~al.}(2008)\citenamefont{Fouche, Robilliard,
  Faure, Rizzo, Mauchain et~al.}}]{Fouche:2008jk}
\bibinfo{author}{\bibfnamefont{M.}~\bibnamefont{Fouche}},
  \bibinfo{author}{\bibfnamefont{C.}~\bibnamefont{Robilliard}},
  \bibinfo{author}{\bibfnamefont{S.}~\bibnamefont{Faure}},
  \bibinfo{author}{\bibfnamefont{C.}~\bibnamefont{Rizzo}},
  \bibinfo{author}{\bibfnamefont{J.}~\bibnamefont{Mauchain}},
  \bibnamefont{et~al.}, \bibinfo{journal}{Phys.Rev.}
  \textbf{\bibinfo{volume}{D78}}, \bibinfo{pages}{032013}
  (\bibinfo{year}{2008}), \eprint{0808.2800}.

\bibitem[{\citenamefont{Afanasev et~al.}(2009)\citenamefont{Afanasev, Baker,
  Beard, Biallas, Boyce et~al.}}]{Afanasev:2008fv}
\bibinfo{author}{\bibfnamefont{A.}~\bibnamefont{Afanasev}},
  \bibinfo{author}{\bibfnamefont{O.}~\bibnamefont{Baker}},
  \bibinfo{author}{\bibfnamefont{K.}~\bibnamefont{Beard}},
  \bibinfo{author}{\bibfnamefont{G.}~\bibnamefont{Biallas}},
  \bibinfo{author}{\bibfnamefont{J.}~\bibnamefont{Boyce}},
  \bibnamefont{et~al.}, \bibinfo{journal}{Phys.Lett.}
  \textbf{\bibinfo{volume}{B679}}, \bibinfo{pages}{317} (\bibinfo{year}{2009}),
  \eprint{0810.4189}.

\bibitem[{\citenamefont{Ehret et~al.}(2009)}]{Ehret:2009sq}
\bibinfo{author}{\bibfnamefont{K.}~\bibnamefont{Ehret}} \bibnamefont{et~al.}
  (\bibinfo{collaboration}{ALPS collaboration}),
  \bibinfo{journal}{Nucl.Instrum.Meth.} \textbf{\bibinfo{volume}{A612}},
  \bibinfo{pages}{83} (\bibinfo{year}{2009}), \eprint{0905.4159}.

\bibitem[{\citenamefont{Andriamonje et~al.}(2007)}]{Andriamonje:2007ew}
\bibinfo{author}{\bibfnamefont{S.}~\bibnamefont{Andriamonje}}
  \bibnamefont{et~al.} (\bibinfo{collaboration}{CAST Collaboration}),
  \bibinfo{journal}{JCAP} \textbf{\bibinfo{volume}{0704}}, \bibinfo{pages}{010}
  (\bibinfo{year}{2007}), \eprint{hep-ex/0702006}.

\bibitem[{\citenamefont{Redondo}(2008)}]{Redondo:2008aa}
\bibinfo{author}{\bibfnamefont{J.}~\bibnamefont{Redondo}},
  \bibinfo{journal}{JCAP} \textbf{\bibinfo{volume}{0807}}, \bibinfo{pages}{008}
  (\bibinfo{year}{2008}), \eprint{0801.1527}.

\bibitem[{\citenamefont{Andreas et~al.}(2012)\citenamefont{Andreas, Niebuhr,
  and Ringwald}}]{Andreas:2012mt}
\bibinfo{author}{\bibfnamefont{S.}~\bibnamefont{Andreas}},
  \bibinfo{author}{\bibfnamefont{C.}~\bibnamefont{Niebuhr}}, \bibnamefont{and}
  \bibinfo{author}{\bibfnamefont{A.}~\bibnamefont{Ringwald}},
  \bibinfo{journal}{Phys.Rev.} \textbf{\bibinfo{volume}{D86}},
  \bibinfo{pages}{095019} (\bibinfo{year}{2012}), \eprint{1209.6083}.

\bibitem[{\citenamefont{Jaeckel et~al.}(2012)\citenamefont{Jaeckel, Jankowiak,
  and Spannowsky}}]{Jaeckel:2012yz}
\bibinfo{author}{\bibfnamefont{J.}~\bibnamefont{Jaeckel}},
  \bibinfo{author}{\bibfnamefont{M.}~\bibnamefont{Jankowiak}},
  \bibnamefont{and}
  \bibinfo{author}{\bibfnamefont{M.}~\bibnamefont{Spannowsky}}
  (\bibinfo{year}{2012}), \eprint{1212.3620}.

\bibitem[{\citenamefont{Gninenko}(2013)}]{Gninenko:2013sr}
\bibinfo{author}{\bibfnamefont{S.}~\bibnamefont{Gninenko}},
  \bibinfo{journal}{Phys. Rev.} \textbf{\bibinfo{volume}{D87}},
  \bibinfo{pages}{035030} (\bibinfo{year}{2013}), \eprint{1301.7555}.

\bibitem[{\citenamefont{An et~al.}(2013)\citenamefont{An, Pospelov, and
  Pradler}}]{An:2013yfc}
\bibinfo{author}{\bibfnamefont{H.}~\bibnamefont{An}},
  \bibinfo{author}{\bibfnamefont{M.}~\bibnamefont{Pospelov}}, \bibnamefont{and}
  \bibinfo{author}{\bibfnamefont{J.}~\bibnamefont{Pradler}}
  (\bibinfo{year}{2013}), \eprint{1302.3884}.

\bibitem[{\citenamefont{Jaeckel and Ringwald}(2008)}]{Jaeckel:2007ch}
\bibinfo{author}{\bibfnamefont{J.}~\bibnamefont{Jaeckel}} \bibnamefont{and}
  \bibinfo{author}{\bibfnamefont{A.}~\bibnamefont{Ringwald}},
  \bibinfo{journal}{Phys.Lett.} \textbf{\bibinfo{volume}{B659}},
  \bibinfo{pages}{509} (\bibinfo{year}{2008}), \eprint{0707.2063}.

\bibitem[{\citenamefont{Gninenko and Redondo}(2008)}]{Gninenko:2008pz}
\bibinfo{author}{\bibfnamefont{S.~N.} \bibnamefont{Gninenko}} \bibnamefont{and}
  \bibinfo{author}{\bibfnamefont{J.}~\bibnamefont{Redondo}},
  \bibinfo{journal}{Phys.Lett.} \textbf{\bibinfo{volume}{B664}},
  \bibinfo{pages}{180} (\bibinfo{year}{2008}), \eprint{0804.3736}.

\bibitem[{\citenamefont{Jaeckel and Redondo}(2008)}]{Jaeckel:2008sz}
\bibinfo{author}{\bibfnamefont{J.}~\bibnamefont{Jaeckel}} \bibnamefont{and}
  \bibinfo{author}{\bibfnamefont{J.}~\bibnamefont{Redondo}},
  \bibinfo{journal}{Europhys.Lett.} \textbf{\bibinfo{volume}{84}},
  \bibinfo{pages}{31002} (\bibinfo{year}{2008}), \eprint{0806.1115}.

\bibitem[{\citenamefont{Caspers et~al.}(2009)\citenamefont{Caspers, Jaeckel,
  and Ringwald}}]{Caspers:2009cj}
\bibinfo{author}{\bibfnamefont{F.}~\bibnamefont{Caspers}},
  \bibinfo{author}{\bibfnamefont{J.}~\bibnamefont{Jaeckel}}, \bibnamefont{and}
  \bibinfo{author}{\bibfnamefont{A.}~\bibnamefont{Ringwald}},
  \bibinfo{journal}{JINST} \textbf{\bibinfo{volume}{4}},
  \bibinfo{pages}{P11013} (\bibinfo{year}{2009}), \eprint{0908.0759}.

\bibitem[{\citenamefont{Bjorken et~al.}(2009)\citenamefont{Bjorken, Essig,
  Schuster, and Toro}}]{Bjorken:2009mm}
\bibinfo{author}{\bibfnamefont{J.~D.} \bibnamefont{Bjorken}},
  \bibinfo{author}{\bibfnamefont{R.}~\bibnamefont{Essig}},
  \bibinfo{author}{\bibfnamefont{P.}~\bibnamefont{Schuster}}, \bibnamefont{and}
  \bibinfo{author}{\bibfnamefont{N.}~\bibnamefont{Toro}},
  \bibinfo{journal}{Phys.Rev.} \textbf{\bibinfo{volume}{D80}},
  \bibinfo{pages}{075018} (\bibinfo{year}{2009}), \eprint{0906.0580}.

\bibitem[{\citenamefont{Essig et~al.}(2011)\citenamefont{Essig, Schuster, Toro,
  and Wojtsekhowski}}]{Essig:2010xa}
\bibinfo{author}{\bibfnamefont{R.}~\bibnamefont{Essig}},
  \bibinfo{author}{\bibfnamefont{P.}~\bibnamefont{Schuster}},
  \bibinfo{author}{\bibfnamefont{N.}~\bibnamefont{Toro}}, \bibnamefont{and}
  \bibinfo{author}{\bibfnamefont{B.}~\bibnamefont{Wojtsekhowski}},
  \bibinfo{journal}{JHEP} \textbf{\bibinfo{volume}{1102}}, \bibinfo{pages}{009}
  (\bibinfo{year}{2011}), \eprint{1001.2557}.

\bibitem[{\citenamefont{Dent et~al.}(2012)\citenamefont{Dent, Ferrer, and
  Krauss}}]{Dent:2012mx}
\bibinfo{author}{\bibfnamefont{J.~B.} \bibnamefont{Dent}},
  \bibinfo{author}{\bibfnamefont{F.}~\bibnamefont{Ferrer}}, \bibnamefont{and}
  \bibinfo{author}{\bibfnamefont{L.~M.} \bibnamefont{Krauss}}
  (\bibinfo{year}{2012}), \eprint{1201.2683}.

\bibitem[{\citenamefont{Jaeckel}(2013)}]{Jaeckel:2013ija}
\bibinfo{author}{\bibfnamefont{J.}~\bibnamefont{Jaeckel}}
  (\bibinfo{year}{2013}), \eprint{1303.1821}.

\bibitem[{\citenamefont{Badertscher et~al.}(2004)\citenamefont{Badertscher,
  Belov, Crivelli, Felcini, Fetscher et~al.}}]{Badertscher:2003rk}
\bibinfo{author}{\bibfnamefont{A.}~\bibnamefont{Badertscher}},
  \bibinfo{author}{\bibfnamefont{A.}~\bibnamefont{Belov}},
  \bibinfo{author}{\bibfnamefont{P.}~\bibnamefont{Crivelli}},
  \bibinfo{author}{\bibfnamefont{M.}~\bibnamefont{Felcini}},
  \bibinfo{author}{\bibfnamefont{W.}~\bibnamefont{Fetscher}},
  \bibnamefont{et~al.}, \bibinfo{journal}{Int.J.Mod.Phys.}
  \textbf{\bibinfo{volume}{A19}}, \bibinfo{pages}{3833} (\bibinfo{year}{2004}),
  \eprint{hep-ex/0311031}.

\end{thebibliography}

\end{document}